\documentclass[reprint,superscriptaddress,amsmath,amssymb,aps,prl]{revtex4-2}
\usepackage{graphicx}
\usepackage{dcolumn}
\usepackage{bm}
\usepackage{color}
\usepackage{units}
\usepackage{wasysym}
\usepackage{MnSymbol}
\usepackage{comment}
\usepackage{times}
\usepackage{upgreek}
\usepackage[mathlines]{lineno}
\usepackage[unicode=true,pdfusetitle,bookmarks=false,colorlinks=true,citecolor=blue,urlcolor=blue,linkcolor=blue]{hyperref}
\begin{document}

\title{Theory of Capillary Tension and Interfacial Dynamics of Motility-Induced Phases}

\author{Luke Langford}
\affiliation{Department of Materials Science and Engineering, University of California, Berkeley, California 94720, USA}

\author{Ahmad K. Omar}
\email{aomar@berkeley.edu}
\affiliation{Department of Materials Science and Engineering, University of California, Berkeley, California 94720, USA}
\affiliation{Materials Sciences Division, Lawrence Berkeley National Laboratory, Berkeley, California 94720, USA}
\begin{abstract}
The statistical mechanics of equilibrium interfaces has been well-established for over a half century. 
In the last decade, a wealth of observations have made increasingly clear that a new perspective is required to describe interfaces arbitrarily far from equilibrium. 
In this work, beginning from microscopic particle dynamics that break time-reversal symmetry, we systematically derive the interfacial dynamics of coexisting motility-induced phases.
Doing so allows us to identify the athermal energy scale that excites interfacial fluctuations and the nonequilibrium surface tension that resists these excitations. 
Our theory identifies that, in contrast to equilibrium fluids, this active surface tension contains contributions arising from nonconservative forces which act to suppress interfacial fluctuations   and, crucially, is distinct from the mechanical surface tension of Kirkwood and Buff.  
 We find that the interfacial stiffness scales linearly with the intrinsic persistence length of the constituent active particle trajectories, in agreement with simulation data.
We demonstrate that at wavelengths much larger than the persistence length, the interface obeys surface area minimizing Boltzmann statistics with our derived nonequilibrium interfacial stiffness playing an identical role to that of equilibrium systems. 
\end{abstract}

\maketitle

\section*{Introduction} 
Nonequilibrium phase transitions are ubiquitous in the living world and are increasingly realized in synthetic systems. 
``Active matter" -- systems with constituents that \textit{locally} consume and convert energy into mechanical motion -- represents a broad class of these out-of-equilibrium systems. 
Whether found in colonies of motile bacteria~\cite{Liu2019Self-DrivenFormation} or suspensions of catalytic colloids~\cite{Palacci2013LivingSurfers}, these active phase transitions are inherently dissipative and thus elude a thermodynamic description. 
The interface separating coexisting active phases has displayed a number of surprising findings, including negative mechanical surface tensions (despite manifestly stable interfaces)~\cite{Bialke2015, Patch2018, Solon2018GeneralizedEnsembles, Hermann2019b, Omar2020, Zakine2020, Lauersdorf2021}, propagating waves~\cite{Adkins2022}, and odd surface flows~\cite{Soni2019TheFluid}. 
Moreover, many of these unique aspects of active interfaces are implicated in the observation of novel droplet coarsening dynamics~\cite{Tjhung2018} and even in determining macroscopic phase behavior~\cite{Aifantis1983TheRule, Solon2018GeneralizedMatter, Omar2023b}.
However, the statistical mechanical framework for understanding fluid interfaces, developed over a half century ago~\cite{Rowlinson1982MolecularCapillarity}, relies on equilibrium Boltzmann statistics. 
It is clear that a new perspective that can describe interfaces arbitrarily far from equilibrium is now required.

The interface associated with motility-induced phase separation (MIPS) -- the phenomenon in which purely repulsive particles phase separate at sufficient activity -- has been crucial in attempts to understand nonequilibrium interfaces. 
Efforts to understand active interfaces have focused on the surface tension, $\upgamma$, the central material property for describing \textit{equilibrium} fluid interfaces. 
Indeed, the equilibrium surface tension controls capillary fluctuations~\cite{Rowlinson1982MolecularCapillarity,Evans1979TheFluids}, the Laplace pressure and shape of finite-sized fluid droplets~\cite{Yang1976}, as well as the nucleation barrier of droplets~\cite{DEBENEDETTI2020MetastableLiquids}. 
Kirkwood and Buff~\cite{Kirkwood1949} provided the mechanical definition of surface tension using reversible work arguments -- a stress-based definition which may appear to be applicable to systems out of equilibrium.
However, when applying this definition to an active interface, Bialk\'e~\textit{et al.}~\cite{Bialke2015} measured a strongly \textit{negative} surface tension. 
Subsequent attempts to alternatively define the surface tension were found to reduce the magnitude of the measured tension with the sign remaining negative~\cite{Omar2020} or relied on equilibrium relations (such as the Gibbs-Duhem relation) that are strictly inapplicable to active systems~\cite{Hermann2019b}.

Despite the ambiguity in defining the surface tension, the MIPS interface displays scaling that is remarkably similar to equilibrium capillary-wave theory (CWT)~\cite{Bialke2015}, with the interfacial height fluctuations of phase-separated active Brownian particles (ABPs) scaling as $\langle |h(k)|^2\rangle \sim k^{-2}$ at low wavevector magnitudes $k$~\cite{Patch2018}. 
In equilibrium, these height fluctuations are proportional to the inverse interfacial stiffness, $\upgamma/k_B T$: thermal energy acts to excite interfacial fluctuations and surface tension acts as a restoring force~\cite{Rowlinson1982MolecularCapillarity}.
The observed consistency of active interfaces~\cite{Bialke2015,Patch2018, DelJunco2019} with CWT raises a number of intriguing questions: what is the theoretical foundation for this $k^{-2}$ scaling?; what plays the role of thermal energy in exciting an athermal active interface?; and, crucially, what is the definition and nature of the surface tension that attenuates these fluctuations? 

Recent progress has been made towards the theory of nonequilibrium interfacial fluctuations. 
Lee~\cite{Lee2017InterfaceSeparations} proposed an interfacial equation-of-motion using kinetic theory but neglected the effects of mass conservation.
Nardini and colleagues~\cite{Nardini2017EntropyMatter,Tjhung2018, Fausti2021CapillarySeparation}, building on the work of Bray~\cite{Bray2002InterfaceShear}, made this approach more concrete by linking the stochastic density field dynamics of the continuum active model B+ (AMB+) to those of the interfacial height field~\cite{Fausti2021CapillarySeparation}.
The stationary height fluctuation spectra that emerged from their equation-of-motion was in agreement with CWT.
This perspective for deriving interfacial height field dynamics \textit{begins} from the stochastic dynamics of a coarse-grained density field (models B~\cite{Hohenberg1977TheoryPhenomena} and AMB+ in Refs.~\cite{Bray2002InterfaceShear} and~\cite{Fausti2021CapillarySeparation}, respectively) and knowledge of the averaged density profile of the macroscopically phase separated state~\cite{Cahn1958FreeEnergy,Wittkowski2014ScalarSeparation,Tjhung2018}.
Importantly, while obtaining closed-form expressions for the fluctuating hydrodynamics of microscopic models of active matter remains an outstanding challenging, the theoretical description of the phase-separated state of active Brownian spheres, including the average density profile, was recently presented in Ref.~\cite{Omar2023b}.
We are now poised to begin developing a physical understanding of how \textit{microscopic} forces contribute to nonequilibrium interfacial dynamics.

In this work, we systematically derive a fluctuating hydrodynamic description of interacting active Brownian particles (ABPs), beginning from microscopic considerations. 
The fluctuating hydrodynamics are then used to solve for the dynamics of the interfacial height field, allowing us to identify the \textit{capillary-wave tension  ($\upgamma_{\rm cw}$)} of MIPS and the \textit{entirely athermal} energy scale ($k_BT^{\rm act}$) that excites the interface. 
Crucially, we show that this surface tension  \textit{is distinct} from the mechanical surface tension and includes effects beyond the interparticle forces that determine equilibrium interfacial tensions, including nematic surface flows that act to stabilize the interface against active fluctuations. 
These contributions are reminiscent of the physical picture proposed by Edwards and Wilkinson~\cite{Edwards1982TheAggregate}, in which a surface tension emerges from preferential accumulation of granular particles at concave surfaces over convex surfaces.
Our derived height field dynamics are found to recover area-minimizing Boltzmann statistics, i.e.,~$P[h(x)] \sim \text{exp}\left[-(\upgamma_{\rm cw}/k_BT^{\rm act})\int d\mathbf{x} |\boldsymbol{\nabla}h|^2\right]$ in the limit of long wavelengths despite the microscopic dynamics remaining far from equilibrium. 
It is precisely in this limit of area-minimizing dynamics that CWT is recovered.
It is our hope that the perspective offered in this work will provide a concrete path forward towards an understanding of interfaces arbitrarily far from equilibrium beginning from microscopic considerations.

\section*{Results}
\subsection*{Fluctuating hydrodynamics of interacting ABPs}

Our aim is to systematically derive the interfacial dynamics of active phase separations beginning from microscopic considerations.
We consider a system of $N$ interacting ABPs in which the time variation of the position $\mathbf{r}_i$ and orientation $\mathbf{q}_i$ of the $i$th particle follow overdamped Langevin equations:
\begin{linenomath*}
\begin{subequations}
  \label{eq:eom}
    \begin{align}
        \Dot{\mathbf{r}}_i &= U_o \mathbf{q}_i + \frac{1}{\zeta}\sum_{j\neq i}^N\mathbf{F}_{ij}\label{eq:rdot}, \\ \Dot{\mathbf{q}}_i &= \mathbf{q}_i\times \bm{\Omega}_i\label{eq:qdot},
    \end{align} 
\end{subequations}
\end{linenomath*}
where $\mathbf{F}_{ij}$ is the (pairwise) interparticle force, $U_o$ is the intrinsic active speed, $\zeta$ is the translational drag coefficient, and $\bm{\Omega}_i$ is a stochastic angular velocity with zero mean and variance $\langle \bm{\Omega}_i (t) \bm{\Omega}_j(t')\rangle = 2D_R\delta_{ij}\delta(t-t')\mathbf{I}$ (where $\mathbf{I}$ is the identity tensor). 
Here, $D_R$ is the rotational diffusivity, which may be athermal in origin, and is inversely related to the orientational relaxation time $\tau_R \equiv D_R^{-1}$.
One can then define an intrinsic run length, $\ell_o = U_o\tau_R$, the typical distance an ideal particle travels before reorienting.

We seek to describe the dynamics of an instantaneous interface that is defined by the coarse-grained density field,  $\rho(\mathbf{r},t) = \sum_{i=1}^{N}\Delta(\mathbf{r}-\mathbf{r}_i(t))$ where $\Delta$ is a kernel of finite spatial width such that $\rho$ is spatially continuous.
Beginning from Eqs.~\eqref{eq:rdot} and~\eqref{eq:qdot}, we derive a fluctuating hydrodynamic~\cite{Dean1996LangevinProcesses} description of the coarse-grained density of ABPs, as detailed in Section 1 of the Supplementary Information~\footnote{See Supplementary Information at [URL], which includes Refs.~\cite{Irving1950,Dean1996LangevinProcesses,Cugliandolo2015StochasticDipoles,Zwanzig2001NonequilibriumMechanics,Omar2023b,vanKampen2007StochasticChemistry,Omar2021,Noll1955,Lehoucq2010,Hardy1982FormulasWaves,Epstein2019,Bray2002InterfaceShear,Fausti2021CapillarySeparation,Takatori2014,Solon2015,Omar2020,Korteweg1904,Yang1976,Evans1979TheFluids,Gardiner2002HandbookMethods,Archer2004DynamicalDeterministic,Grun2006Thin-FilmNoise,Delfau2016PatternEquation,vanKampen1981ItoStratonovich,Bray1994TheoryKinetics,Kirkwood1949,Arfken2012MathematicalGuide,Aifantis1983TheRule,Solon2018GeneralizedMatter,Rowlinson1982MolecularCapillarity,Bialke2015,Weeks1971,Anderson2020,Willard2010,Patch2018}, for detailed derivations of the fluctuating hydrodynamic equations and interfacial dynamics, numerical details, and supplemental media.}.  
The evolution equation for the density satisfies the continuity equation $\partial \rho/\partial t = -\boldsymbol{\nabla} \cdot \mathbf{J}$ where the density flux, $\mathbf{J} =  \left( \boldsymbol{\nabla} \cdot \bm{\sigma}^C + \zeta U_o \mathbf{m} \right)/\zeta$, contains a diffusive contribution from the interparticle conservative stress, $\bm{\sigma}^C$, and a term proportional to the polar order density, $\mathbf{m}(\mathbf{r},t) = \sum_{i=1}^{N}\mathbf{q}_i(t)\Delta(\mathbf{r}-\mathbf{r}_i(t))$. 

An evolution equation for $\mathbf{m}(\mathbf{r}, t)$ and all other one-body orientational fields is required for a full spatial and temporal description of the density field. 
However, the inclusion of higher order orientational fields simply acts to increase the temporal and spatial resolution~\cite{Yan2015a}.
To describe MIPS, inclusion of the traceless nematic order density $\mathbf{Q}' = \mathbf{Q} - \rho/d\mathbf{I}$ [where $\mathbf{Q}(\mathbf{r},t) = \sum_{i=1}^{N}\mathbf{q}_i(t)\mathbf{q}_i(t)\Delta(\mathbf{r}-\mathbf{r}_i(t))$], is necessary~\cite{Omar2023b}.
We therefore close the hierarchy of equations by assuming $\mathbf{B}(\mathbf{r},t) = \sum_{i=1}^{N}\mathbf{q}_i(t)\mathbf{q}_i(t)\mathbf{q}_i(t)\Delta(\mathbf{r}-\mathbf{r}_i(t))$ is isotropic, as demonstrated in SI Section 1.2~\cite{Note1, Omar2023b}. 
We further note that the polar and nematic order relax on faster timescales than the density field, allowing us to neglect their time variation and assume quasi-stationary density relaxation.
Under these closures, the \emph{deterministic} component of the polarization force density takes the form $\frac{\zeta \ell_o}{d-1}\boldsymbol{\nabla}\cdot \left (U \mathbf{Q}\right) \ \equiv \boldsymbol{\nabla} \cdot \bm{\sigma}^{\rm act}$, where $U$ is the effective speed of the particles (reduced from $U_o$ due to interparticle interactions) and $\bm{\sigma}^{\rm act}$ is an active \textit{effective} stress~\cite{Omar2020, Fily2012, Takatori2014, Solon2015}.
The nematic field can be directly related to the density field, allowing for a closed set of equations solely in terms of the density with:
\begin{linenomath*}
\begin{subequations}
\label{eq:abphydro}
\begin{align}
    \frac{\partial\rho}{\partial t} &= -\boldsymbol{\nabla} \cdot \mathbf{J},\  \label{eq:continuity} \\ \mathbf{J} &= \frac{1}{\zeta}\boldsymbol{\nabla}\cdot \left(\bm{\sigma}^C+\bm{\sigma}^{\rm{act}}\right) + \bm{\eta}^{\rm act},\ \\  
    \bm{\sigma}^C &= \left[-p_C(\rho) + \kappa_1(\rho)\nabla^2\rho + \kappa_2(\rho)|\boldsymbol{\nabla}\rho|^2\right]\mathbf{I} \nonumber \\  &\quad + \kappa_3(\rho)\boldsymbol{\nabla}\rho\boldsymbol{\nabla}\rho + \kappa_4(\rho)\boldsymbol{\nabla}\boldsymbol{\nabla}\rho,\  \label{eq:cstress}\\ \bm{\sigma}^{\rm{act}} &= \left[-\frac{\ell_o U_o \zeta \overline{U}\rho}{d(d-1)} + {a(\rho)}\nabla^2\rho\right]\mathbf{I} + {b(\rho)}\boldsymbol{\nabla}\rho\boldsymbol{\nabla}\rho,\ \label{eq:astress} \\ a(\rho) &= \frac{3 \ell_o^2}{2d(d-1)(d+2)}\overline{U}^2\frac{\text{d} p_C}{\text{d}\rho}, \ \label{eq:beta} \\ b(\rho) &= \frac{3\ell_o^2\overline{U}}{2d(d-1)(d+2)}\frac{d}{d\rho}\left[\overline{U}\frac{\text{d}p_C}{\text{d}\rho}\right],\ \label{eq:lambda}
\end{align}
\end{subequations}
\end{linenomath*}
where $p_C(\rho)$ is the conservative interaction pressure~\cite{Omar2023b},  $\overline{U}(\rho) \equiv U(\rho)/U_o$ is the dimensionless effective speed~\cite{Omar2023b}, $d(>1)$ is the system dimensionality. 
Here $\{\kappa_i(\rho)\}$ are expansion coefficients of the conservative stress~\footnote{In equilibrium, $\kappa_1(\rho) = \rho\kappa(\rho)$, $\kappa_2(\rho) = (\kappa(\rho)+\rho\kappa'(\rho)/2)$, $\kappa_3(\rho) = -\kappa(\rho)$, $\kappa_4(\rho) = 0$, where $\kappa(\rho)$ is related to a moment of the direct correlation function and pairwise interaction force~\cite{Evans1979TheFluids}.} in the reversible limit as first derived by Korteweg~\cite{Korteweg1904, Yang1976, Evans1979TheFluids}.
%, the precise forms of which are well established  and depend on the details of the pairwise interaction force
We note that the derivation of Eq.~\eqref{eq:abphydro} made use of a quasi-stationary approximation (i.e.,~$|\mathbf{J}|$ is small) which is valid for dynamics near the flux-free states (e.g.,~stationary phase coexistence) under consideration in this work.

Equation~\eqref{eq:abphydro} introduces an entirely athermal stochastic contribution to the flux, $\bm{\eta}^{\rm act}$, with zero mean and a variance of:
\begin{linenomath*}
    
\begin{align}
     \langle \bm{\eta}^{\rm act}(\mathbf{r},t)\bm{\eta}^{\rm act}(\mathbf{r}',t') \rangle = &2\frac{k_BT^{\rm act}}{\zeta}\left(\rho\mathbf{I} - \frac{d}{d-1}\mathbf{Q}'\right)\nonumber \\ &\times\delta(t-t')\delta(\mathbf{r}-\mathbf{r}') \ \label{eq:etavar},
\end{align}
\end{linenomath*}
where $k_BT^{\rm{act}} \equiv \ell_o \zeta U_o /d(d-1)$ is the active energy scale.
We reemphasize that, within our closure, the traceless nematic field $\mathbf{Q}'$ can be expressed fully in terms of the density field [see SI Eq.~1.32].
To arrive at these noise statistics we \textit{approximate} the coarse-graining kernel as $\Delta(\mathbf{r}-\mathbf{r}_i(t)) \approx \delta(\mathbf{r}-\mathbf{r}_i(t))$~\cite{Archer2004DynamicalDeterministic,Barre2015Motility-InducedAlignment,Bouchet2016PerturbativeExample,Tjhung2018}, allowing us to make use of the properties of delta functions~\cite{Dean1996LangevinProcesses} and express Eq.~\eqref{eq:etavar} solely in terms of one-body fields, as shown in SI Section 1.2.
We emphasize that this approximation was only necessary for finding the noise variance.
These statistics are similar to those found by Cugliandolo~\textit{et al.}~\cite{Cugliandolo2015StochasticDipoles} in the context of passive dipolar particles~\footnote{We are not aware of the presence of Eq.~\eqref{eq:etavar} in the \emph{active} literature, likely due to the emphasis on 2d systems with orientations restricted to be in the plane~\cite{Farrell2012PatternMotility}.}. 
%In this case Eq.~\eqref{eq:qdot} can be further simplified.}. 
Unlike the fluctuating hydrodynamic description of passive systems, the noise derived here is anisotropic due to the presence of the nematic order.
In a system with no spontaneous nematic order (i.e.,~$\mathbf{Q'} = \mathbf{0}$), the noise reduces to a form identical to that of a passive system~\cite{Dean1996LangevinProcesses} with $k_BT^{\rm{act}}$ playing the same role as thermal energy.
However, even in this scenario where the fluctuation-dissipation theorem is satisfied, the dynamics described in Eq.~\ref{eq:abphydro} remain distinctly active as the nematic flows encapsulated in $\bm{\sigma}^{\rm{act}}$ cannot be derived from a variational principle.  
As expected, \textit{deterministic} terms in Eq.~\eqref{eq:abphydro} are identical to those of Ref.~\cite{Omar2023b}.
However, the route to deriving an expression for the density evolution which preserves the \textit{stochastic} fluctuations is distinct from the route used in Ref.~\cite{Omar2023b} which averaged over all noise and focused on stationary states.

\begin{figure}
	\centering
	\includegraphics[width=.48\textwidth]{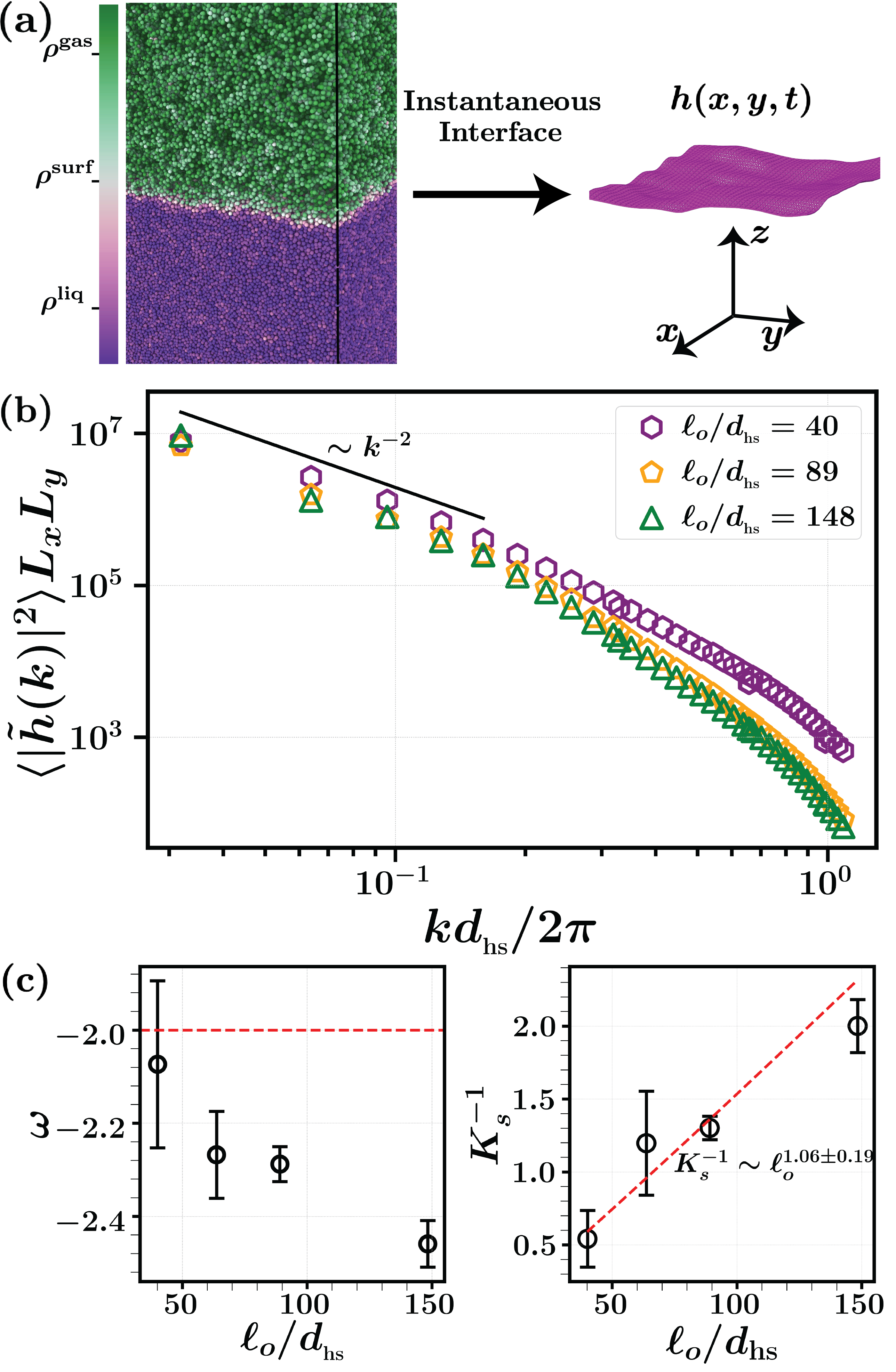}
	\caption{\protect\small{{(a) Visualization~\cite{Stukowski2010VisualizationTool} of instantaneous interface construction obtained through Brownian dynamics. See SI Section 5} for videos of interfacial dynamics. (b) Average interfacial fluctuation spectrum (units of $d_{\rm hs}^4$). Statistical uncertainty is provided in SI Section 4.7. (c) Values of $\omega$ (dimensionless) and $K_{s}^{-1}$ (units of $d_{\rm hs}^{-2}$) obtained from power law fits $\langle |\tilde{h}(k)|^2 \rangle = K_{s}k^{\omega}$.}}
	\label{fig:figure1}
\end{figure}

\subsection{System details}

The above fluctuating hydrodynamics may be applied to any system with particles obeying Eq.~\eqref{eq:eom} -- pairwise interactions of different forms will simply alter the constitutive relations present in Eq.~\eqref{eq:abphydro}.
Here, we focus on three dimensional (3d) active Brownian hard spheres as: (i) the order parameter distinguishing the two coexisting phases is simply the volume fraction; (ii) the phase diagram is well characterized~\cite{Omar2021} and entirely described by two dimensionless geometric parameters, the run length $\ell_o/d_{\rm hs}$ (where $d_{\rm hs}$ is the hard-sphere diameter) and volume fraction; and (iii) expressions for the required constitutive relations have been recently developed~\cite{Omar2023b}.
Furthermore, for active hard spheres, MIPS occurs at an appreciable level of activity [$\ell_o/d_{\rm hs} \ge \mathcal{O}(10)$] resulting in $\{\kappa_i(\rho)\} \ll a(\rho), b(\rho)$~\cite{Omar2023b} and allowing us to neglect the interaction interfacial stresses present in Eq.~\eqref{eq:cstress}. 
This approximation has recently been justified numerically through multiple approaches Refs.~\cite{Omar2020,Evans2023}.
We conduct particle-based Brownian dynamics simulations~\cite{Anderson2020} at activities and volume fractions within the MIPS binodal, with simulation details in SI Section 4.7.
The chosen simulation geometry results in a planar interface with its normal (on average) in the $z$-direction~\cite{Omar2021}. 
The remaining coordinates are tangent to the interface and are denoted by the vector $\mathbf{x}$, i.e.,~$\mathbf{r}\to (\mathbf{x},z)$. 
The height field $h(\mathbf{x})$ associated with the instantaneous interface~\cite{Willard2010} is defined as the location in space where $\rho$ is halfway between the binodal densities [see Fig.~\ref{fig:figure1}(a)], i.e.,~$\rho(\mathbf{x}, z=h) = \left(\rho^{\rm liq} + \rho^{\rm gas}\right)/2 = \rho^{\rm surf}$.
The discrete Fourier transform ($\mathbf{x}\to\mathbf{k}$) of $h(\mathbf{x})$ is denoted by $\tilde{h}(k)$.

As we later establish, (see SI Section 4.7), the computational cost for observing the relaxation of capillary-waves associated with $k$ scales as $\left(1/k\right)^{d+3}$, limiting the numerical accessibility of the mesoscopic limit where CWT might apply. 
We therefore conduct, to our knowledge, the largest scale simulation of 3d ABPs currently in the literature, with $631444$ particles for a duration of $89000~d_{\rm hs}/U_o$~\cite{Anderson2020}.
The calculated fluctuation spectrum is shown in  Fig.~\ref{fig:figure1}(b). 
Power law fits (i.e.,~$\langle |h(k)|^2\rangle = K_{s} k^{\omega}$) to the first decade of wavevectors reveal that the height fluctuations approach a $k^{-2}$ scaling in the low-$k$ limit [see Fig.~\ref{fig:figure1}(c)], consistent with the 2d data of Patch~\textit{et al.}~\cite{Patch2018} and corroborating the apparent applicability of CWT to these driven interfaces. 
 These fits also demonstrate the interfacial stiffness, $K_{s}^{-1}$ [see Fig.~\ref{fig:figure1}(c)], scales linearly with run length, again consistent with 2d simulations~\cite{Patch2018}. 

\subsection{Fluctuations of MIPS interfaces} 

In order to connect the dynamics of the density field to those of an interfacial height field we introduce the ansatz proposed by Bray~\textit{et al.}~\cite{Bray2002InterfaceShear}, which states that the instantaneous stochastic density field is, within a displacement of the interfacial height field $h$ in the $z$-dimension, equal to the noise-averaged stationary density field, $\upvarphi$:
\begin{equation}
    \rho(\mathbf{r},t) = \upvarphi\left[z - h(\mathbf{x},t)\right].
    \label{eq:brayapprox}
\end{equation}
This ansatz implies that $|\partial h/\partial \mathbf{x}|^2 << 1$~\cite{Bray2002InterfaceShear}; it is equivalent to stating the normal vector to  contours of $\rho$ negligibly deviate from the $z$-direction~\footnote{Note that the stochastic density evolution equation has been derived using It$\hat{\text{o}}$'s discretization scheme. However, we proceed as if it could be interpreted in the Stratonovich sense with no modification. In applying Bray's ansatz, we switch to a Stratonovich scheme and neglect the spurious drift term, which has been shown to vanish when considering conserved order parameters with a stochastic flux~\cite{Gardiner2002HandbookMethods,Grun2006Thin-FilmNoise,Delfau2016PatternEquation}.}.

As recently demonstrated by Fausti~\textit{et al.}~\cite{Fausti2021CapillarySeparation}, substitution of the ansatz [Eq.~\eqref{eq:brayapprox}] into the continuity equation [Eq.~\eqref{eq:abphydro}] results in nonlinear terms which coincide with those found in the quasi-1D flux balance used to derive the generalized Maxwell construction pseudovariable {needed to calculate the binodals}~\cite{Aifantis1983TheRule, Solon2018GeneralizedMatter, Omar2023b}.
The use of the pseudovariable allows us to incorporate the effects of the nematic flows (nonlinear terms in the fluctuating hydrodynamics) to the interfacial tension (linear term in the height field dynamics), which we demonstrate in SI Section 4.6.
This pseduovariable, $\mathcal{E}$, satisfies the differential equation $a\partial^2 \mathcal{E}/\partial \rho^2 = (2b - \partial a/\partial \rho) \partial \mathcal{E}/\partial \rho $ and is found to be $\mathcal{E} \sim p_C$ for the present system~\cite{Omar2023b}. 
 
The ansatz [Eq.~\eqref{eq:brayapprox}] coupled with our fluctuating hydrodynamic theory allow for the determination of a \textit{linear} equation-of-motion for $h(k)$, the continuous Fourier transform of $h(\mathbf{x})$. 
The resulting Langevin equation (see SI  Section 2 for a detailed derivation) for $h(k)$ is:
\begin{linenomath*}
\begin{subequations}
\label{eq:hdynamics}
\begin{align}
    & \zeta_{\rm{eff}} \frac{\partial h}{\partial t}  = -k^3\upgamma_{\rm{cw}} h +  \chi^{\rm{iso}} + \chi^{\rm{aniso}}\ \label{eq:interfacelangevin}, \\ &\zeta_{\rm{eff}}(k) = \frac{\zeta A^2(k)}{2B(k)\rho^{\rm surf}} \label{eq:effdrag}, \\ \upgamma_{\rm{cw}}(k) &= \frac{A(k)}{ \rho^{\rm surf}B(k)}\biggl[\int_{-\infty}^{\infty}\text{d}u \frac{\partial \mathcal{E}}{\partial u} \frac{\partial\upvarphi}{\partial u} a(\upvarphi) \nonumber \\ + \int_{-\infty}^{\infty}\text{d}u \frac{\partial \mathcal{E}}{\partial u}&\int_{-\infty}^{\infty} \text{d}z'\text{sgn}(u-z')e^{-k|u-z'|}b(\upvarphi)\left(\frac{\partial\upvarphi}{\partial z'}\right)^2 \biggr]\ \label{eq:capwavetension},
\end{align}
\end{subequations}
\end{linenomath*}
where the integration coordinate is defined as $u = z -h$. 
$\upgamma_{\rm cw}$ is the capillary-wave tension and has physical dimensions of energy per unit area (length) in 3d (2d). 
$\zeta_{\rm eff}$ is an effective drag coefficient, and $\chi^{\rm{iso}}(\mathbf{k},t)$ and $\chi^{\rm{aniso}}(\mathbf{k},t)$ are independent stochastic forces with zero mean that, respectively, originate from the isotropic and anisotropic contributions to $\bm{\eta}^{\rm act}$.
The variance of $\chi^{\rm{iso}}$ is:
\begin{linenomath*}
\begin{align}
    \langle \chi^{\rm{iso}}(\mathbf{k},t)\chi^{\rm{iso}}(\mathbf{k}',t')\rangle = &2k(k_BT^{\rm{act}})\zeta_{\rm{eff}}\nonumber \\ &\times\delta(t-t')\delta(\mathbf{k}+\mathbf{k}')(2\pi)^{(d-1)}\label{eq:chivariso},
\end{align}
\end{linenomath*}
while the variance of $\chi^{\rm aniso}$ and integral expressions for $A(k),B(k)$ (which solely depend on $\upvarphi$ and $\mathcal{E}$) are provided in SI Section 2.1.

The derived capillary-wave tension $\upgamma_{\rm{cw}}(k)$ is distinct from the mechanical surface tension, as shown in SI Section 4.
For active hard spheres, the active force and diameter must set the scale of the surface tension with $\upgamma_{\rm cw} \sim \zeta U_o/d_{\rm hs}$ while the run length dependence is nontrivial.
Defining a reduced run length, $\lambda \equiv \ell_o/\ell_o^{\rm crit} - 1$ (where $\lambda > 0$ results in the phase-separated states of interest), the surface tension has the form $\upgamma_{\rm cw}d_{\rm hs}/\zeta U_o = g(\lambda,k)$ where $g(\lambda,k)$ is a dimensionless function that must approach zero as the critical point is approached (i.e.,~$\lambda \rightarrow 0$). 
In this limit of vanishing capillary tension, the characteristic timescale that naturally emerges from Eq.~\eqref{eq:interfacelangevin}, $\tau(k) \equiv \zeta_{\rm eff}/\upgamma_{\rm cw}k^3 \rightarrow \infty$. 
These increasingly long timescales with decreasing $\lambda$ are consistent with the increased statistical uncertainty numerically observed at low activities in Fig.~\ref{fig:figure1}. 
We note that the reduction of $\upgamma_{\rm cw}$ with decreasing $\lambda$ implies a vanishing stabilizing force (with $\upgamma_{\rm cw} > 0$) in Eq.~\eqref{eq:interfacelangevin} as the critical point is approached -- in this regime, the assumptions underlying the ansatz [Eq.~\eqref{eq:brayapprox}] no longer hold.

\begin{figure}
	\centering
	\includegraphics[width=.48\textwidth]{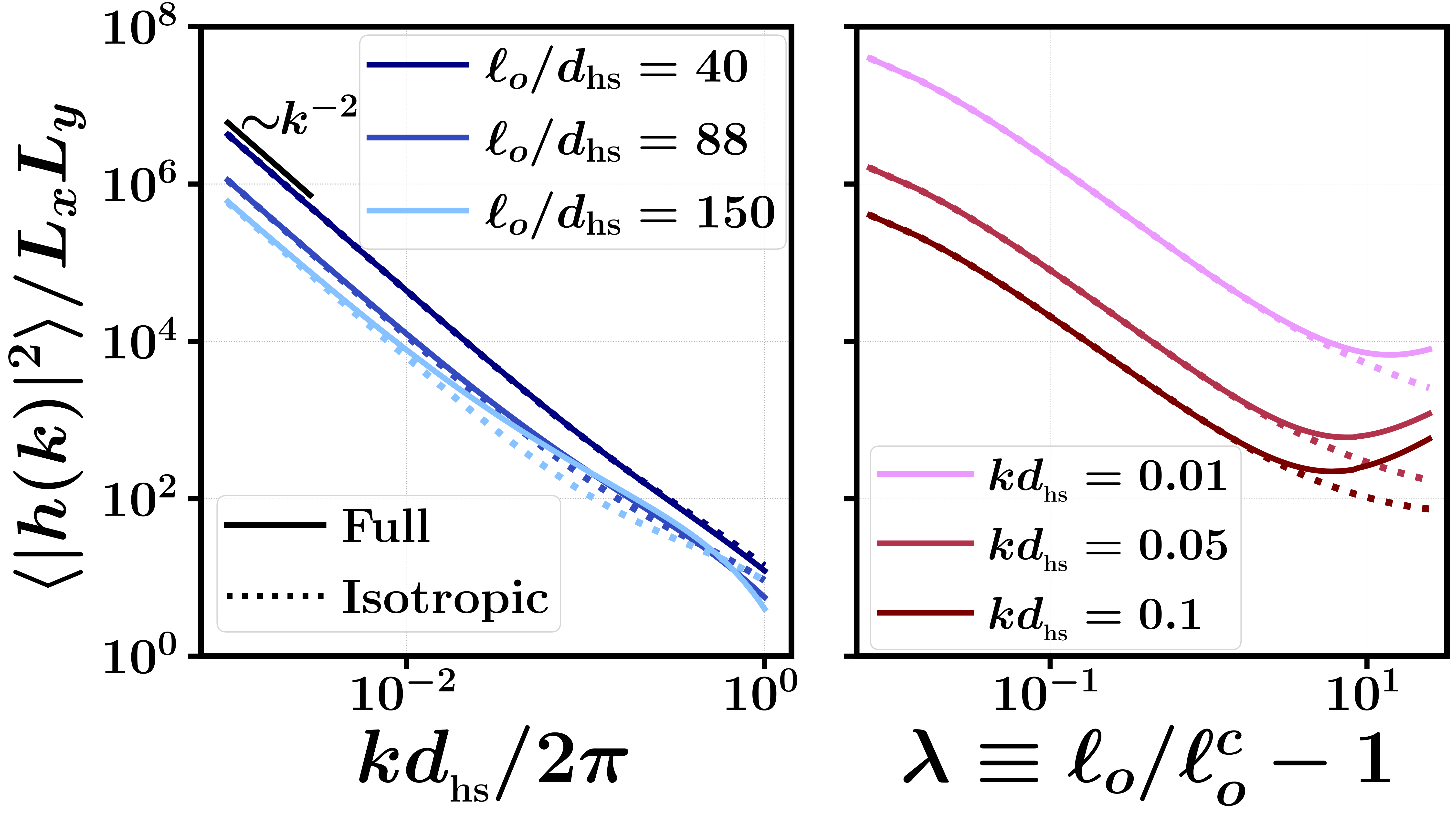}
	\caption{\protect\small{{Stationary fluctuations of the interfacial height field (units of $d_{\rm hs}^4$) as given by Eq.~\eqref{eq:fluctfull} (solid lines) and Eq.~\eqref{eq:fluctefftemp} (dotted lines)}.
    The critical activity is $\ell_o^c/d_{\rm hs} \approx 16.2$.  The anisotropic terms of Eq.~\eqref{eq:fluctfull} result in a non-monotonic dependence of the stationary fluctuations with activity.}}
	\label{fig:figure3}
\end{figure}

To determine the natural energy scale of $\upgamma_{\rm cw}$ we use Eq.~\eqref{eq:interfacelangevin} to determine the stationary fluctuations of the height field with:
\begin{linenomath*}
\begin{align}
        \langle |h(k)|^2 \rangle=\frac{L^{(d-1)}}{\upgamma_{\rm{cw}}} \biggl[ \frac{k_BT^{\rm{act}}}{k^2} + \frac{C(k) + D(k)}{2k^3 (d-1)\rho^{\rm surf}B(k)}\biggr],
        \label{eq:fluctfull}
\end{align}
\end{linenomath*}
where $L^{(d-1)}$ is the projected area (length) of the interface for 3d (2d) systems.
The first term in Eq.~\eqref{eq:fluctfull} arises entirely due to $\chi^{\rm{iso}}$ and is consistent with CWT while the second term emerges from $\chi^{\rm{aniso}}$ ($C(k)$ and $D(k)$ are provided in SI Section 2.2).
In the low-$k$ limit, the scaling of the second term of Eq.~\eqref{eq:fluctfull} with $k$ can be identified as $1/k$ by inspecting SI Eqs.~(4.14-4.17).
Using the equations-of-state of 3d ABPs~\cite{Omar2023b} and solving for $\upvarphi$ (see SI Section 4.2), we calculate the fluctuations with and without anisotropic contributions, as shown in Fig.~\ref{fig:figure3}. 
While the anisotropic contributions can result in a non-monotonic dependence of the height fluctuations with activity at finite wavelengths, these contributions are negligible at wavelengths much larger than the run length, i.e.,~$\ell_o k << 1$, and at low activities. 
As a result, CWT is precisely recovered at low $k$ with: 
\begin{linenomath*}
\begin{align}
    \langle |h(k)|^2 \rangle = \frac{L^{(d-1)}k_BT^{\rm{act}}}{\upgamma_{\rm{cw}}k^2} + \mathcal{O}\left(\frac{1}{k}\right).
    \label{eq:fluctefftemp}
\end{align}
\end{linenomath*}

While in this large wavelength limit, the variance of the height fluctuations are consistent with CWT, it can be shown that \textit{all statistical moments} of $h(k)$ will have the same form of those of an equilibrium interface.
We demonstrate this by solving the steady-state Fokker-Planck equation associated with Eq.~\eqref{eq:interfacelangevin} in SI Section 2.5, finding a Boltzmann distribution of interfacial shapes ${P\left[h(\mathbf{x})\right] \propto \text{exp}\left[-(\upgamma_{\rm cw} / k_BT^{\rm act})\int d\mathbf{x} |\boldsymbol{\nabla} h|^2\right]}$.
This distribution implies that at long wavelengths our interfacial dynamics can be exactly mapped onto that of an equilibrium system~\cite{Bray2002InterfaceShear} with an interfacial energy given by the product of $\upgamma_{\rm cw}$ with the surface area and a temperature of $k_BT^{\rm act}$.
Thus, when the anisotropic fluctuations can be ignored, \textit{an area-minimizing principle is recovered for the MIPS interface}, despite the nonequilibrium origins of the interfacial tension and athermal excitations.

\begin{figure}
	\centering
	\includegraphics[width=.48\textwidth]{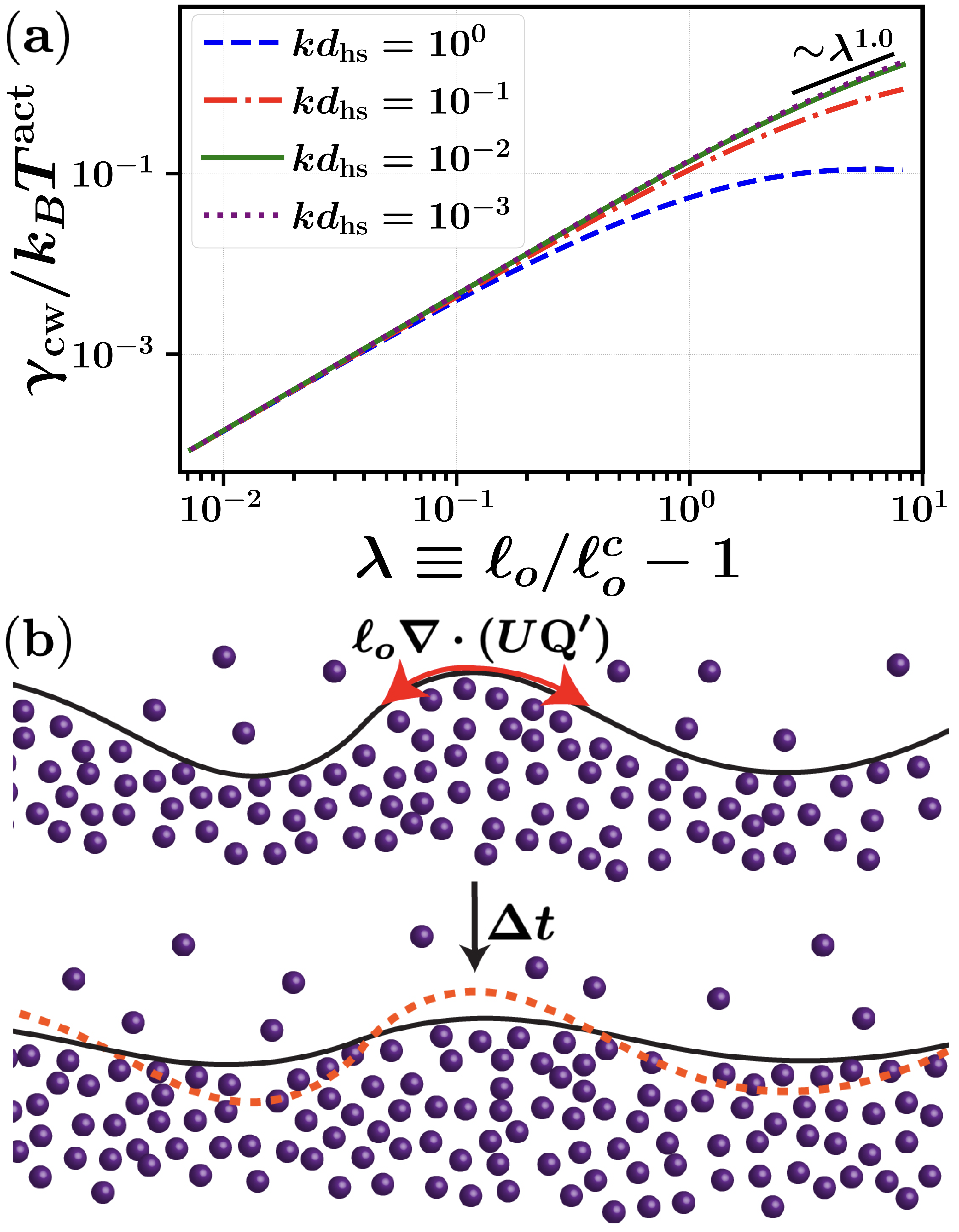}
	\caption{\protect\small{{(a) Interfacial stiffness (units of $d_{\rm hs}^{-2}$) as a function of the reduced activity, $\lambda$, and $k$, calculated using Eq.~\eqref{eq:capwavetension}. (b) Schematic demonstrating the impact of tangential nematic flows that are encapsulated in $\upgamma_{\rm{cw}}$. Following time evolution $\Delta t\sim\tau(k)\equiv\zeta_{\rm{eff}}/k^3\upgamma_{\rm cw}$, nematic flows lead to particle escape from convex hills and accumulation at convex troughs.} }}
	\label{fig:figure2}
\end{figure}

From Eq.~\eqref{eq:fluctefftemp} it is clear that interfacial fluctuations are excited by the athermal active energy scale and suppressed by the capillary-wave tension: the active interfacial stiffness is $\upgamma_{\rm cw}/k_BT^{\rm act}$.
We numerically evaluate this stiffness, shown in Fig.~\ref{fig:figure2}. 
The capillary-wave tension is positive for all activities and increases towards a constant in the macroscopic (i.e.,~$k\rightarrow0$) limit. 
In the limit of large activity, we find that the interfacial stiffness scales linearly with run length, consistent with the observed stiffness scaling reported by Patch~\textit{et al}.~\cite{Patch2018}.
Our theory now allows us to determine the origin of this scaling by identifying the precise form of $\upgamma_{\rm cw}$ [Eq.~\eqref{eq:capwavetension}] and the interfacial stiffness. 
While the active energy scale has a trivial scaling with reduced activity, $k_BT^{\rm act}\sim\lambda$, the scaling of the capillary-wave tension with activity requires examining the $\lambda$ dependence of the density profile ($\upvarphi$), interfacial coefficients ($a$ and $b$), and the pseudovariable ($\mathcal{E}$).
For the present system, $\mathcal{E}$ is independent of activity while the density profile is also largely independent of activity at large $\lambda$.
The latter can be appreciated by considering the relatively constant binodal densities of the MIPS phase diagram at large $\lambda$~\cite{Omar2021, Omar2023b}.
The asymptotic activity dependence of $\upgamma_{\rm cw}$ thus closely follows that of the active interfacial stress coefficients, $a$ and $b$, which generate nematic flows in response to density gradients and scale as $\lambda^2$.
In SI Section 4.6, we provide numerical evidence that use of the pseudovariable $\mathcal{E}$ further encapsulates the effects of nematic flows into the capillary-wave tension.
The physical origin for the linear dependence of the stiffness on activity is therefore a result of stabilizing nematic flows competing with the active energy scale. 
Fig.~\ref{fig:figure2}(b) provides a schematic of nematic flows which result in particles escaping from hills and accumulating in troughs.

\begin{figure}
    \centering
    \includegraphics[width = 0.48\textwidth]{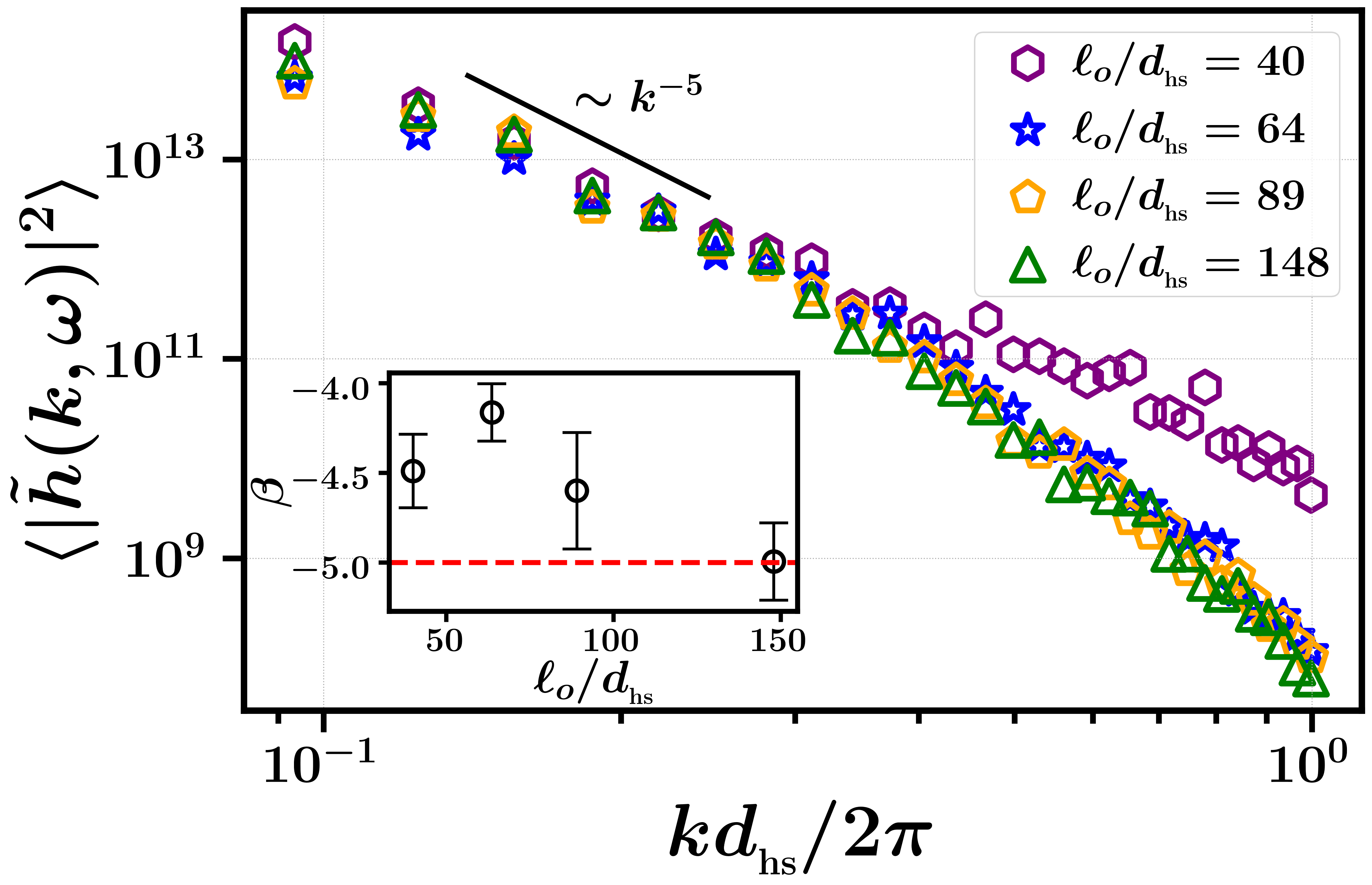}
    \caption{\protect\small{ Power spectra $\langle |h(k,\omega)|^2 \rangle$ at each simulated activity as a function of $k$ with $\omega d_{\rm hs}/ 2\pi U_o = 2.82\times 10^{-4}$ fixed. (Inset) Power law scaling obtained from logarithmic fits to $\langle |h(k,\omega)|^2 \rangle \sim k^{\beta}$ for low values of $k$ (i.e.,~$kd_{\rm hs}/2\pi < 0.5$).}}
    \label{fig:figure4}
\end{figure}

While we have focused on the stationary properties of the interface, our theory also makes predictions for dynamic properties.
Starting from Eq.~\eqref{eq:interfacelangevin}, we solve for the interfacial power spectra $\langle |h(k,\omega)|^2 \rangle$ (as detailed in SI Section 2.3):
\begin{equation}
    \langle |h(k,\omega)|^2 \rangle = \frac{2kL^{d-1}\Gamma k_BT^{\rm act}}{\zeta_{\rm eff}\left(\omega^2 + \tau(k)^{-2}\right)},
    \label{eq:powerspec}
\end{equation}
where $\Gamma$ is the total simulation time.
We note that Eq.~\eqref{eq:powerspec} assumes values of $k$ such that anisotropic fluctuations are negligible (i.e.,~$\ell_o k << 1$).
At these low values of $k$ and in the low frequency limit (i.e.,~$\omega \to 0$), we predict the power spectra to scale as $k^{-5}$.
Figure~\ref{fig:figure4} displays the power spectra measured from Brownian dynamics simulation.
We note that despite the extensive size and duration of our computer simulations, our analysis is limited to less than a single decade of $k$ values, and thus a robust power-law fit cannot be established.
Nevertheless, it is encouraging that the fit obtained approaches the theoretically predicted scaling of $k^{-5}$ with increasing activity, agreeing well with this prediction at the highest activity.
A potential source of error for the low run length dynamics are nonlinear corrections~\cite{Besse2023InterfaceSystems,Toner2023RougheningSystems} to Eq.~\eqref{eq:interfacelangevin}, neglected in the present study through use of Eq.~\eqref{eq:brayapprox}.
These nonlinear corrections are anticipated to be more important closer to the critical point due to the larger magnitude of interfacial fluctuations [see Fig.~\ref{fig:figure1}(c) or SI Section 5].
At higher activities, error may also come from the significance of anisotropic noise as the identified parameter $k\ell_o$ grows. 
These competing dependencies of our theory's error may offer an explanation for the non-monotonic behavior of the inset of Fig.~\ref{fig:figure4}.
We hope that this finding will encourage additional numerical studies of the dynamics of active interfaces.

\section*{Discussion}
Starting from microscopic particle dynamics, we systematically derive a mesoscopic Langevin equation for the interface of MIPS.
From this equation-of-motion, we identify the capillary-wave tension governing the dynamics of the interface and identify the athermal source of interfacial excitations. 
Interestingly, the derived energy scale for interfacial excitations is consistent with predictions made utilizing stochastic thermodynamics~\cite{Bialke2015, Speck2016}.  
Our capillary-wave tension encapsulates the effects of active nematic flows on the height field dynamics and is rooted in mass transport generated by nonconservative forces.
The origins of this tension are similar to the physical picture proposed by Edwards and Wilkinson~\cite{Edwards1982TheAggregate}, in which surface tension emerged from deposited particles preferentially settling in local minima, and perhaps points to a general phenomena of dynamically-stabilized interfaces.
Our interfacial Langevin equation recovers a surface-area minimizing Boltzmann distribution at low activities and low wavevectors, despite the microscopic particle dynamics breaking time-reversal symmetry. 
In this limit, an active analog to CWT is recovered with an interfacial stiffness that scales linearly with run length.

The fluctuating hydrodynamics derived in this Article may be applied to understand other nonequilibrium phenomena requiring a stochastic description, including active nucleation~\cite{Richard2016NucleationParticles,Redner2016,Du2020Rod-assistedSuspensions,Cates2023ClassicalSeparation}.
Moreover, it is our hope that by deriving interfacial dynamics beginning from microscopic equations of motion, the procedure outline here could provide a way forward for examining the interfacial dynamics of chiral active matter, which can exhibit odd surface flows~\cite{Soni2019TheFluid}.
This work might also be extended to include multiple order parameters, shedding light on phenomena such as active crystallization~\cite{Omar2021} or multicomponent driven systems~\cite{DelJunco2018,DelJunco2019,Saha2020,You2020,Chiu2023}, such as biomolecular condensates~\cite{Brangwynne2009GermlineDissolution/Condensation,Banani2017BiomolecularBiochemistry,Shin2017LiquidDisease}.
Finally, while our derived interfacial dynamics are linear, the inclusion of nonlinear contributions could allow for the determination of the universality class of the MIPS interface~\cite{Besse2023InterfaceSystems,Caballero2018StrongClass,Toner2023RougheningSystems}.
It is our hope that the theoretical explanations for puzzling observations in active matter provided here will assist in setting a foundation for further investigation of nonequilibrium interfacial phenomena.

\section*{Data Availability}
All the relevant data of this study are included in the paper and supplementary information file, and raw data are available from the corresponding author upon request.

\section*{Code Availability}
All code used for this work are available from the corresponding author upon request.

\section*{Acknowledgements} 
We thank Hyeongjoo Row and Adrianne Zhong for helpful discussions. 
L.L. was supported in part by the Department of Defense (DoD) through the National Defense Science \& Engineering Graduate (NDSEG) Fellowship Program. 
A.K.O. was supported by the Laboratory Directed Research and Development Program of Lawrence Berkeley National Laboratory under U.S. Department of Energy Contract No. DE-AC02-05CH11231 and the UC Berkeley College of Engineering.
This research used the Savio computational cluster resource provided by the Berkeley Research Computing program. 

\section*{Author Contributions}
Both authors contributed extensively to the work presented in this paper.

%\bibliography{submission4/interfacerefs_edit}
%apsrev4-2.bst 2019-01-14 (MD) hand-edited version of apsrev4-1.bst
%Control: key (0)
%Control: author (72) initials jnrlst
%Control: editor formatted (1) identically to author
%Control: production of article title (-1) disabled
%Control: page (0) single
%Control: year (1) truncated
%Control: production of eprint (0) enabled
%

\end{document}

\maketitle

\pagenumbering{roman}

\setcounter{secnumdepth}{2}
\setcounter{tocdepth}{2}
{
	\hypersetup{
		linkcolor=Black,
		citecolor=Black
	}
	\vspace{-30pt}
	\tableofcontents
}

\noindent\rule{5.0cm}{0.4pt}
	{\footnotesize
	$\\$
	{
	    {$^\dag \,$}aomar@berkeley.edu}
	}

\newpage 
\pagenumbering{arabic}
\section{\label{sec:fluctuating_hydro}Fluctuating Hydrodynamics of Active Brownian Particles}
In this Section, we derive the fluctuating hydrodynamics of ABPs that serve as the starting point for our interfacial analysis discussed in Section~\ref{sec:interface_langevin} as well as the main text.
Our derivation begins from the microscopic particle dynamics and systematically derives the evolution of the coarse-grained density field. 
This procedure resembles that of Irving and Kirkwood~\cite{Irving1950}, which is used to derive the dynamics of fields or collective variables \textit{averaged} over the relevant $N$-body microscopic distribution.  
A key distinction here is that we do not average over a statistical ensemble in order to obtain \textit{stochastic} evolution equations, as demonstrated by Dean in the case of passive systems~\cite{Dean1996LangevinProcesses}. 
The absence of averaging over a statistical distribution introduces an additional term to the dynamical operator needed to determine the evolution equation of arbitrary fields, as we discuss below.
In the case of athermal ABPs, the stochastic particle orientations give rise to a contribution to the dynamical operator that is identical to that of passive systems with orientational degrees of freedom~\cite{Cugliandolo2015StochasticDipoles}.
We arrive at a fluctuating hydrodynamic description of the coarse-grained density field that solely depends on the stochastic density and the prescribed activity. 

\subsection{\label{sec:microhydro}Exact Coarse-Grained Density Evolution}
Our derivation of the fluctuating hydrodynamics for active Brownian particles begins with the overdamped Langevin equation for the position $\mathbf{r}_i$ and orientation $\mathbf{q}_i$ of particle $i$:

\begin{subequations}
    \label{seq:particleeom}
    \begin{align}
        \Dot{\mathbf{r}}_i &= U_o \mathbf{q}_i + \frac{1}{\zeta}\sum_{j\neq i}^N\mathbf{F}_{ij}\label{seq:rdot}, \\ \Dot{\mathbf{q}}_i &= \mathbf{q}_i\times \bm{\Omega}_i, \label{seq:qdot}
    \end{align}
where $\dot{\mathbf{a}}$ denotes the time variation of $\mathbf{a}$, $\mathbf{F}_{ij}$ is the interparticle force, $U_o$ is the intrinsic active speed, $\zeta$ is the translational drag coefficient, $\bm{\Omega}_i$ is a Brownian angular velocity with zero mean and a variance of $\langle \bm{\Omega}_i (t) \bm{\Omega}_j(t')\rangle = 2D_R\delta_{ij}\delta(t-t')\mathbf{I}$, and $D_R$ is the rotational diffusivity.
We note that, as presented, Eq.~\eqref{seq:qdot} must be interpreted in the Stratonovich convention to conserve the magnitude of $\mathbf{q}_i$~\cite{Cugliandolo2015StochasticDipoles}.
By adding the relevant drift term (see Section~\ref{sec:stochcalc}) we may convert Eq.~\eqref{seq:qdot} to an equivalent statement interpreted with the It\^{o} convention:
\begin{align}
    \Dot{\mathbf{q}}_i &= \mathbf{q}_i\times \bm{\Omega}_i - 2D_R\mathbf{q}_i. \label{seq:qdotito}
\end{align}
\end{subequations}

We aim to describe the fluctuations of the coarse-grained density field, $\rho(\mathbf{r}; t) = \sum_{i=1}^N\Delta(\mathbf{r} - \mathbf{r}_i)$, where $\Delta(\mathbf{r}-\mathbf{r}_i)$ is a kernel of finite spatial width such that $\rho(\mathbf{r}; t)$ is spatially continuous. 
To do this, we first define the dynamical operator required to describe the time evolution of an observable (interpreted with the It\^{o} convention), $\mathcal{O}$, which can arbitrarily depend on the microscopic degrees of freedom, $(\mathbf{r}^N, \mathbf{q}^N)$\footnote{We note that while we have denoted the time derivative $\mathcal{O}$ as a partial derivative, it is in fact a total derivative as the only time dependence of $\mathcal{O}$ is implicit through the microscopic degrees of freedom~\cite{Zwanzig2001NonequilibriumMechanics}.
We choose to express it as a partial derivative in order to avoid confusion with a material derivative later on.}:
\begin{equation}
    \frac{\partial \mathcal{O}}{\partial t} = \sum_{i=1}^{N}\left[\dot{\mathbf{r}}_i\cdot\frac{\partial \mathcal{O}}{\partial \mathbf{r}_i} + \dot{\mathbf{q}}_i\cdot\frac{\partial \mathcal{O}}{\partial \mathbf{q}_i} + \mathcal{I}\right].
    \label{seq:itostart}
\end{equation}
$\mathcal{I}$ represents terms emerging from the the It\^{o} chain rule (see Section~\ref{sec:stochcalc} for details):
\begin{equation}
    \mathcal{I}  =D_R\boldsymbol{\nabla}^2_{\mathbf{q}_i}\mathcal{O} + 2D_R\mathbf{q}_i\cdot\frac{\partial\mathcal{O}}{\partial \mathbf{q}_i},
    \label{seq:driftidentity}
\end{equation}
where $\boldsymbol{\nabla}_{\mathbf{q}_i}\equiv \mathbf{q}_i\times\partial/\partial\mathbf{q}_i$ is the rotational gradient operator.
Substitution of Eq.~\eqref{seq:driftidentity} and Eq.~\eqref{seq:qdotito} into Eq.~\eqref{seq:itostart} results in:
\begin{equation}
    \frac{\partial \mathcal{O}}{\partial t} = \sum_{i=1}^{N}\left[\dot{\mathbf{r}}_i\cdot\frac{\partial \mathcal{O}}{\partial \mathbf{r}_i} + D_R\boldsymbol{\nabla}^2_{\mathbf{q}_i}\mathcal{O} + \left(\mathbf{q}_i\times\bm{\Omega}_i\right)\cdot\frac{\partial \mathcal{O}}{\partial \mathbf{q}_i}\right].
    \label{seq:langito}
\end{equation}
We now define the dynamical operator $\mathcal{L}$ that evolves an arbitrary observable in time as:
\begin{equation}
    \label{seq:doperator}
    \mathcal{L} =   \sum_{i=1}^N\bigg[\underbrace{\dot{\mathbf{r}}_i\cdot\frac{\partial }{\partial \mathbf{r}_i} + D_R \boldsymbol{\nabla_{\mathbf{q}_i}}^2}_{\substack{\text{deterministic}}} + \underbrace{\left(\mathbf{q}_i\times\bm{\Omega}_i\right)\cdot\frac{\partial }{\partial \mathbf{q}_i}}_{\substack{\text{stochastic}}} \bigg],
\end{equation}
where we note that the first two terms are deterministic (i.e.,~solely depending on the microscopic configuration) while the last term is stochastic through the explicit dependence on $\Omega_i$. 
This distinction allows us to further appreciate that the dynamical operator can be split into two contributions with:
\begin{subequations}
\label{seq:opbreakdown}
\begin{align}
    \mathcal{L} &= \mathcal{L}_{\rm FP}^* + \mathcal{L}_S,\ \label{seq:addoperators}\\
    \mathcal{L}_{\rm FP}^* &= \sum_{i=1}^N\bigg[\dot{\mathbf{r}}_i\cdot\frac{\partial }{\partial \mathbf{r}_i} + D_R \boldsymbol{\nabla_{\mathbf{q}_i}}^2 \bigg],\ \label{seq:FPop}\\
    \mathcal{L}_S &= \sum_{i=1}^N\bigg[\left(\mathbf{q}_i\times\bm{\Omega}_i\right)\cdot\frac{\partial }{\partial \mathbf{q}_i} \bigg], \label{seq:stochop}
\end{align}
\end{subequations}
where we have recognized that the deterministic portion of the operator is \textit{precisely} the adjoint of the Fokker-Planck operator for ABPs (see Ref.~\cite{Omar2023b}). 
The dynamics of an arbitrary observable are thus given by:
\begin{equation}
\label{seq:Odynamics}
\frac{\partial}{\partial t}\mathcal{O} = \mathcal{L}\mathcal{O} = \mathcal{L}_{\rm FP}^*\mathcal{O} + \mathcal{L}_S\mathcal{O}.
\end{equation}
An alternative and equivalent derivation of these dynamics, following Refs.~\cite{vanKampen2007StochasticChemistry, Cugliandolo2015StochasticDipoles}, is described in Section~\ref{sec:stochcalc}.

In what follows, we will proceed to derive the dynamics of the density and its one-body orientational moments as derived by Ref.~\cite{Omar2023b}.
There, exact expressions for $\mathcal{L}_{\rm FP}^*\mathcal{O}$ were derived and ultimately an expectation over the noise-averaged $N$-body distribution was performed.
The latter operation eliminated the need to evaluate $\mathcal{L}_S\mathcal{O}$ which vanishes upon averaging over the noise distribution. 
Here, for completeness, we will recapitulate the expressions found in Ref.~\cite{Omar2023b} for  $\mathcal{L}_{\rm FP}^*\mathcal{O}$\footnote{We note that, in this work, we are considering the dynamics of coarse-grained variables while Ref.~\cite{Omar2023b} examined the dynamics of microscopic variables (averaged over the distribution). 
The use of coarse-grained field variables, however, will simply introduce the coarse-graining kernel in place of the Dirac delta function employed in the definition of microscopic field variables.} while now also determining the stochastic contributions to our field variables. 

The primary field variable of interest is the density, which acts as the unambiguous order parameter in three dimensional motility-induced phase separation~\cite{Omar2021}. 
In this Subsection, we will first focus on the \textit{exact} dynamics of the coarse-grained density field before introducing the closures and approximations necessary to ultimately describe these dynamics solely in terms of the density field itself in the following Subsection. 

The coarse-grained density at position $\mathbf{r}$ is defined as:
\begin{equation}
    \label{seq:densitycoarsegrained}
     \rho(\mathbf{r}, t) = \sum_{i=1}^{N}\Delta(\mathbf{r}-\mathbf{r}_i).
\end{equation}
Substituting $\mathcal{O} = \rho$ into Eq.~\eqref{seq:Odynamics} results in simply $\partial \rho/\partial t = \mathcal{L}_{\rm FP}^*\rho$ (and is thus fully derived in Ref.~\cite{Omar2023b}) as the density contains no explicit dependence on the particle orientations. 
The exact evolution equation for the coarse-grained density follows as:
\begin{subequations}
\label{seq:rho_evolution}
\begin{equation}
    \frac{\partial \rho}{\partial t} =  - \boldsymbol{\nabla} \cdot \mathbf{J}, \label{seq:rho_eqn}
\end{equation}
where $\mathbf{J} = \sum_{i=1}^N\dot{\mathbf{r}}_i\Delta (\mathbf{r} - \mathbf{r}_i)$ is the flux of density and is given by:
\begin{equation}
    \mathbf{J} = U_o \mathbf{m} + \frac{1}{\zeta} \boldsymbol{\nabla} \cdot \bm{\sigma}^{C}, \label{seq:rho_flux}
\end{equation}
\end{subequations}
where $\mathbf{m} = \sum_{i=1}^N \mathbf{q}_i \Delta (\mathbf{r} - \mathbf{r}_i)$ and $\bm{\sigma}^{C}$ are, respectively, the polar order and conservative interaction stress.
For the pairwise interactions under consideration in this work, the interaction stress is:
\begin{equation}
\label{seq:cstress}
\bm{\sigma}^C =  -\frac{1}{2}\sum_{i=1}^{N}\sum_{j\neq i}^{N}\mathbf{F}_{ij}\mathbf{r}_{ij}b_{ij},
\end{equation}
where  $b_{ij}(\mathbf{r}; \mathbf{r}_{i}, \mathbf{r}_{j})  = \int_{0}^{1} \Delta(\mathbf{r} -  \mathbf{r}_{j} - \lambda \mathbf{r}_{ij}) \ d\lambda$ is the bond function~\cite{Noll1955, Lehoucq2010,Hardy1982FormulasWaves,Epstein2019} and $\mathbf{r}_{ij} = \mathbf{r}_i - \mathbf{r}_j$. 

The density flux [Eq.~\eqref{seq:rho_flux}] naturally introduces the polar order field, the dynamics of which we now consider.
The time evolution of the polarization is given by $\partial \mathbf{m} / \partial t = \mathcal{L}_{\rm FP}^*\mathbf{m} + \mathcal{L}_S\mathbf{m}$ where we now must consider the stochastic operator. 
The resulting evolution equation follows as:
\begin{subequations}
\label{seq:polarization_evolution}
\begin{equation}
    \frac{\partial \mathbf{m}}{\partial t} = - \boldsymbol{\nabla} \cdot \mathbf{J^m} - (d-1)D_R\mathbf{m} + \bm{\eta}^\mathbf{m}, \label{seq:m_eqn}
\end{equation}
where we have defined a polarization noise vector as $\bm{\eta}^\mathbf{m} = \mathcal{L}_S\mathbf{m}$.
The polarization flux $\mathbf{J^m} = \sum_{i=1}^N\dot{\mathbf{r}}_i\mathbf{q}_i\Delta (\mathbf{r} - \mathbf{r}_i)$ (a second rank tensor) is given by:
\begin{equation}
    \mathbf{J^m}= U_o \mathbf{Q} +  \frac{1}{\zeta} \boldsymbol{\kappa}^{\mathbf{m}} 
    + \frac{1}{\zeta} \boldsymbol{\nabla} \cdot \boldsymbol{\Sigma}^{\mathbf{m}}. \label{seq:m_flux}
\end{equation}
\end{subequations}
The polarization flux consists of three contributions: convection at the ideal active speed by the nematic density tensor $\mathbf{Q} = \sum_{i=1}^N \mathbf{q}_i\mathbf{q}_i \Delta (\mathbf{r} - \mathbf{r}_i)$, forcing from a ``body-force-like'' term $\boldsymbol{\kappa}^{\mathbf{m}}$ (dimensions of force density) and forcing from a ``stress-like'' term $\boldsymbol{\Sigma}^{\mathbf{m}}$ (dimensions of stress). 
These terms emerge from the Fokker-Planck portion of the operator and are defined as: 
\begin{subequations}
\label{seq:mbodystress}
    \begin{align}    
    \boldsymbol{\kappa}^{\mathbf{m}}
    &=  
    \frac{1}{2} \sum_i^N \sum_{j \neq i}^N  \mathbf{F}_{ij} (\mathbf{q}_{i} - \mathbf{q}_{j})b_{ij}, \label{seq:kappa_m} \\
    \boldsymbol{\Sigma}^{\mathbf{m}} &= 
    -\frac{1}{2} \sum_i^N \sum_{j \neq i}^N  \mathbf{r}_{ij} \mathbf{F}_{ij} \mathbf{c}_{ij}, \label{seq:Sig_m}
\end{align}
\end{subequations}
where $\mathbf{c}_{ij} = \int_{0}^1 \text{d}\lambda \left(\mathbf{q}_j - \lambda\mathbf{q}_{ij}\right)\Delta\left(\mathbf{r}-\mathbf{r}_{j} - \lambda\mathbf{r}_{ij}\right)$ and $\mathbf{q}_{ij} = \mathbf{q}_i - \mathbf{q}_j$.
The physical interpretation of these terms are discussed in Ref.~\cite{Omar2023b}.
Here, we note that $\bm{\kappa}^{\mathbf{m}}$ vanishes for two particles with the same orientation and is maximal for two particles with \emph{opposite} orientation. 
One can thus interpret $\bm{\kappa}^{\mathbf{m}}$ as a force that resists the nematic convection and reduces the effective convective speed. 
While the first two terms in Eq.~\eqref{seq:m_flux} are associated with convection, the final term, captures the flux of polar generated by interactions across surfaces. 
Constitutive equations for Eq.~\eqref{seq:mbodystress} will later be introduced. 

We now consider the stochastic term, $\bm{\eta}^\mathbf{m}$, appearing in Eq.~\eqref{seq:m_eqn}. 
Applying the stochastic operator on the polarization results in an expression for the polarization noise with:
\begin{equation}
    \label{seq:mnoise}
    \bm{\eta}^\mathbf{m} = \sum_{i=1}^N\mathbf{q}_i\times\bm{\Omega}_i\Delta(\mathbf{r} - \mathbf{r}_i).
\end{equation}
We can immediately recognize that this noise is nonlocal in space for finite coarse-graining width.
Furthermore, describing the variance of this noise in terms of field variables is also not readily possible. 
An approximation that allows this noise to be expressed solely in terms of the one-body orientional fields will be made in Section~\ref{sec:approx}.

The time evolution of the nematic order is now required. 
Again, like the polar order (and unlike the density), the stochastic contribution cannot be neglected. 
The nematic dynamics are found to be: 
\begin{subequations}\label{seq:Qtilde_evolution}
\begin{equation}
    \frac{\partial \mathbf{Q}}{\partial t} = - \boldsymbol{\nabla} \cdot \mathbf{J}^{\mathbf{Q}} - 2dD_R \left(\mathbf{Q} - \frac{1}{d} \rho \mathbf{I} \right) + \bm{\eta^{\mathbf{Q}}}, \label{seq:Qtilde_eqn}
\end{equation}
where we have defined a polarization noise tensor as $\bm{\eta}^\mathbf{Q} = \mathcal{L}_S\mathbf{Q}$.
The nematic flux $\mathbf{J^Q} = \sum_{i=1}^N\dot{\mathbf{r}}_i\mathbf{q}_i\mathbf{q}_i\Delta (\mathbf{r} - \mathbf{r}_i)$ (a third rank tensor) is given by:
\begin{equation}
    \mathbf{J}^{\mathbf{Q}}= U_o \mathbf{B}
    + \frac{1}{\zeta} \boldsymbol{\kappa}^{\mathbf{Q}} 
    + \frac{1}{\zeta} \boldsymbol{\nabla} \cdot \boldsymbol{\Sigma}^{\mathbf{Q}}. \label{seq:Qtilde_flux}
\end{equation}
\end{subequations}
Just as in the case of the polar order flux, the nematic flux consists of three contributions: convection at the ideal active speed by $\mathbf{B} = \sum_{i=1}^N \mathbf{q}_i\mathbf{q}_i\mathbf{q}_i \Delta (\mathbf{r} - \mathbf{r}_i)$, forcing from a ``body-force-like'' term $\boldsymbol{\kappa}^{\mathbf{Q}}$ (dimensions of force density) and forcing from a ``stress-like'' term $\boldsymbol{\Sigma}^{\mathbf{Q}}$ (dimensions of stress). 
Expressions for the latter two contributions have the following form: 
\begin{subequations}
\label{seq:Qbodystress}
    \begin{align}
    \boldsymbol{\kappa}^{\mathbf{Q}} &= \frac{1}{2} \sum_i^N \sum_{j \neq i}^N  \mathbf{F}_{ij} (\mathbf{q}_{i}\mathbf{q}_{i} - \mathbf{q}_{j}\mathbf{q}_{j})b_{ij}, \label{seq:kappa_Q} \\
    \boldsymbol{\Sigma}^{\mathbf{Q}} &= -\frac{1}{2} \sum_i^N \sum_{j \neq i}^N  \mathbf{r}_{ij} \mathbf{F}_{ij} \mathbf{d}_{ij}, \label{seq:Sig_Q}
    \end{align}
\end{subequations}
where $\mathbf{d}_{ij} = \int_0^{1}\text{d}\lambda \left(\mathbf{q}_i\mathbf{q}_i - \lambda\left(\mathbf{q}_i\mathbf{q}_i - \mathbf{q}_j\mathbf{q}_j\right)\right)\Delta(\mathbf{r} - \mathbf{r}_j - \lambda\mathbf{r}_{ij})$. 
Despite the change of the tensorial rank of these quantities from those found in the polar order, the physical interpretations remains the same. 
The nematic noise vector takes the following form:
\begin{equation}
    \label{seq:Qnoise}
    \bm{\eta}^\mathbf{Q} = \sum_{i=1}^N2\mathbf{q}_i\times\bm{\Omega}_i\mathbf{q}_i\Delta(\mathbf{r} - \mathbf{r}_i),
\end{equation}
and, as with $\bm{\eta}^{\mathbf{m}}$, will have statistics approximately found in Section~\ref{sec:approx}.

We can similarly obtain an evolution equation for the \textit{traceless} nematic order, defined as: 
\begin{subequations}\label{seq:Q_evolution}
\begin{equation}
\mathbf{Q'} = \mathbf{Q} - \frac{\rho}{d}\mathbf{I}. \label{seq:Qdef}
\end{equation}
Inserting this definition into Eq.~\eqref{seq:Qtilde_evolution} results in:
\begin{equation}
    \frac{\partial \mathbf{Q'}}{\partial t} = - \boldsymbol{\nabla} \cdot \mathbf{J}^{\mathbf{Q'}} - 2dD_R\mathbf{Q'} + \bm{\eta^{\mathbf{Q}}}, \label{seq:Q_eqn}
\end{equation}
where the traceless nematic field flux is simply $\mathbf{J}^{\mathbf{Q'}} = \mathbf{J}^{\mathbf{Q}} - \frac{1}{d}\mathbf{J}\mathbf{I}$ with:
\begin{equation}
    \mathbf{J}^{\mathbf{Q'}}= U_o \mathbf{B} + \frac{1}{\zeta} \boldsymbol{\kappa}^{\mathbf{Q}} - \frac{1}{d} \mathbf{J}\mathbf{I} + \frac{1}{\zeta} \boldsymbol{\nabla} \cdot \boldsymbol{\Sigma}^{\mathbf{Q}}. \label{seq:Q_flux}
\end{equation}
\end{subequations}

In principle we could continue deriving the evolution equation of $\mathbf{B}$ and the infinite hierarchy of one-body orientation fields. 
However, as discussed in Section~\ref{sec:approx}, obtaining the \textit{linear} height field dynamics of the interface does not require the consideration of orientation fields beyond the nematic tensor. 
We thus conclude our derivation of the coarse-grained density dynamics of athermal ABPs and now begin to simplify our expressions with approximations consistent with obtaining a long wavelength capillary theory.

\subsection{\label{sec:approx}Approximations and Closures}

Our aim is to derive to stochastic height field dynamics associated with ABP interfaces.
Doing so requires a stochastic description of the field variable used to define the location of the interface.
In our case, this is the coarse-grained density. 
In Sec.~\ref{sec:microhydro}, we derived the \textit{exact} stochastic dynamics of the coarse-grained density field and, in doing so, found that the density field dynamics are coupled to the dynamics of the other one-body orientational fields. 
An exact description of the height field dynamics would require solving the coupled dynamics of the density, polarization and nematic order \textit{and} an ansatz that introduces less error than that of Bray's~\cite{Bray2002InterfaceShear} (see Section~\ref{sec:interface_langevin}). 
However, as we are interested in obtaining a theory for long wavelength capillary fluctuations, we will make use of Bray's ansatz and other simplifying approximations in our fluctuating hydrodynamics consistent with this aim.
These assumptions reduce our fluctuating hydrodynamic description of the density field to \textit{solely} consist of a single field variable, the density field itself.
This proves convenient in the derivation of the interfacial dynamics, as a single field was assumed in Bray's treatment and more recently by Fausti~\textit{et al.}~\cite{Fausti2021CapillarySeparation}.

We first assume that the relaxation dynamics of the polar and nematic order are faster than those of the density field. 
The faster relaxation dynamics of higher order one-body orientational moments can be appreciated from the conservation equations derived in Sec.~\ref{sec:microhydro}. 
Comparison of Eqs.~\eqref{seq:m_eqn} and~\eqref{seq:Q_eqn} in the absence of a spatial gradient reveals that the nematic order will exhibit a temporal decay that is a factor of $2d/(d-1)$ faster than that of the polar order. 
We neglect the time variation of both the polar and nematic order in Eqs.~\eqref{seq:m_eqn} and~\eqref{seq:Q_eqn}, respectively, as they are anticipated to be faster than those of the density field. 
Doing so allows us to identify that the polar order can be expressed as $\mathbf{m} = \frac{\tau_R}{d-1}\left(\boldsymbol{\nabla} \cdot \mathbf{J^\mathbf{m}} - \bm{\eta}^{\mathbf{m}} \right)$. 
Substitution of this into the density flux [Eq.~\eqref{seq:m_flux}] allows us to define several key mechanical terms:
\begin{subequations}
\label{seq:rho_flux_simplified}
\begin{equation}
    \mathbf{J} = \frac{1}{\zeta} \boldsymbol{\nabla} \cdot \bm{\Sigma} + \frac{\ell_o}{d-1}\bm{\eta}^{\mathbf{m}} , \label{seq:rho_flux2}
\end{equation}
where we have defined the \textit{dynamic} stress tensor: 
\begin{equation}
\bm{\Sigma} = \bm{\sigma}^{\rm{act}} + \bm{\sigma}^C, \label{seq:micro_dynstress}
\end{equation}
recognizing that the body force generated by the polarization appears to take the form of an \textit{effective} stress upon neglecting the temporal variation of the polar order. 
This effective active (or ``swim''~\cite{Takatori2014}) stress is defined as~\cite{Solon2015,Omar2020}: 
\begin{equation}
\bm{\sigma}^{\rm{act}} = \frac{\zeta \ell_o}{d-1} \mathbf{J^m}. \label{seq:micro_activestress}
\end{equation}
\end{subequations}

We now find the statistics of $\bm{\eta}^{\mathbf{m}}$, which is given by:
\begin{equation}
    \label{seq:mnoisemicro}
    \bm{\eta}^\mathbf{m} = \sum_{i=1}^N\mathbf{q}_i\times\bm{\Omega}_i\delta(\mathbf{r} - \mathbf{r}_i).
\end{equation}
While the mean is clearly zero, the variance is given by:
\begin{equation}
    \label{seq:mvar1}
    \left \langle \bm{\eta}^\mathbf{m}(\mathbf{r}, t)\bm{\eta}^\mathbf{m}(\mathbf{r'}, t') \right \rangle = \sum_{i=1}^N\sum_{j=1}^N\left\langle (\mathbf{q}_i\times\bm{\Omega}_i)(\mathbf{q}_j\times\bm{\Omega}_j)\right \rangle \Delta(\mathbf{r} - \mathbf{r}_i)\Delta(\mathbf{r'} - \mathbf{r}_j),
\end{equation} 
where $\langle \cdot\cdot\cdot \rangle$ denotes an expectation over the distribution of the stochastic angular velocities. 
Straightforward manipulation allows us to express ${\left\langle (\mathbf{q}_i\times\bm{\Omega}_i)(\mathbf{q}_j\times\bm{\Omega}_j)\right \rangle = 2D_R\left[(\mathbf{q}_i\cdot\mathbf{q}_j)\mathbf{I} - \mathbf{q}_i\mathbf{q}_j\right]\delta_{ij}\delta(t-t')}$, substitution of which into Eq.~\eqref{seq:mvar1} results in:
\begin{equation}
    \label{seq:mvar2}
    \left \langle \bm{\eta}^\mathbf{m}(\mathbf{r}, t)\bm{\eta}^\mathbf{m}(\mathbf{r'}, t') \right \rangle = \sum_{i=1}^N2D_R\left(\mathbf{I} - \mathbf{q}_i\mathbf{q}_i \right)\Delta(\mathbf{r} - \mathbf{r}_i)\Delta(\mathbf{r'} - \mathbf{r}_i)\delta(t-t'),
\end{equation}
where the independence of the variance of angular velocities between different particles has eliminated one of the particle summations and ensured that the noise is a one-body property.
We now approximate the statistics of $\bm{\eta}^{\mathbf{m}}$ with the statistics of $\hat{\bm{\eta}}^{\mathbf{m}}$, which is the noise associated with the \textit{microscopic} polar order flux and has a variance of:
\begin{equation}
    \label{seq:mvar2micro}
    \left \langle \hat{\bm{\eta}}^\mathbf{m}(\mathbf{r}, t)\hat{\bm{\eta}}^\mathbf{m}(\mathbf{r'}, t') \right \rangle = \sum_{i=1}^N2D_R\left(\mathbf{I} - \mathbf{q}_i\mathbf{q}_i \right)\delta(\mathbf{r} - \mathbf{r}_i)\delta(\mathbf{r'} - \mathbf{r}_i)\delta(t-t').
\end{equation}
Using the identity $\delta(\mathbf{r} - \mathbf{r}_i)\delta(\mathbf{r'} - \mathbf{r}_i)= \delta(\mathbf{r} - \mathbf{r'})\delta(\mathbf{r} - \mathbf{r}_i)$, which is the only component of our approximation that cannot be applied to arbitrary $\Delta$, and invoking the definitions of the microscopic density and nematic order, we arrive at our final expression for the microscopic polarization noise statistics:
\begin{subequations}
\label{seq:etamstatisticsmicro}
\begin{align}
    \left \langle \hat{\bm{\eta}}^\mathbf{m}(\mathbf{r}, t) \right \rangle &= \mathbf{0}, \label{seq:etamavg} \\ 
    \left \langle \hat{\bm{\eta}}^\mathbf{m}(\mathbf{r}, t)\hat{\bm{\eta}}^\mathbf{m}(\mathbf{r'}, t') \right \rangle &= 2D_R\left(\hat{\rho}\mathbf{I} - \mathbf{\hat{Q}} \right)\delta(\mathbf{r} - \mathbf{r'})\delta(t-t'), \label{seq:etamvarmicro}
\end{align}
\end{subequations}
where $\hat{\rho}$ and $\mathbf{\hat{Q}}$ can be evaluated at $\mathbf{r}$ or $\mathbf{r'}$. 
This stochastic polarization term is identical to that found by Cugliandolo~\textit{et al.}~\cite{Cugliandolo2015StochasticDipoles} in describing the fluctuating hydrodynamics of passive dipoles with the only physical distinction being the (generally) athermal origins of $D_R$ for active systems. 
By approximating the statistics of $\bm{\eta}^{\mathbf{m}}$ with the statistics of $\hat{\bm{\eta}}^{\mathbf{m}}$ we find:
\begin{subequations}
\label{seq:etamstatistics}
\begin{align}
    \left \langle \bm{\eta}^\mathbf{m}(\mathbf{r}, t) \right \rangle &= \mathbf{0}, \\ 
    \left \langle \bm{\eta}^\mathbf{m}(\mathbf{r}, t)\bm{\eta}^\mathbf{m}(\mathbf{r'}, t') \right \rangle &= 2D_R\left(\rho\mathbf{I} - \mathbf{Q} \right)\delta(\mathbf{r} - \mathbf{r'})\delta(t-t'), \label{seq:etamvar}
\end{align}
\end{subequations}
Furthermore, by absorbing the factor of $\ell_o/(d-1)$ present in Eq.~\eqref{seq:rho_flux2} into $\bm{\eta}^{\mathbf{m}}$, we define $\bm{\eta}^{\rm act} = \ell_o/(d-1)\bm{\eta}^{\mathbf{m}}$ which has statistics following directly from those of the polar order noise [Eq.~\eqref{seq:etamstatistics}]: 
\begin{subequations}
\label{seq:etaactivestatistics}
\begin{align}
    \left \langle \bm{\eta}^{\rm act}(\mathbf{r}, t) \right \rangle &= \mathbf{0}, \label{seq:etaactavg} \\ 
       \left \langle \bm{\eta}^{\rm act}(\mathbf{r},t)\bm{\eta}^{\rm act}(\mathbf{r}',t') \right \rangle &= 2\frac{k_BT^{\rm act}}{\zeta}\left(\rho\mathbf{I} - \frac{d}{d-1}\mathbf{Q}'\right) \delta(t-t')\delta(\mathbf{r}-\mathbf{r}') \ \label{seq:etaactvar},
\end{align}
where we have defined the active energy scale as $k_BT^{\rm{act}} \equiv \ell_o \zeta U_o /d(d-1)$. 
\end{subequations}
The statistics of $\bm{\eta}^{\mathbf{Q}}$ can be approximated analogously to those of the polarization flux with:
\begin{subequations}
\label{seq:etaQstatistics}
\begin{align}
    \left \langle \bm{\eta}^\mathbf{Q}(\mathbf{r}, t) \right \rangle &= \mathbf{0}, \label{seq:etaQavg} \\ 
    \left \langle \bm{\eta}^\mathbf{Q}(\mathbf{r}, t)\bm{\eta}^\mathbf{Q}(\mathbf{r'}, t') \right \rangle &= 8D_R\left(\mathbf{Q}\mathbf{I} - \mathbf{C} \right)\delta(\mathbf{r} - \mathbf{r'})\delta(t-t'), \label{seq:etaQvar}
\end{align}
\end{subequations}
where $\mathbf{C} = \sum_{i=1}^N \mathbf{q}_i\mathbf{q}_i\mathbf{q}_i \mathbf{q}_i\Delta (\mathbf{r} - \mathbf{r}_i)$.

At this point, the unknown quantities required to describe our stochastic density flux are constitutive relations for the conservative stress [Eq.~\eqref{seq:cstress}] as well as the body-force-like and stress-like terms [Eqs.~\eqref{seq:mbodystress} and~\eqref{seq:Qbodystress}] and the remaining one-body orientational moments, $\mathbf{Q}'$, $\mathbf{B}$ and $\mathbf{C}$. 
Before proposing constitutive equations and closure relations we recall that predicting states of coexistence \textit{requires} the retention of terms that are at least third order in spatial gradient of the density in our flux expression [Eq.~\eqref{seq:rho_flux2}] and thus second order in our dynamic stress [Eq.~\eqref{seq:micro_dynstress}].

We first focus on the required constitutive equations, beginning with the conservative stress.
In the case of a homogeneous density, the conservative stress would simply return the isotropic interaction pressure, $\bm{\sigma}^C = -p_C\mathbf{I}$, an equation of state that depends on the bulk density and activity. 
A general second-order gradient expansion of the conservative stress has the form:
\begin{equation}
\bm{\sigma}^C = \left[-p_C + \kappa_1\nabla^2\rho + \kappa_2|\boldsymbol{\nabla}\rho|^2\right]\mathbf{I}  + \kappa_3\boldsymbol{\nabla}\rho\boldsymbol{\nabla}\rho + \kappa_4\boldsymbol{\nabla}\boldsymbol{\nabla}\rho,\  \label{seq:expandcstress}
\end{equation}
where the linear term vanishes due to inversion symmetry. 
In the reversible limit, which occurs as $\ell_o \rightarrow 0$, the coefficients $\{\kappa_i\}$ take the form of the Korteweg stress~\cite{Korteweg1904, Yang1976, Evans1979TheFluids} as noted in the main text. 
We forgo providing general microscopic expressions for $\{\kappa_i\}$ that are valid for all activities as, for active hard spheres, there is likely only a narrow region of activity where these coefficients both depart significantly from the reversible Korteweg stress and are comparable in scale to the gradient terms that will arise from the active stress. 

The physical interpretation of the body-force-like terms is clear: they arise from interparticle interactions and \textit{oppose} free convection at the intrinsic active speed $U_o$. 
Following Ref.~\cite{Omar2023b}, we therefore propose the following constitutive relation for 
$\bm{\kappa}^{\mathbf{m}}$:
\begin{equation}
    \bm{\kappa}^{\mathbf{{m}}} = -\zeta \left(U_o-U\right)\mathbf{Q},
    \label{seq:kappaconst}
\end{equation}
where $U$ is the \textit{effective} active speed and is bound between $0$ and $U_o$.
We adopt a similar constitutive equation for the body-force-like term appearing in the nematic flux with: 
\begin{equation}
    \bm{\kappa}^{\mathbf{{Q}}} = -\zeta \left(U_o-U\right)\mathbf{B},
    \label{seq:kappaconstQ}
\end{equation}
where we have assumed that the same effective active speed appears here as in Eq.~\eqref{seq:kappaconst}. 
The polarization flux also requires a constitutive relation for the stress-like term, $\bm{\Sigma}^{\mathbf{m}}$, which cannot be \textit{a priori} discarded on the basis of spatial gradients. 
However, here we neglect this contribution as it was shown to have a marginal quantitative impact on the ABP hard sphere/disk phase diagrams in Ref.~\cite{Omar2023b}. 

With Eq.~\eqref{seq:kappaconst} and $\bm{\Sigma}^{\mathbf{m}} \approx \mathbf{0}$, the simplified polarization flux results in the active stress taking the following form:
\begin{subequations}
\label{seq:activestress2}
\begin{equation}
\bm{\sigma}^{\rm{act}} = -p_{\rm act}\mathbf{I} + \frac{\zeta \ell_o}{d-1} \left(U \mathbf{Q'} \right),
\label{seq:micro_activestressnematic}
\end{equation}
where the active pressure is defined as:
\begin{equation}
\label{seq:activepressure}
p_{\rm act} = \frac{\zeta \ell_o U}{d(d-1)}\rho.
\end{equation}
\end{subequations}
From Eq.~\eqref{seq:activestress2} it is clear that the gradient contributions to the active stress will be generated by the convection of the traceless nematic order and equation of state for $U$ (which will depend on the nature of the particle interactions) in addition to $p_C$. 

We now turn to the traceless nematic field. 
Neglecting the time evolution of $\mathbf{Q}'$ in Eq.~\eqref{seq:Q_eqn} allows us to express $\mathbf{Q'} = \frac{\tau_R}{2d}\left(\boldsymbol{\nabla} \cdot \mathbf{J^\mathbf{Q'}} - \bm{\eta}^{\mathbf{Q}} \right)$.
A description of the nematic order requires expressions for $\bm{\eta}^{\mathbf{Q}}$ (and thus, $\mathbf{C}$), $\bm{B}$, and $\bm{\Sigma}^{\mathbf{Q}}$ [recall, we have already assumed the form of $\bm{\kappa}^{\mathbf{Q}}$ in Eq.~\eqref{seq:kappaconstQ}].
We can now appreciate that each orientational field contains contributions related to the divergence of the next higher order field (e.g.,~$\mathbf{m} \sim \boldsymbol{\nabla} \cdot \mathbf{Q}$) and therefore introduces higher order spatial gradients. 
We thus can safely assume that $\mathbf{B}$ is isotropic:
\begin{equation}
\label{seq:tracelessB}
    \mathbf{B} = \frac{1}{d+2}\bm{\alpha}\cdot\mathbf{m},
\end{equation}
where $\bm{\alpha}$ is the fourth-order identity tensor given in Einstein notation as $\alpha_{ijkl}=\delta_{ij}\delta_{kl} + \delta_{ik}\delta_{jl} + \delta_{il}\delta_{jk}$.
Further, $\bm{\Sigma}^{\mathbf{Q}}$ can be neglected as its contribution to the dynamic stress will be higher than second order in spatial gradient. 
We will also discard $\bm{\eta}^{\mathbf{Q}}$ as its variance (at lowest order) will enter into the flux at the level of $\boldsymbol{\nabla}\cdot\boldsymbol{\nabla}\cdot\mathbf{Q}\mathbf{I}$.
Finally, the fast relaxation of the polar order and nematic field implies that the density flux can be approximated to be vanishingly small such that $\zeta U_o \mathbf{m} \approx -\boldsymbol{\nabla}\cdot\bm{\sigma}^C$ [see Eq.~\eqref{seq:rho_flux}].
Under these simplifications, the traceless nematic field takes the following form:
\begin{equation}
    \mathbf{Q}' = \frac{\tau_R}{2\zeta U_o d(d+2)}\boldsymbol{\nabla}\cdot \left[U \left(\boldsymbol{\nabla}\cdot\bm{\sigma}^C\right)\cdot\bm{\alpha} \right].
    \label{seq:tracenematiccompact}
\end{equation}

With expressions for the bulk and gradient contributions for the conservative and active stress and the active noise statistics, we now have an expression for the density flux that \textit{solely} depends on the density field itself. 
Evaluating these expressions will require equations of state for the effective active speed, bulk interaction pressure, and any relevant interfacial coefficients in the conservative stress. 

\subsection{\label{sec:summary}Summary of Coarse-Grained Density Fluctuating Hydrodynamics}
We now summarize the fluctuating hydrodynamic equations for the coarse-grained density field, $\rho$.
It is convenient to reorganize Eq.~\eqref{seq:tracenematiccompact} under the simplification (consistent with our gradient theory) of $\bm{\sigma}^C \approx -p_C\mathbf{I}$:
\begin{equation}
    \mathbf{Q}' = \frac{3\ell_o}{2\zeta d(d+2)} \biggl[-\overline{U}(\rho) \frac{\partial p_C}{\partial \rho}\nabla^2\rho \mathbf{I} -\frac{\partial }{\partial \rho}\left[\overline{U}(\rho)\frac{\partial p_C}{\partial \rho}\right]\boldsymbol{\nabla}\rho\boldsymbol{\nabla}\rho.\biggr],
    \label{seq:tracenematicexpand}
\end{equation}
where we have introduced the dimensionless effective active speed, $\overline{U} \equiv U/U_o$. 
Substitution of Eq.~\eqref{seq:tracenematicexpand} into the active stress allows us to straightforwardly identify the active stress gradient terms. 

The fluctuating hydrodynamics of athermal ABPs can now be summarized:
\begin{subequations}
\label{seq:abphydro}
\begin{align}
    \frac{\partial\rho}{\partial t} &= -\boldsymbol{\nabla} \cdot \mathbf{J},\  \label{seq:continuity} \\ \mathbf{J} &= \frac{1}{\zeta}\boldsymbol{\nabla}\cdot \bm{\Sigma} + \bm{\eta}^{\rm act},\ \\ \bm{\Sigma} &= \bm{\sigma}^C + \bm{\sigma}^{\rm act},\ \label{seq:dynstressagain} \\  
    \bm{\sigma}^C &= \left[-p_C(\rho) + \kappa_1(\rho)\nabla^2\rho + \kappa_2(\rho)|\boldsymbol{\nabla}\rho|^2\right]\mathbf{I}  + \kappa_3(\rho)\boldsymbol{\nabla}\rho\boldsymbol{\nabla}\rho + \kappa_4(\rho)\boldsymbol{\nabla}\boldsymbol{\nabla}\rho,\  \label{seq:cstressagain}\\ \bm{\sigma}^{\rm{act}} &= \left[-p_{\rm act} + {a(\rho)}\nabla^2\rho\right]\mathbf{I} + {b(\rho)}\boldsymbol{\nabla}\rho\boldsymbol{\nabla}\rho,\ \label{seq:astress} \\ a(\rho) &= \frac{3 \ell_o^2}{2d(d-1)(d+2)}\overline{U}^2\frac{\text{d} p_C}{\text{d}\rho}, \ \label{seq:beta} \\ b(\rho) &= \frac{3\ell_o^2\overline{U}}{2d(d-1)(d+2)}\frac{d}{d\rho}\left[\overline{U}\frac{\text{d}p_C}{\text{d}\rho}\right],\ \label{seq:lambda} 
\end{align}
where the statistics of the noise $\bm{\eta}^{\rm act}$ are:
\begin{align}
        \langle \bm{\eta}^{\rm act}\rangle &= \mathbf{0}, \\ \langle \bm{\eta}^{\rm act}(\mathbf{r},t)\bm{\eta}^{\rm act}(\mathbf{r}',t') \rangle &= 2\frac{k_BT^{\rm act}}{\zeta}\left(\rho\mathbf{I} - \frac{d}{d-1}\mathbf{Q}'\right)\delta(t-t')\delta(\mathbf{r}-\mathbf{r}'),
\end{align}
\end{subequations}
and the simplified traceless nematic tensor is provided in Eq.~\eqref{seq:tracenematicexpand} and its slightly more general form provided in Eq.~\eqref{seq:tracenematiccompact}. 
As stated in Section~\ref{sec:microhydro}, our derived Langevin equations for observables such as Eq.~\eqref{seq:abphydro} are to be interpreted in the It\^{o} convention. 
However, as the noise appearing in Eq.~\eqref{seq:abphydro} is \textit{conserved}, the drift term associated with switching to the Stratonovich convention vanishes~\cite{Gardiner2002HandbookMethods,Archer2004DynamicalDeterministic,Grun2006Thin-FilmNoise,Delfau2016PatternEquation}.

The derived fluctuating hydrodynamics are general to athermal ABPs that obey the equations-of-motion described at the outset of this Section [Eqs.~\eqref{seq:particleeom}].
As input, these fluctuating hydrodynamics require equations of state that will certainly depend on the precise form of the conservative particle interactions.
Specifically, the density and activity dependence of $\overline{U}$, $p_C$, and the gradient coefficients $\{\kappa_i\}$ must be provided. 
For hard spheres, the active stress gradient coefficients scale as $\ell_o^2$ [see Eqs.~\eqref{seq:beta} and~\eqref{seq:lambda}] while those of the conservative interaction stress ($\{\kappa_i\}$) can only scale with powers of the hard sphere diameter $d_{\rm hs}$.
We therefore can safely discard $\{\kappa_i\}$ as we will exclusively focus on phase-separated hard spheres where the run length is at least an order of magnitude larger than the hard sphere diameter [$\ell_o/d_{\rm hs} \ge \mathcal{O}(10)$].

\subsection{\label{sec:stochcalc}Supplemental Stochastic Calculus Details}
Here, we provide further details on the derivation of the dynamical operator $\mathcal{L}$ which evolves an arbitrary observable $\mathcal{O}$ [see Eqs.~\eqref{seq:opbreakdown}]. 
We begin by detailing the procedure used to find the drift term necessary to define Eq.~\eqref{seq:qdotito} and the It\^{o} chain rule term necessary to find Eq.~\eqref{seq:langito}. 
Within this section we will make frequent use of Einstein notation.
While doing so, particle indices will remain as Latin subscripts while tensor components will be given by Greek superscripts.

To convert from the Stratonovich convention to the It\^{o} convention, it is convenient to express Eq.~\eqref{seq:qdot} in the following form:
\begin{equation}
    \dot{\mathbf{q}}_i = \mathbf{g}_i(\mathbf{q}_i) \cdot \bm{\Lambda}_i(t),
\end{equation}
where $\bm{\Lambda}_i(t)$ is a unit Gaussian white noise vector with correlations $\langle \bm{\Lambda}_i(t) \bm{\Lambda}_j(t') \rangle = \delta_{ij}\delta(t-t')\mathbf{I}$ and $\mathbf{g}_i$ is a second rank tensor.
The variance of $\mathbf{q}_i\times \bm{\Omega}_i$, found in Section~\ref{sec:approx}, allows us to immediately identify $\mathbf{g}_i$ as:
\begin{equation}
    \mathbf{g}_i = \left(2D_R\left(\mathbf{I}-\mathbf{q}_i\mathbf{q}_i\right)\right)^{1/2}.
    \label{seq:gnaive}
\end{equation}
We simplify  Eq.~\eqref{seq:gnaive} by noting that:
\begin{align}
    \left(\mathbf{I} - \mathbf{q}_i\mathbf{q}_i\right)\left(\mathbf{I} - \mathbf{q}_i\mathbf{q}_i\right) &= \left(\delta^{\alpha\beta} - q_i^{\alpha}q_i^{\beta}\right)\left(\delta^{\beta\gamma} - q_i^{\beta}q_i^{\gamma}\right)\nonumber \\ &= \delta^{\alpha \gamma} - q_i^{\alpha}q_i^{\gamma} - q_i^{\alpha}q_i^{\gamma} + q_i^{\alpha}q_i^{\gamma} \nonumber \\ &= \left(\mathbf{I} - \mathbf{q}_i\mathbf{q}_i\right),\label{seq:gproperties}
\end{align}
where we have used $|\mathbf{q}_i| = 1$. 
$\mathbf{g}_i$ thus takes the form:
\begin{equation}
    \mathbf{g}_i = \sqrt{2D_R}\left(\mathbf{I}-\mathbf{q}_i\mathbf{q}_i\right).
    \label{seq:gfinal}
\end{equation}
As outlined by van Kampen~\cite{vanKampen1981ItoStratonovich,vanKampen2007StochasticChemistry}, we may convert Stratonovich interpreted dynamics to It\^{o} interpreted dynamics by addition of the drift term given by:
\begin{align}
    \frac{1}{2}\frac{\partial g^{\lambda \mu}}{\partial q_i^{\kappa}}g^{\kappa\mu} &= D_R\left(\frac{\partial}{\partial q_i^{\kappa}}\left(\delta^{\lambda \mu} - q_i^{\lambda}q_i^{\mu}\right)\right)\left(\delta^{\kappa\mu} - q_i^{\kappa}q_i^{\mu}\right)\nonumber \\ &= -D_R\left(\delta^{\lambda\kappa}q_i^{\mu} + q^{\lambda}\delta^{\kappa \mu}\right)\left(\delta^{\kappa\mu} - q_i^{\kappa}q_i^{\mu}\right) \nonumber \\ &= -2D_R\mathbf{q}_i.\label{seq:itodrifttermsolution}
\end{align}
Thus the It\^{o} equivalent dynamics to Eq.~\eqref{seq:qdot} are:
\begin{equation}
    \dot{\mathbf{q}}_i = \mathbf{q}_i\times\bm{\Omega}_i - 2D_R\mathbf{q}_i,
\end{equation}
completing our derivation of Eq.~\eqref{seq:qdotito}. 

We now solve for the associated It\^{o} chain rule term when taking a time derivative of $\mathcal{O}$. 
The form of such a term (see Refs.~\cite{vanKampen1981ItoStratonovich,vanKampen2007StochasticChemistry}) is:
\begin{equation}
    \mathcal{I}  = \frac{1}{2}\frac{\partial^2\mathcal{O}}{\partial q_i^{\alpha}\partial q_i^{\beta}}g^{\alpha\gamma}g^{\gamma\beta} = D_R \frac{\partial^2\mathcal{O}}{\partial q_{i}^{\alpha}\partial q_i^{\beta}} \left(\delta^{\alpha\beta}-q_i^{\alpha}q_i^{\beta}\right).\label{seq:itochainrulesolve}
\end{equation}
Let's now consider the effect of a rotational Laplace operator acting on $\mathcal{O}$:
\begin{align}
    \boldsymbol{\nabla}^{2}_{\mathbf{q}_i} \mathcal{O} &= \epsilon^{\alpha\beta\gamma}q_i^{\beta}\frac{\partial}{\partial q_i^{\gamma}}\left(\epsilon^{\alpha \mu\nu}q_i^{\mu}\frac{\partial\mathcal{O}}{\partial q_i^{\nu}}\right)\nonumber \\ &= \epsilon^{\alpha\beta\gamma}\epsilon^{\alpha\mu\nu}q_i^{\beta}\left(\delta^{\gamma\mu}\frac{\partial\mathcal{O}}{\partial q_i^{\nu}} + q_i^{\mu}\frac{\partial^2\mathcal{O}}{\partial q_i^{\gamma}\partial q_i^{\nu}}\right)\nonumber \\ &= \left(\delta^{\beta\mu}\delta^{\gamma\nu} - \delta^{\beta\nu}\delta^{\gamma \mu}\right)q_i^{\beta}\left(\delta^{\gamma\mu}\frac{\partial\mathcal{O}}{\partial q_i^{\nu}} + q_i^{\mu}\frac{\partial^2\mathcal{O}}{\partial q_i^{\gamma}\partial q_i^{\nu}}\right)\nonumber \\ &= \frac{\partial^2\mathcal{O}}{\partial q_i^{\beta}\partial q_i^{\gamma}}\left(\delta^{\beta\gamma} - q_i^{\beta}q_i^{\gamma}\right) - 2q_i^{\beta}\frac{\partial\mathcal{O}}{\partial q_i^{\beta}}.\label{seq:rotlaplaceO}
\end{align}
Eq.~\eqref{seq:rotlaplaceO} together with Eq.~\eqref{seq:itochainrulesolve} implies that:
\begin{equation}
    \mathcal{I} = D_R\boldsymbol{\nabla}^{2}_{\mathbf{q}_i}\mathcal{O} + 2D_R\mathbf{q}_i\cdot\frac{\partial\mathcal{O}}{\partial \mathbf{q}_i},
\end{equation}
which concludes our derivation of the It\^{o} chain rule term appearing in Eq.~\eqref{seq:itostart}.

We now present an alternative derivation for the evolution of $\mathcal{O}$ that starts from the Stratonovich expression of the orientational dynamics Eq.~\eqref{seq:qdot} using the Kramers-Moyal expansion as outlined in Refs.~\cite{vanKampen2007StochasticChemistry,Cugliandolo2015StochasticDipoles}. 
Beginning with Eq.~\eqref{seq:qdot}, the Stratonovich interpreted time evolution of an arbitrary observable $\mathcal{O}$ is given by:
\begin{align}
    \frac{\partial}{\partial t}\mathcal{O} =  \sum_{i=1}^N\left[\left(U_o \mathbf{q}_i + \frac{1}{\zeta}\sum_{j\neq i}^N\mathbf{F}_{ij}\right)\cdot\frac{\partial \mathcal{O}}{\partial \mathbf{r}_i} + \left(\mathbf{q}_i\times \bm{\Omega}_i\right)\cdot\frac{\partial \mathcal{O}}{\partial \mathbf{q}_i} \right].
    \label{seq:stratomicro}
\end{align}
Comparison with the It\^{o} description [Eq.~\eqref{seq:langito}], the rotary diffusion term is absent. 
To make use of the connection to the Fokker-Planck adjoint operator, it is convenient to recover this diffusion operator. 
We therefore solve for a Langevin equation equivalent to Eq.~\eqref{seq:stratomicro} but interpreted in the It\^{o} convention, which will contain the desired diffusion term. 
We do so by solving for the Kramers-Moyal expansion coefficients, where the $k$th coefficient is given by:
\begin{align}
    \lim_{\Delta t \to 0}\frac{\langle \left(\Delta \mathcal{O}\right)^k \rangle}{\Delta t},
\end{align}
where $\langle \cdot\cdot\cdot \rangle$ denotes an expectation over the distribution of the stochastic angular velocities and we have defined:
\begin{align}
    \Delta \mathcal{O} \equiv \mathcal{O}(t+\Delta t) - \mathcal{O}(t) = \int_{t}^{t+\Delta t}d\tau \frac{\partial \mathcal{O}}{\partial \tau}.
\end{align}
As outlined in Refs.~\cite{vanKampen2007StochasticChemistry,Cugliandolo2015StochasticDipoles}, a Langevin equation for $\mathcal{O}$ in the It\^{o} convention has deterministic terms given by the $k=1$ Kramers-Moyal expansion coefficient and a stochastic term with variance given by the $k=2$ coefficient.
We solve for the $k=1$ expansion coefficient:
\begin{align}
    \lim_{\Delta t \to 0}\frac{\langle \Delta \mathcal{O} \rangle}{\Delta t} = \sum_{i=1}^{N}\left[\dot{\mathbf{r}_i}\cdot\frac{\partial\mathcal{O}}{\partial \mathbf{r}_i} + D_{R}\boldsymbol{\nabla}_{\mathbf{q}_i}^2\mathcal{O}\right],
\end{align}
where the term proportional to the rotational Laplace operator arises due to the noise averaging of the stochastic term.
The $k=2$ terms are found to be:
\begin{align}
    \lim_{\Delta t \to 0}\frac{\langle \left(\Delta \mathcal{O} \right)^2\rangle}{\Delta t} = &\frac{1}{\Delta t}\left\langle\left[\int_{t}^{t+\Delta t}dt'\sum_{i=1}^{N}\dot{\mathbf{r}_i}\cdot\frac{\partial \mathcal{O}}{\partial \mathbf{r}_i}\right]^2\right\rangle \nonumber \\ &+\frac{2}{\Delta t}\int_{t}^{t+\Delta t}dt'\int_{t}^{t+\Delta t}dt''\left\langle \sum_{i=1}^{N}\sum_{j=1}^{N}\dot{\mathbf{r}}_i\cdot\frac{\partial\mathcal{O}}{\partial \mathbf{r}_i}(\mathbf{q}_j\times\bm{\Omega_j})\cdot\frac{\partial \mathcal{O}}{\partial \mathbf{q}_j} \right\rangle \nonumber \\ & + \frac{1}{\Delta t}\int_{t}^{t+\Delta t}dt'\int_{t}^{t+\Delta t}dt''\left\langle\sum_{i=1}^{N}\sum_{j=1}^{N} (\mathbf{q}_i\times\bm{\Omega_i})\cdot\frac{\partial \mathcal{O}}{\partial \mathbf{q}_i}(\mathbf{q}_j\times\bm{\Omega_j})\cdot\frac{\partial \mathcal{O}}{\partial \mathbf{q}_j}\right\rangle.
    \label{seq:kmcoeff2}
\end{align}
As argued by van Kampen~\cite{vanKampen2007StochasticChemistry}, the only non-negligible term in Eq.~\eqref{seq:kmcoeff2} is the last line. 
Thus we may write the Langevin equation for $\mathcal{O}$ as
\begin{equation}
    \frac{\partial \mathcal{O}}{\partial t} = \sum_{i=1}^{N}\left[\dot{\mathbf{r}_i}\cdot\frac{\partial\mathcal{O}}{\partial \mathbf{r}_i} + D_{R}\boldsymbol{\nabla}_{\mathbf{q}_i}^2\mathcal{O}\right] + \eta^{\mathcal{O}}(\mathbf{r},\mathbf{q},t),
    \label{seq:lang1}
\end{equation}
where we have defined $\eta^{\mathcal{O}}$ as a noise with zero mean and correlations:
\begin{equation}
    \langle \eta^{\mathcal{O}}(\mathbf{r},\mathbf{q},t)\eta^{\mathcal{O}}(\mathbf{r}',\mathbf{q}',t') \rangle = \left\langle\sum_{i=1}^{N}\sum_{j=1}^{N} (\mathbf{q}_i\times\bm{\Omega_i})\cdot\frac{\partial \mathcal{O}}{\partial \mathbf{q}_i}(\mathbf{q}_j\times\bm{\Omega_j})\cdot\frac{\partial \mathcal{O}}{\partial \mathbf{q}_j}\right\rangle.
\end{equation}
We may therefore equivalently express $\eta^{\mathcal{O}}$ as:
\begin{equation}
    \eta^{\mathcal{O}}(\mathbf{r},\mathbf{q},t) = \sum_{i=1}^{N}(\mathbf{q}_i\times\bm{\Omega}_i)\cdot\frac{\partial \mathcal{O}}{\partial q_i}.
    \label{seq:etaostat}
\end{equation}
Substitution of Eq.~\eqref{seq:etaostat} into Eq.~\eqref{seq:lang1} results in:
\begin{equation}
    \frac{\partial \mathcal{O}}{\partial t} = \sum_{i=1}^{N}\left[\dot{\mathbf{r}_i}\cdot\frac{\partial\mathcal{O}}{\partial \mathbf{r}_i} + D_{R}\boldsymbol{\nabla}_{\mathbf{q}_i}^2\mathcal{O} + \left(\mathbf{q}_i\times\bm{\Omega}_i\right) \cdot \frac{\partial \mathcal{O}}{\partial \mathbf{q}_i}\right] .
    \label{seq:lang2}
\end{equation}
Eq.~\eqref{seq:lang2} is identical to Eq.~\eqref{seq:langito} and both are interpreted in the It\^{o} sense.  \newpage
\section{\label{sec:interface_langevin}Langevin Dynamics of an Active Interface}

In this Section, we introduce the ansatz proposed by Bray~\textit{et al.}~\cite{Bray2002InterfaceShear,Bray1994TheoryKinetics} to connect the stochastic density field dynamics to a Langevin equation for the interfacial height field up to linear order in the height. 
From this Langevin equation, we define a capillary-wave tension as the coefficient of the linear term.
This capillary tension is defined such that it has dimensions of energy per area (length) in 3d (2d) and correctly accounts for the effects of nonconservative forces on height fluctuations, in contrast to the mechanical tension of Kirkwood and Buff~\cite{Kirkwood1949}.
The athermal noise derived in our fluctuating hydrodynamics [see Eq.~\eqref{seq:etaactvar}] breaks detailed balance and will thus result in height field noise statistics that generally violate the fluctuation dissipation theorem (FDT). 
We derive these noise statistics in detail and discuss the limit of when the FDT is effectively satisfied in Section~\ref{sec:isoerror} in addition to the main text. 
Using the derived interfacial equation of motion and noise statistics, we then solve for the stationary height fluctuations $\langle |h(k)|^2 \rangle$ and the capillary relaxation time scale. 
A dimensional analysis of all derived terms is subsequently given for reference. 
Finally, we demonstrate the conditions required for an interfacial equation to recover a Boltzmann distribution related to the area of the interface.

In the remainder of this Supplemental Material, we will use a prime ($'$) symbol to denote two distinct operations.  
If a prime follows an integration variable, e.g., $\int \text{d}z'$, the prime is simply meant to indicate that the variable is a dummy variable. 
If the prime follows a function, then the prime is meant to indicate a derivative with respect to the argument of that function (e.g., $a'(\rho) = \partial a(\rho)/\partial \rho, \upvarphi'(z) = \partial \upvarphi/\partial z$).

\subsection{\label{sec:heightfieldevo}Height Field Evolution}
We transform the density evolution to a height field evolution using the ansatz proposed by  Bray~\textit{et al.}, $\rho(\mathbf{r},t) = \upvarphi(z - h(\mathbf{x},t))$~\cite{Bray2002InterfaceShear},
where $\upvarphi$ is the noise-averaged \textit{stationary} density field.
We assume that the conditions for a planar phase-separated state are met such that $\upvarphi(z)$ is simply a function of $z$, the normal direction to the interface. 
Furthermore, $\upvarphi(z)$ is spatially constant (at the binodal densities) at nearly all points except for within the interface, where the density transitions between the two binodal densities. 
The ansatz implies the following chain-rule relations:
\begin{subequations}
\begin{align}
    \frac{\partial \rho}{\partial t} &= -\upvarphi'\frac{\partial h}{\partial t}\label{seq:brayevolve} \\ |\boldsymbol{\nabla} \rho|^2 &= \left(1 + |\nabla_x h|^2\right)(\upvarphi')^2 \\ \nabla^2\rho &= \left(1 +|\nabla_xh|^2\right)\upvarphi'' - \nabla_x^2 h\upvarphi',
\end{align}
\end{subequations}
where we denote the gradient in all directions other than $z$ with $\nabla_x$.
Substitution of the continuity equation and our derived constitutive equation for the density flux [Eq.\eqref{seq:abphydro}] into Eq.~\eqref{seq:brayevolve} results in:
\begin{align}
    -\upvarphi'\frac{\partial h}{\partial t} &= -\frac{1}{\zeta}\boldsymbol{\nabla} \cdot\boldsymbol{\nabla}\cdot \bm{\Sigma} - \boldsymbol{\nabla}\cdot\bm{\eta}^{\rm act}.\label{seq:doubledivmotiv}
\end{align}
We now express the double divergence of the dynamic stress using the above chain rule relations,
\begin{align}
    \boldsymbol{\nabla}\cdot\boldsymbol{\nabla}\cdot\bm{\Sigma} = &\nabla^2\bigl[-\mathcal{P}(\upvarphi) + \left(1 +|\nabla_xh|^2\right)a(\upvarphi)\upvarphi'' -\nabla_x^2 h a(\upvarphi)\upvarphi' \bigr] + \boldsymbol{\nabla}\cdot\boldsymbol{\nabla}\cdot\left(b(\rho)\nabla\rho\nabla\rho\right),
    \label{seq:divdivstressinit}
\end{align}
where $\mathcal{P}\equiv p_{C} + p_{\rm act}$ is the dynamic pressure. 
It is straightforward to show:
\begin{align}
    \boldsymbol{\nabla}\cdot\boldsymbol{\nabla} \cdot \left(b(\rho)\boldsymbol{\nabla}\rho \boldsymbol{\nabla}\rho\right) = -\frac{\partial}{\partial z}\left[b(\upvarphi)(\upvarphi')^2\right]\nabla_x^2 h + \nabla^2\left[b(\upvarphi)(\upvarphi')^2\left(1 + |\nabla_xh|^2\right)\right]. \label{seq:doubledivbterm}
\end{align}
Substituting Eq.~\eqref{seq:doubledivbterm} into Eq.~\eqref{seq:divdivstressinit} gives
\begin{align}
    \boldsymbol{\nabla}\cdot\boldsymbol{\nabla}\cdot\bm{\Sigma} =  &\nabla^2\bigl[-\mathcal{P}(\upvarphi)-\nabla_x^2 h a(\upvarphi)\upvarphi'  + \left(1 +|\nabla_xh|^2\right)\left(a(\upvarphi)\upvarphi'' + b(\upvarphi)(\upvarphi')^2\right) \bigr] \nonumber \\ &-\frac{\partial}{\partial z}\left[b(\upvarphi)(\upvarphi')^2\right]\nabla_x^2 h.
\end{align}
The height evolution is then given by
\begin{align}
    -\zeta\upvarphi'\frac{\partial h}{\partial t} =&-\nabla^2\bigl[-\mathcal{P}(\upvarphi)-\nabla_x^2 h a(\upvarphi)\upvarphi' + \left(1 +|\nabla_xh|^2\right)\left(a(\upvarphi)\upvarphi'' + b(\upvarphi)(\upvarphi')^2\right) \bigr] \nonumber \\ &+\frac{\partial}{\partial z}\left[b(\upvarphi)(\upvarphi')^2\right]\nabla_x^2 h - \zeta \boldsymbol{\nabla}\cdot\bm{\eta}^{\rm{act}}.
    \label{seq:hevoinit}
\end{align}

Throughout the SM and the main text, we will denote the continuous Fourier transform of $h(\mathbf{x},t)$ as $h(\mathbf{k},t)$ (dimensions of $\left[\rm{length}\right]^d$), defined as
\begin{align}
    h(\mathbf{k},t) = \int d\mathbf{x} e^{i\mathbf{k}\cdot\mathbf{x}}h(\mathbf{x},t),
\end{align}
while we will denote the discrete (fast) Fourier transform as $\Tilde{h}(\mathbf{k},t)$ (dimensions of $\left[\rm{length}\right]$).
The continuous Fourier transform of Eq.~\eqref{seq:hevoinit} in the $\mathbf{x}$ direction (i.e. all directions except $z$) results in
\begin{align}
    -\zeta\upvarphi'\frac{\partial h}{\partial t} =&-\left[\frac{\partial^2}{\partial z^2 }- k^2\right]\bigl[-\mathcal{P}(\upvarphi)\delta(k)+k^2 h a(\upvarphi)\upvarphi'  + \mathcal{F}\left[1 +|\nabla_xh|^2\right]\left(a(\upvarphi)\upvarphi'' + b(\upvarphi)(\upvarphi')^2\right) \bigr] \nonumber \\ &-\frac{\partial}{\partial z}\left[b(\upvarphi)(\upvarphi')^2\right]|k^2 h - \zeta\mathcal{F}\left[\boldsymbol{\nabla}\cdot\bm{\eta}^{\rm{act}}(\mathbf{r},t)\right],
    \label{seq:heightevofourier}
\end{align}
where we have defined the wavevector magnitude as $k = |\mathbf{k}|$ and the Fourier transform of $f(\mathbf{x)}$ as $\mathcal{F}\left[f(\mathbf{x})\right]$. 
The statistics of the Fourier transformed noise, as well as all subsequent transformations to the noise, will be derived in Section~\ref{sec:noise_stat}.

The Green's function of a Laplace operator that has been Fourier transformed in all directions but the $z$-direction is defined by~\cite{Arfken2012MathematicalGuide}:
\begin{align}
    \underbrace{\left[\frac{\partial^2}{\partial z^2 }- k^2\right]}_{\substack{\text{Laplacian}}} \underbrace{\left[\frac{1}{2k}e^{-k|z-z'|}\right]}_{\substack{\text{Green's Function}}} = -\delta(z-z').
\end{align}
We thus multiply Eq.~\eqref{seq:heightevofourier} by the Green's function and integrate across all $z$ to find:
\begin{align}
    \zeta\frac{\partial h}{\partial t} \int \text{d}z' &e^{-k|z-z'|}\upvarphi'(z')= -2k^3 h a(\upvarphi)\upvarphi' - 2k\mathcal{F}\left[|\nabla_xh|^2\right]\left(a(\upvarphi)\upvarphi'' + b(\upvarphi)(\upvarphi')^2\right)  \nonumber \\ &+k^2 h \int \text{d}z' e^{-k||z-z'|}\frac{\partial}{\partial z'}\left[b(\upvarphi)(\upvarphi')^2\right]|  +\zeta\int \text{d}z' e^{-k|z-z'|}\mathcal{F}\left[\boldsymbol{\nabla}\cdot\bm{\eta}^{\rm{act}}(\mathbf{r},t)\right],
    \label{seq:hevogf}
\end{align}
where we have discarded terms proportional to $k\delta(k)=0$. 
At this point, Eq.~\eqref{seq:hevogf} retains a dependence on the $z$-dimension and contains nonlinear terms (i.e., $\mathcal{F}\left[|\nabla_xh|^2\right]$) that are outside of the scope of the ansatz.
Fausti~\textit{et al.}~\cite{Fausti2021CapillarySeparation} recognized that both the $z$-dependence and the nonlinear terms can be eliminated by multiplying Eq.~\eqref{seq:hevogf} by the $z$-derivative of a density-dependent pseudovariable $\mathcal{E}$ and integrating across the $z$-dimension.
Under the ansatz, this pseudovariable coincides with the same pseudovariable used in equal-area constructions to determine nonequilibrium binodals~\cite{Aifantis1983TheRule, Solon2018GeneralizedMatter, Omar2023b}. 
Here $\mathcal{E}$, first introduced by Aifantis and Serrin~\cite{Aifantis1983TheRule}, is defined by:
\begin{align}
    \frac{\partial^2\mathcal{E}}{\partial \rho^2} = \frac{2b(\rho) - a'(\rho)}{a(\rho)}\frac{\partial \mathcal{E}}{\partial \rho}.
    \label{seq:pseudodef}
\end{align}
We therefore multiply Eq.~\eqref{seq:hevogf} by the $z$-derivative of the pseudovariable, $\mathcal{E}$, that depends \textit{solely} on density (and is nonzero only within the interface) and integrate over all $u = z- h$. 
We note that while integrating over $u$, the error incurred by approximating this integral as solely over $z$ (i.e., neglecting $h$) has been shown to introduce error on the order of $k^2h^3$~\cite{Bray2002InterfaceShear}, which is higher order than the error introduced by using the ansatz. 
Such a treatment also implies that we may freely switch between  $\upvarphi$ and $\rho$ in all integrals with respect to $u$.
The physical interpretation of multiplying by the pseudovariable is further discussed in Section~\ref{sec:pseudoeffect}.
Use of the pseudovariable results in:
\begin{align}
    \int \text{d}u \frac{\partial \mathcal{E}}{\partial u}\left[a(\upvarphi)\upvarphi'' + b(\upvarphi)(\upvarphi')^2\right] = 0.\label{seq:pseudobyparts}
\end{align}
Up to a constant of integration, $\partial\mathcal{E}/\partial z$ can be solved for using the definitions of $b(\upvarphi)$ and $a(\upvarphi)$ as
\begin{align}
    \frac{\partial \mathcal{E}}{\partial z} = \mathcal{C}\frac{\partial \rho}{\partial z}\frac{\left(p'_C(\rho)\overline{U}(\rho)\right)^2}{a(\rho)} \thicksim \frac{\partial \rho}{\partial z}\frac{\partial p_C}{\partial \rho},
\end{align}
which is in agreement with the pseudovariable for active Brownian particles found in Ref.~\cite{Omar2023b}. 
We note that in the equilibrium limit, $\ell_o/d_{\rm{hs}}\to 0$, the coefficients associated with Eq.~\eqref{seq:pseudodef} are dictated by the Korteweg expansion~\cite{Korteweg1904} and the resulting solution becomes $\mathcal{E} \thicksim 1/\rho$. 

Multiplication of Eq.~\eqref{seq:hevogf} by $\partial\mathcal{E}/\partial u$ and integrating over all $u = z-h$ results in:
\begin{align}
    \zeta A(k)\frac{\partial h}{\partial t} = &-2k^3 h \int \text{d} u a(\upvarphi)\upvarphi'\frac{\partial \mathcal{E}}{\partial u} +k^2 h \int \text{d}u \frac{\partial \mathcal{E}}{\partial u}\int\text{d}z' e^{-k|u-z'|}\frac{\partial}{\partial z'}\left[b(\upvarphi)(\upvarphi')^2\right] \nonumber \\ &+\zeta\int \text{d}u \frac{\partial \mathcal{E}}{\partial u}\int \text{d}z' e^{-k|u-z'|}\mathcal{F}\left[\boldsymbol{\nabla}\cdot\bm{\eta}^{\rm{act}}(\mathbf{r},t)\right]\label{seq:firstafterpseudo},
\end{align}
where $A(k)$ is defined as
\begin{align}
    A(k) = \int \int \text{d}u\text{d}z e^{-k|u-z|}\frac{\partial \mathcal{E}}{\partial u}(u)\upvarphi'(z).
\end{align}

Consider the second integral on the RHS of Eq.~\eqref{seq:firstafterpseudo}. 
Integrating by parts gives
\begin{align}
    \int\text{d}z' e^{-k|u-z'|}\frac{\partial}{\partial z'}\left[b(\upvarphi)(\upvarphi')^2\right] = -\int \text{d}z' k\text{sgn}(u-z')e^{-k|u-z'|}b(\upvarphi)(\upvarphi')^2,
\end{align}
where $\rm{sgn}$ denotes the sign function which returns $1$ ($-1$) for any positive (negative) argument.
Then the equation of motion for the height field simplifies to 
\begin{align}
    \zeta_{\rm{eff}} \frac{\partial h}{\partial t}  &= -k^3\upgamma_{\rm{cw}} h + \chi(k,t) + O\left(k^3h^2\right),
    \label{seq:interfacelangevin}  
\end{align}
where the error is due to both the ansatz and approximating integrals with respect to $u$ as if they were with respect to $z$. 
The effective drag coefficient is defined as:
\begin{align}
    \zeta_{\rm{eff}} = \frac{\zeta A^2(k)}{2B(k)\rho^{\rm surf}}.
\end{align}
$\upgamma_{\rm{cw}}$ is the capillary-wave tension:
\begin{align}
    \upgamma_{\rm{cw}}(k) = \frac{A(k)}{
    \rho^{\rm surf}B(k)}\biggl[\int_{-\infty}^{\infty}\text{d}u \frac{\partial \mathcal{E}}{\partial u} \upvarphi' a(\upvarphi)  + \int_{-\infty}^{\infty}\text{d}u \frac{\partial \mathcal{E}}{\partial u}\int_{-\infty}^{\infty} \text{d}z'\text{sgn}(u-z')e^{-k|u-z'|}b(\upvarphi)(\upvarphi')^2 \biggr],
\end{align}
where $\rho^{\rm surf} \equiv \left(\rho^{\rm liq} + \rho^{\rm gas}\right)/2$.
The noise $\chi$ has the form:
\begin{align}
    \chi(k,t) = \frac{A(k)\zeta }{B(k)2\rho^{\rm surf}} \int \text{d}u \frac{\partial \mathcal{E}}{\partial u}\int \text{d}z' e^{-k|u-z'|}\mathcal{F}\left[\nabla\cdot\bm{\eta}^{\rm act}\right],
    \label{seq:chidef}
\end{align}
where the function $B(k)$ is defined as:
\begin{align}
    B(k) = \int \int \text{d}u\text{d}u' e^{-k|u-u'|}\frac{\partial \mathcal{E}}{\partial u}(u)\frac{\partial \mathcal{E}}{\partial u'}(u').
\end{align}
The emergence of $\rho^{\rm surf}$ and $B(k)$ occurs when solving for the statistics of $\chi$, as detailed in the next Subsection.

\subsection{\label{sec:noise_stat}Noise Statistics}
We first solve for the statistics of $\xi(\mathbf{k},z,t) = \mathcal{F}\left[\boldsymbol{\nabla}\cdot\bm{\eta}^{\rm{act}}(\mathbf{r},t)\right]$. 
The variance of $\xi$ is given by
\begin{align}
    \langle \xi(\mathbf{k},z,t) \xi (\mathbf{k}',z',t')\rangle =  \int \int \text{d}\mathbf{x}\text{d}\mathbf{x}' e^{i\mathbf{x}\cdot\mathbf{k}}e^{i\mathbf{x}'\cdot\mathbf{k}'} \boldsymbol{\nabla}\cdot \boldsymbol{\nabla}'\cdot \langle\bm{\eta}^{\rm{act}}(\mathbf{r},t)\bm{\eta}^{\rm{act}}(\mathbf{r}',t') \rangle. \label{seq:xivarinit}
\end{align}
Then by substituting Eq.~\eqref{seq:etaactvar} into Eq.~\eqref{seq:xivarinit} we find:
\begin{align}
    \langle \xi (\mathbf{k},z,t) &\xi (\mathbf{k}',z',t')\rangle = \nonumber \\ &-2\frac{k_BT^{\rm{act}}}{\zeta}\delta(t-t')\int \int\text{d}\mathbf{x}\text{d}\mathbf{x}'  e^{i\mathbf{x}\cdot\mathbf{k}}e^{i\mathbf{x}'\cdot\mathbf{k}'}   \boldsymbol{\nabla}\cdot  \left(\left(\rho\mathbf{I} - \frac{d}{d-1}\mathbf{Q}'\right)\cdot\boldsymbol{\nabla}\delta(\mathbf{r}-\mathbf{r}')\right).
\end{align}
Using Eq.~\eqref{seq:tracenematicexpand}, and the definitions of $a(\rho)$ and $b(\rho)$,
\begin{align}
    \left(\rho\mathbf{I} - \frac{d}{d-1}\mathbf{Q}'\right) = \rho\mathbf{I}+ \frac{d}{\ell_o^2 D_R \zeta \overline{U}(\rho)}\left[a(\rho)\nabla^2\rho \mathbf{I} + b(\rho)\boldsymbol{\nabla}\rho\boldsymbol{\nabla}\rho\right].
    \label{seq:tracenemconst}
\end{align}
Then the statistics of $\xi$ can be split into three contributions
\begin{align}
    \langle \xi(\mathbf{k},z,t) \xi (\mathbf{k}',z',t')\rangle = \sum_{i=1}^3\langle \xi(\mathbf{k},z,t) \xi (\mathbf{k}',z',t')\rangle_i,
\end{align}
with the three contributions defined as:
\begin{subequations}
    \begin{align}
        \langle \xi (\mathbf{k},z,t) \xi (\mathbf{k}',z',t') \rangle_1 = &-2\frac{k_BT^{\rm{act}}}{\zeta}\delta(t-t')\int\int\text{d}\mathbf{x}\text{d}\mathbf{x}' e^{i\mathbf{x}\cdot\mathbf{k}}e^{i\mathbf{x}'\cdot\mathbf{k}}\boldsymbol{\nabla}\cdot\left(\rho \boldsymbol{\nabla}\delta(\mathbf{r}-\mathbf{r}')\right), \label{seq:xione} \\ \langle \xi (\mathbf{k},z,t) \xi (\mathbf{k}',z',t') \rangle_2 = &-\frac{2k_BT^{\rm{act}}d}{\ell_o^2 \zeta^2 D_R }\delta(t-t')\int\int\text{d}\mathbf{x}\text{d}\mathbf{x}' \nonumber \\ & \times e^{i\mathbf{x}\cdot\mathbf{k}}e^{i\mathbf{x}'\cdot\mathbf{k}}\boldsymbol{\nabla}\cdot\left(\frac{1}{\overline{U}(\rho)}a(\rho)\nabla^2\rho \boldsymbol{\nabla}\delta(\mathbf{r}-\mathbf{r}')\right), \label{seq:xitwo} \\ \langle \xi (\mathbf{k},z,t) \xi (\mathbf{k}',z',t') \rangle_3 = &-\frac{2k_BT^{\rm{act}}d}{\ell_o^2 \zeta^2 D_R }\delta(t-t')\int\int\text{d}\mathbf{x}\text{d}\mathbf{x}'\nonumber \\ &\times  e^{i\mathbf{x}\cdot\mathbf{k}}e^{i\mathbf{x}'\cdot\mathbf{k}}\boldsymbol{\nabla}\cdot\left(\frac{1}{\overline{U}(\rho)}b(\rho)\boldsymbol{\nabla}\rho\boldsymbol{\nabla}\rho\cdot \boldsymbol{\nabla}\delta(\mathbf{r}-\mathbf{r}')\right).\label{seq:xithree} 
    \end{align}
\end{subequations}
We evaluate these terms separately to lowest order in $k$, beginning with the first. 
Consider the term:
\begin{align}
    \boldsymbol{\nabla}\cdot \left(\rho\boldsymbol{\nabla}\delta(\mathbf{r}-\mathbf{r}') \right) = \frac{\partial }{\partial z}\left(\rho \frac{\partial}{\partial z}\delta(\mathbf{r}-\mathbf{r}')\right) + \nabla_x\cdot\left(\rho\nabla_x\delta(\mathbf{r}-\mathbf{r}')\right).
    \label{seq:graddeltas}
\end{align}
Upon substitution of Eq.~\eqref{seq:graddeltas} into Eq.~\eqref{seq:xione}, and ignoring the dependence of $\upvarphi$ on $h$, we integrate by parts to find that
\begin{align}
    \langle \xi (\mathbf{k},z,t) \xi (\mathbf{k}',z',t') \rangle_1 = &2\frac{k_BT^{\rm{act}}}{\zeta}(2\pi)^{d-1}\delta(\mathbf{k} + \mathbf{k}')\delta(t-t')\nonumber \\ & \times\left(k^2\upvarphi\delta(z-z') - \frac{\partial}{\partial z}\left(\upvarphi\frac{\partial}{\partial z}\delta(z-z')\right)\right).
\end{align}
The same lines of argument may be used for Eq.~\eqref{seq:xitwo} to show:
\begin{align}
    \langle \xi (\mathbf{k},z,t) \xi (\mathbf{k}',z',t') \rangle_2 = &\frac{2k_BT^{\rm{act}}d}{\ell_o^2 \zeta^2 D_R }(2\pi)^{d-1}\delta(\mathbf{k} + \mathbf{k}')\delta(t-t')\nonumber \\ &\times\left(\frac{k^2a(\upvarphi)}{\overline{U}(\upvarphi)}\upvarphi''\delta(z-z') - \frac{\partial}{\partial z}\left(\frac{a(\upvarphi)\upvarphi''}{\overline{U}(\upvarphi)}\frac{\partial}{\partial z}\delta(z-z')\right)\right).
\end{align}
The lowest order in $k$ contribution from Eq.~\eqref{seq:xithree} will be the term where all derivatives are in the $z$-direction, giving
\begin{align}
    \langle \xi (\mathbf{k},z,t) \xi &(\mathbf{k}',z',t') \rangle_3 = -\frac{2k_BT^{\rm{act}}d}{\ell_o^2 \zeta^2 D_R }(2\pi)^{d-1}\delta(\mathbf{k} + \mathbf{k}')\delta(t-t')\frac{\partial}{\partial z}\left(\frac{b(\rho)}{\overline{U}(\rho)}(\upvarphi')^2\frac{\partial}{\partial z}\delta(z-z')\right).
\end{align}
Now we introduce $L(\mathbf{k},t)$:
\begin{align}
    L(\mathbf{k},t) = \int \text{d}u \frac{\partial \mathcal{E}}{\partial u} \int \text{d}z' e^{-k|u-z'|}\xi(\mathbf{k},z',t).
\end{align}
The variance of $L$ is given by:
\begin{align}
    \langle L(\mathbf{k},z,t)L(\mathbf{k}',z',t') \rangle = &\int \text{d}u \frac{\partial \mathcal{E}}{\partial u} \int \text{d}u'\frac{\partial \mathcal{E}}{\partial u'} \int\int \text{d}z'\text{d}z''  e^{-k|u-z'|}e^{-k|u'-z''|}\langle\xi(\mathbf{k},z',t)\xi(\mathbf{k}',z'',t') \rangle.
\end{align}
Then, as with $\xi$, we split up the variance of $L$ into three contributions,
\begin{subequations}
    \begin{align}
    \langle L(\mathbf{k},t)L(\mathbf{k}',t') \rangle_1 = \int \text{d}u \frac{\partial \mathcal{E}}{\partial u} \int \text{d}u'\frac{\partial \mathcal{E}}{\partial u'} \int\int \text{d}z'\text{d}z''   e^{-k|u-z'|}e^{-k|u'-z''|}\langle\xi(\mathbf{k},z',t)\xi(\mathbf{k}',z'',t') \rangle_1, \label{seq:chione} \\ \langle L(\mathbf{k},t)L(\mathbf{k}',t') \rangle_2 = \int \text{d}u \frac{\partial \mathcal{E}}{\partial u} \int \text{d}u' \frac{\partial \mathcal{E}}{\partial u'}\int\int \text{d}z'\text{d}z''  e^{-k|u-z'|}e^{-k|u'-z''|}\langle\xi(\mathbf{k},z',t)\xi(\mathbf{k}',z'',t') \rangle_2, \label{seq:chitwo} \\ \langle L(\mathbf{k},t)L(\mathbf{k}',t') \rangle_3 = \int \text{d}u \frac{\partial \mathcal{E}}{\partial u} \int \text{d}u' \frac{\partial \mathcal{E}}{\partial u'}\int\int \text{d}z'\text{d}z''  e^{-k|u-z'|}e^{-k|u'-z''|}\langle\xi(\mathbf{k},z',t)\xi(\mathbf{k}',z'',t') \rangle_3.\label{seq:chithree} 
\end{align}
\end{subequations}
We begin with the first contribution, Eq.~\eqref{seq:chione}:
\begin{align}
   \langle L(\mathbf{k},t)L(\mathbf{k}',t') \rangle_1 = &2\frac{k_BT^{\rm{act}}}{\zeta}(2\pi)^{d-1}\delta(\mathbf{k} + \mathbf{k}') \delta(t-t')\int \text{d}u \frac{\partial \mathcal{E}}{\partial u} \int \text{d}u'\frac{\partial \mathcal{E}}{\partial u'} \int\int \text{d}z'\text{d}z''  \nonumber \\ &\times e^{-k|u-z'|}e^{-k|u'-z''|}\left(k^2\upvarphi\delta(z'-z'') - \frac{\partial}{\partial z'}\left(\upvarphi\frac{\partial}{\partial z'}\delta(z'-z'')\right)\right).
\end{align}
The first contribution may be separated into two subcontributions,
\begin{subequations}
    \begin{align}
        \langle L(\mathbf{k},t)L(\mathbf{k}',t') \rangle_1^a = &2\frac{k_BT^{\rm{act}}}{\zeta}(2\pi)^{d-1}\int \text{d}u \frac{\partial \mathcal{E}}{\partial u} \int \text{d}u'\frac{\partial \mathcal{E}}{\partial u'} \int\int \text{d}z'\text{d}z''  \nonumber \\ &\times e^{-k|u-z'|}e^{-k|u'-z''|} \delta(\mathbf{k} + \mathbf{k}')\delta(t-t')\left(k^2\upvarphi\delta(z'-z'')\right), \label{seq:acontribute}\\ \langle L(\mathbf{k},t)L(\mathbf{k}',t') \rangle_1^b = &- 2\frac{k_BT^{\rm{act}}}{\zeta}(2\pi)^{d-1} \int \text{d}u \frac{\partial \mathcal{E}}{\partial u} \int \text{d}u' \frac{\partial \mathcal{E}}{\partial u'}\int\int \text{d}z'\text{d}z''  \nonumber \\ &\times e^{-k|u-z'|}e^{-k|u'-z''|}\delta(\mathbf{k} + \mathbf{k}')\delta(t-t')\frac{\partial}{\partial z'}\left(\upvarphi\frac{\partial}{\partial z'}\delta(z'-z'')\right).\label{seq:bcontribute}
    \end{align}
\end{subequations}
In Eq.~\eqref{seq:acontribute}, the delta function immediately eliminates one of the integrals, leaving us with
\begin{align}
    \langle L(\mathbf{k},t)L(\mathbf{k}',t') \rangle_1^a = &\frac{2D_R(d-1)}{d}(2\pi)^{d-1}  \int \text{d}u \frac{\partial \mathcal{E}}{\partial u} \int \text{d}u' \frac{\partial \mathcal{E}}{\partial u'} \int \text{d}z' \nonumber \\ &\times e^{-k|u-z'|}e^{-k|u'-z'|} \delta(\mathbf{k} + \mathbf{k}')\delta(t-t')k^2\upvarphi.
\end{align}
Because $\partial\mathcal{E}/\partial z$ is zero everywhere except within the interface, we will approximate $\upvarphi$ as its value halfway between the binodal densities, i.e., $\upvarphi \approx (\rho^{\rm liq} + \rho^{\rm gas})/2 =\rho^{\rm surf}$. 
In addition, by splitting up the integral over $z'$ into its contributions when $z'$ is less than, greater than, or between $u,u'$ one can show that
\begin{align}
    \int \text{d}z'e^{-k|u-z'|}e^{-k|u'-z'|} = \frac{1}{k}e^{-k|u-u'|}.
\end{align}
As a result, Eq.~\eqref{seq:acontribute} simplifies to:
\begin{align}
      \langle L(\mathbf{k},t)L(\mathbf{k}',t') \rangle_1^a = \frac{2k_BT^{\rm{act}}\rho^{\rm surf}}{\zeta}(2\pi)^{d-1}k \nonumber \delta(\mathbf{k} + \mathbf{k}')\delta(t-t')B(k).
\end{align}
Next we simplify Eq.~\eqref{seq:bcontribute}. 
Integrating by part results in:
\begin{align}
     \langle L(\mathbf{k},t)L(\mathbf{k}',t') \rangle_1^b = &2\frac{k_BT^{\rm{act}}}{\zeta}(2\pi)^{d-1}\int \text{d}u \frac{\partial \mathcal{E}}{\partial u} \int \text{d}u' \frac{\partial \mathcal{E}}{\partial u'}\int\int \text{d}z'\text{d}z''   \nonumber \\ &\times e^{-k|u-z'|}e^{-k|u'-z''|}\text{sgn}(u-z')k\delta(\mathbf{k} + \mathbf{k}')\delta(t-t')\upvarphi\frac{\partial}{\partial z'}\delta(z'-z''),
\end{align}
followed by another integration by parts:
\begin{align}
     \langle L(\mathbf{k},t)L(\mathbf{k}',t') \rangle_1^b = &2\frac{k_BT^{\rm{act}}}{\zeta}(2\pi)^{d-1}\int \text{d}u \frac{\partial \mathcal{E}}{\partial u} \int \text{d}u' \frac{\partial \mathcal{E}}{\partial u'}\int \text{d}z'\nonumber \\ &\times  e^{-k|u-z'|}e^{-k|u'-z'|}  \text{sgn}(u-z')\text{sgn}(u'-z')k^2\delta(\mathbf{k} + \mathbf{k}')\delta(t-t')\upvarphi.
\end{align}
Because $u'$ and $u$ will only be evaluated where their magnitudes are very small, and $z'$ will go from negative infinity to infinity, we will approximate $\text{sgn}(u-z')\text{sgn}(u'-z')$ as one.
Then, via the same lines of argument, $b$ simplifies to the same contribution as $a$, 
\begin{align}
      \langle L(\mathbf{k},t)L(\mathbf{k}',t') \rangle_1^b = \frac{2k_BT^{\rm{act}}\rho^{\rm surf}}{\zeta}(2\pi)^{d-1}k \delta(\mathbf{k} + \mathbf{k}')\delta(t-t')B(k).
\end{align}
Which results in
\begin{align}
      \langle L(\mathbf{k},t)L(\mathbf{k}',t') \rangle_1 = \frac{4k_BT^{\rm{act}}\rho^{\rm surf}}{\zeta}(2\pi)^{d-1}k \delta(\mathbf{k} + \mathbf{k}')\delta(t-t')B(k).
\end{align}
For Eq.~\eqref{seq:chitwo}, it is convenient to define the quantity $C(k)$, 
\begin{align}
    C(k) = &\int \int \text{d}u \text{d}u' \frac{\partial \mathcal{E}}{\partial u}(u)\frac{\partial \mathcal{E}}{\partial u'}(u')\int\int \text{d}z' \text{d}z''\nonumber \\  \times &e^{-k|u-z'|}e^{-k|u'-z''|} \biggl(k^2 \frac{a(\rho)}{ \overline{U}} \upvarphi''(z) \delta(z-z') - \frac{\partial}{\partial z}\left( \frac{a(\rho)}{ \overline{U}} \upvarphi''(z) \frac{\partial}{\partial z}\delta(z-z')\right)\biggr).
    \label{seq:ceekayy}
\end{align}
Then Eq.~\eqref{seq:chitwo} can be rewritten as
\begin{align}
    \langle L(\mathbf{k},t)L(\mathbf{k}',t')\rangle_2  =\frac{2k_BT^{\rm{act}}d}{\ell_o^2 \zeta^2 D_R }\delta(t-t')\delta(\mathbf{k}+\mathbf{k}')(2\pi)^{(d-1)} C(k).
\end{align}
Similarly, Eq.~\eqref{seq:chithree} can be rewritten by defining the quantity $D(k)$,
\begin{align}
    D(k) = &\int\int \text{d}u \text{d}u'\frac{\partial \mathcal{E}}{\partial u}(u)\frac{\partial \mathcal{E}}{\partial u'}(u')\int\int \text{d}z'\text{d}z'' \nonumber \\ &\times e^{-k|u-z'|}e^{-k|u'-z''|}  \frac{\partial}{\partial z'}\left[\frac{b(\rho)}{\overline{U}}(\upvarphi'(z'))^2\frac{\partial}{\partial z'}\delta(z-z')\right],
    \label{seq:deekayy}
\end{align}
which results in
\begin{align}
    \langle L(\mathbf{k},t)L(\mathbf{k}',t')\rangle_3 = \frac{2k_BT^{\rm{act}}d}{\ell_o^2 \zeta^2 D_R }\delta(t-t')\delta(\mathbf{k}+\mathbf{k}')(2\pi)^{(d-1)}D(k).
\end{align}
Altogether, we find the full noise correlator as
\begin{align}
    \langle L(\mathbf{k},t)L(\mathbf{k}',t')\rangle = &2\frac{k_BT^{\rm{act}}}{\zeta}\delta(t-t')\delta(\mathbf{k}+\mathbf{k}')(2\pi)^{(d-1)}\nonumber \\ & \times\left[2\rho^{\rm surf}kB(k) + \frac{d}{\ell_o^2\zeta D_R}C(k) +  \frac{d}{\ell_o^2\zeta D_R}D(k)\right],
\end{align}
where $B(k)$, $C(k)$, and $D(k)$ are all to be evaluated numerically. 
Finally, we define $\chi(\mathbf{k},t)$ with the coefficients as defined in Eq.~\eqref{seq:chidef}. 
We then break up the noise into two independent contributions, $\chi = \chi^{\rm{iso}} + \chi^{\rm{aniso}}$ which have zero mean and variances:
\begin{subequations}
    \begin{align}
        \langle \chi^{\rm{iso}}(\mathbf{k},t)\chi^{\rm{iso}}(\mathbf{k}',t')\rangle &= 2k(k_BT)^{\rm{act}}\zeta_{\rm{eff}}\delta(t-t')\delta(\mathbf{k}+\mathbf{k}')(2\pi)^{(d-1)},\label{seq:chivariso} \\ \langle \chi^{\rm{aniso}}(\mathbf{k},t)\chi^{\rm{aniso}}(\mathbf{k}',t')\rangle &= \frac{\zeta_{\rm{eff}}(C(k) + D(k))}{(d-1)\rho^{\rm surf}B(k)} \delta(t-t')\delta(\mathbf{k}+\mathbf{k}')(2\pi)^{(d-1)}.
        \label{seq:chivaraniso}
    \end{align}
\end{subequations}
\subsection{\label{sec:capillarystats}Height Correlations, Relaxation Timescale, and Power Spectrum}
Now with the complete height-field Langevin equation, we proceed to evaluating the capillary fluctuations of the interface, i.e., evaluate $\langle h(\mathbf{k}) h(-\mathbf{k})\rangle$.
If we multiply Eq.~\eqref{seq:interfacelangevin} by $h(-\mathbf{k},t)$, average over the noise, and consider only the steady state, we find
\begin{align}
    k^3 \upgamma_{\rm{cw}} \langle h(\mathbf{k})h(-\mathbf{k}) \rangle = \langle\chi(\mathbf{k},t)h(-\mathbf{k},t)\rangle.
    \label{seq:initcap}
\end{align}
Determining the stationary fluctuations thus require evaluating the noise-averaged correlation of the height field with the noise. 
The implicit solution for $h$ is given by,
\begin{align}
    h(-\mathbf{k},t) = h(-\mathbf{k},0)\text{exp}\left(-\frac{t}{\tau(k)}\right) + \frac{1}{\zeta_{\rm eff}}\int_0^t\text{d}t'\text{exp}\left(-\frac{t-t'}{\tau(k)}\right)\chi(-\mathbf{k},t'),\label{seq:intfactors}
\end{align}
where the timescale for the relaxation of a capillary wave $\tau$ is given by 
\begin{align}
    \tau(k) \equiv \frac{\zeta_{\rm{eff}}}{k^3\upgamma_{\rm{cw}}}.
    \label{seq:relaxtime}
\end{align}
We now multiply Eq.~\eqref{seq:intfactors} by $\chi(\mathbf{k},t)$ and take an expectation:
\begin{align}
    \langle \chi(\mathbf{k},t)h(-\mathbf{k})\rangle = \frac{1}{\zeta_{\rm eff}}\int_0^t \text{d}t'\text{exp}\left(-\frac{t-t'}{\tau(k)}\right)\langle\chi(-\mathbf{k},t')\chi(\mathbf{k},t)\rangle.
\end{align}
We substitute in the statistics of $\chi$, Eq.~\eqref{seq:chivariso}, and Eq.~\eqref{seq:chivaraniso} to find
\begin{align}
    \langle \chi(\mathbf{k},t)h(-\mathbf{k})\rangle = &\frac{1}{\zeta_{\rm eff}}\int_0^t \text{d}t'\text{exp}\left(-\frac{t-t'}{\tau(k)}\right)\delta(t-t')L^{d-1}\nonumber \\ &\times\biggl[2\zeta_{\rm eff}k(k_BT^{\rm act}) + \frac{\zeta_{\rm eff}\left(C(k) + D(k)\right)}{(d-1)\rho^{\rm surf} B(k)} \biggr],
\end{align}
where we have noted that $(2\pi)^{d-1}\delta(\mathbf{k}=\mathbf{0}) = L^{d-1}$ for a system of finite size (in three dimensions this is the area projected by the interface onto the $(x,y)$ plane).
The stationary fluctuations can now be determined:
\begin{align}
    \langle h(\mathbf{k})h(-\mathbf{k}) \rangle = \frac{L^{(d-1)}}{\upgamma_{\rm{cw}}} \biggl[ \frac{k_BT^{\rm{act}}}{k^2} + \frac{C(k) + D(k)}{k^3 (d-1)2\rho^{\rm surf}B(k)}\biggr].
    \label{seq:capfluctsfinal}
\end{align}
As shown in Fig.~2 of the main text, the anisotropic contributions to the fluctuations can be safely ignored at low $k$ and low run length. 
Under these conditions we are left with
\begin{align}
    \langle h(\mathbf{k})h(-\mathbf{k}) \rangle = \frac{L^{(d-1)}k_BT^{\rm{act}}}{\upgamma_{\rm{cw}}k^2}.
    \label{seq:capfluctslowk}
\end{align}
We can also Fourier transform Eq.~\eqref{seq:interfacelangevin} in time to find:
\begin{equation}
    i\omega \zeta_{\rm eff} h(\mathbf{k},\omega) = -k^3\upgamma_{\rm cw}h(\mathbf{k},\omega) + \chi(\mathbf{k},\omega).
    \label{seq:interfacelangevinomega}
\end{equation}
After rearranging Eq.~\eqref{seq:interfacelangevinomega} and multiplying both sides of the equation by its own complex conjugate we find:
\begin{equation}
    h(\mathbf{k},\omega)h(-\mathbf{k},-\omega) = \frac{1}{\zeta_{\rm eff}^2\omega^2 + k^6\upgamma_{\rm cw}^2}\chi(\mathbf{k},\omega)\chi(-\mathbf{k},\omega).
    \label{seq:interfacelangevinomegamod}
\end{equation}
We can then average over the noise to extract the power spectra:
\begin{equation}
    \langle |h(\mathbf{k},\omega)|^2 \rangle = \frac{L^{d-1}\Gamma \zeta_{\rm eff}}{\zeta_{\rm eff}^2\omega^2 + k^6\upgamma_{\rm cw}^2} \left[2k(k_BT)^{\rm act} + \frac{C(k) + D(k)}{(d-1)\rho^{\rm surf}B(k)}\right],
\end{equation}
where we have used $2\pi\delta(\omega =0) = \Gamma$, the total time of a finite duration trajectory.
At low $k$ and low run length, we ignore the anisotropic contributions to the power spectra and find:
\begin{equation}
    \langle |h(k,\omega)|^2 \rangle = \frac{2kL^{d-1}\Gamma k_BT^{\rm act}}{\zeta_{\rm eff}\left(\omega^2 + \tau(k)^{-2}\right)}.
\end{equation}

\subsection{\label{sec:dimanal}Dimensional Analysis}
In order to make the dependencies and scaling of the low-$k$ fluctuations as explicit as possible in 3d, we pick a natural system of units with $d_{\rm{hs}}$ as unit length and $\zeta U_o$ as unit force. 
Then the non-dimensionalized run length is given by $\ell_o = \bar{\ell}_o d_{\rm{hs}}$, the non-dimensionalized wave vector is given by $k = \bar{k}d_{\rm{hs}}^{-1}$, the non-dimensionalized system length $L = \bar{L} d_{\rm{hs}}$, and the non-dimensionalized density $\rho = \bar{\rho}d_{\rm{hs}}^{-3}$. 
From these quantities we can non-dimensionalize more complicated quantities such as:
\begin{align}
    a(\rho) &= \frac{3\ell_o^2}{2 d(d-1)(d+2)}\overline{U}(\rho)\overline{U}(\rho) \frac{\partial p_C}{\partial \rho} \nonumber \\ &= \frac{3\bar{\ell}_o^2d_{\rm{hs}}^2}{2 d(d-1)(d+2)}\overline{U}(\rho)\overline{U}(\rho) \frac{\partial \bar{p}_C}{\partial \bar{\rho}} \frac{\zeta U_o}{d_{\rm{hs}}^2} d_{\rm{hs}}^3\nonumber \\ &= \bar{a}(\bar{\rho}) \zeta U_o d_{\rm{hs}}^3,
\end{align}
\begin{align}
    b(\rho) &= \frac{3\ell_o^2}{2 d(d-1)(d+2)}\overline{U}(\rho)\frac{\partial }{\partial \rho}\left[\overline{U}(\rho)\frac{\partial p_C}{\partial \rho}\right] \nonumber \\ &=\frac{3\bar{\ell}_o^2d_{\rm{hs}}^2}{2 d(d-1)(d+2)}\overline{U}(\rho)\frac{\partial }{\partial \rho}\left[\overline{U}(\rho)\frac{\partial \bar{p}_C}{\partial \bar{\rho}}\right]\frac{\zeta U_o}{d_{\rm{hs}}^2}d_{hs}^6 \nonumber \\ &=\bar{b}(\bar{\rho})\zeta U_o d_{\rm{hs}}^6 ,
\end{align}
\begin{align}
    A(k) &= \mathcal{C}\int \text{d}z_1 \upvarphi'\int \text{d}z_2 \frac{\left(p'_C \overline{U}\right)^2}{a(\upvarphi)}\upvarphi'e^{-k|z_1 - z_2|}\nonumber \\ &= \mathcal{C}\int \text{d}\bar{z}_1 \bar{\upvarphi}' d_{\rm{hs}}^{-3}\int \text{d}\bar{z}_2 \frac{\left(\bar{p}'_C \overline{U}\zeta U_o d_{\rm{hs}}\right)^2}{\bar{a}(\bar{\upvarphi})\zeta U_o \rm{d_{hs}^3}}  \bar{\upvarphi}'d_{\rm{hs}}^{-3}e^{-k|\bar{z}_1 - \bar{z}_2|}\nonumber \\ &= \bar{A}(k)\mathcal{C}\zeta U_o d_{\rm{hs}}^{-7},
\end{align}
\begin{align}
    B(k) &= \mathcal{C}^2\int \text{d}z_1 \upvarphi'\frac{\left(p'_C \overline{U}\right)^2}{a(\upvarphi)}\int \text{d}z_2 \frac{\left(p'_C \overline{U}\right)^2}{a(\upvarphi)}\upvarphi'e^{-k|z_1 - z_2|}\nonumber \\ &=\mathcal{C}^2\int \text{d}\bar{z}_1 \bar{\upvarphi}'d_{\rm{hs}}^{-3}\frac{\left(\bar{p}'_C \overline{U} \zeta U_o d_{\rm{hs}}\right)^2}{\bar{a}(\bar{\upvarphi})\zeta U_o d_{\rm{hs}}^3}\int \text{d}\bar{z}_2 \frac{\left(\bar{p}'_C \overline{U}\zeta U_o d_{\rm{hs}}\right)^2}{\bar{a}(\bar{\upvarphi})\zeta U_o d_{\rm{hs}}^3}\bar{\upvarphi}'d_{\rm{hs}}^{-3}e^{-k|z_1 - z_2|} \nonumber \\ & = \bar{B}(k) \mathcal{C}^2 \left(\zeta U_o\right)^2 d_{\rm{hs}}^{-8},
\end{align}
where $\mathcal{C}$ is an arbitrary constant of integration resulting from the definition of the pseudovariable $\mathcal{E}$ and
\begin{align}
    \upgamma_{\rm{cw}}(k) &= \frac{A(k)}{B(k)\rho^{\rm surf}}\biggl[\int_{-\infty}^{\infty}\text{d}u \upvarphi' \upvarphi'\left(\mathcal{E}p'_C\overline{U}\right)^2 \nonumber \\ &+ \int_{-\infty}^{\infty}\text{d}u \frac{\left(\mathcal{C}p'_C\overline{U}\right)^2}{a(\upvarphi)}\upvarphi'\int_{-\infty}^{\infty} \text{d}z'\text{sgn}(u-z')e^{-\bar{k}|u-z'|}b(\upvarphi)(\upvarphi')^2 \biggr]\nonumber \\ &= \frac{d_{\rm{hs}}^4}{\mathcal{C}\zeta U_o\bar{\rho}^{\rm surf}}\biggl[\int_{-\infty}^{\infty}\text{d}\bar{u} \bar{\upvarphi}' \bar{\upvarphi}'d_{\rm{hs}}^{-7}\left(\mathcal{C}\bar{p}'_C\overline{U}\zeta U_od_{\rm{hs}}\right)^2 \nonumber \\ &+ \int_{-\infty}^{\infty}\text{d}\bar{u} d_{\rm{hs}} \frac{\left(\mathcal{C}p'_C\overline{U} \zeta U_o d_{\rm{hs}}^3\right)^2}{\bar{a}(\bar{\upvarphi}) U_o d_{\rm{hs}}^{7}}\bar{\upvarphi}'d_{\rm{hs}}^{-4}\int_{-\infty}^{\infty} \text{d}\bar{z}'d_{\rm{hs}}\text{sgn}(\bar{u}-\bar{z}')e^{-\bar{k}|\bar{u}-\bar{z}'|}\bar{b}(\bar{\upvarphi}) U_o (\bar{\upvarphi}')^2\rm{d_{hs}^{-2}} \biggr],
\end{align}
Cancelling the like terms in the above expression results in a dimensionless capillary-wave tension:
\begin{align}
    \upgamma_{\rm{cw}}(k) = \bar{\upgamma}_{\rm{cw}}(k) \zeta U_o d_{\rm{hs}}^{-1},
\end{align}
and we can identify the dimensionless function $g(\lambda,\bar{k}) = \bar{\upgamma}_{\rm cw} = \upgamma_{\rm cw}d_{\rm hs}/\zeta U_o$, where $\lambda\equiv \ell_o/\ell_o^c - 1$, as
\begin{align}
    g(\lambda,k) = &\frac{1}{\bar{\rho}^{\rm surf}}\biggl[\int_{-\infty}^{\infty}\text{d}\bar{u} \bar{\upvarphi}' \bar{\upvarphi}'\left(\bar{p}'_C\overline{U}\right)^2 \nonumber \\ &+ \int_{-\infty}^{\infty}\text{d}\bar{u} d_{\rm{hs}} \frac{\left(p'_C\overline{U}  \right)^2}{\bar{a}(\bar{\upvarphi}) }\bar{\upvarphi}'\int_{-\infty}^{\infty} \text{d}\bar{z}'\text{sgn}(\bar{u}-\bar{z}')e^{-\bar{k}|\bar{u}-\bar{z}'|}\bar{b}(\bar{\upvarphi})(\bar{\upvarphi}')^2 \biggr], 
\end{align}
where the dependence on $\lambda$ is inherited from the dependence of the active speed $\overline{U}$ on activity.
The low-$k$ capillary fluctuations in $d=3$ should then be given by
\begin{align}
    \langle h(\mathbf{k})h(-\mathbf{k}) \rangle &=\frac{L^{2}\ell_oU_o\zeta}{6\upgamma_{\rm{cw}}k^2}\nonumber \\ &= \frac{d_{\rm{hs}}^{6}\bar{L}^2 \bar{\ell}_o}{6\bar{\upgamma}_{\rm{cw}}\bar{k}^2}.
\end{align}
A similar procedure can be carried out with $d=2$ in order to find the units of all quantities in two dimensions.
\subsection{\label{sec:fluctdiss}Area Minimization}
We wish to find a condition such that interfacial dynamics (at steady state) recover a Boltzmann distribution \textit{weighted by the interfacial area}. 
In other words,
\begin{equation}
    P\left[h\right] \sim \text{exp}\left[-C \int \text{d}\mathbf{x} \sqrt{1 + |\nabla_x h|^2}\right],
    \label{seq:boltzdistr}
\end{equation}
where $C$ is a physical constant.
For a weakly fluctuating interface where $|\nabla_xh|^2 << 1$, this area may be Taylor expanded as:
\begin{equation}
    P\left[h\right] \sim \text{exp}\left[-C \int \text{d}\mathbf{x} \left(1 + \frac{1}{2}|\nabla_x h|^2\right)\right].
    \label{seq:boltzdistrexp}
\end{equation}
The value of $\int \text{d}\mathbf{x}\left[ 1\right]$ is simply the area of a flat interface, which is an inconsequential constant:
\begin{equation}
    P\left[h\right] \sim \text{exp}\left[-\frac{C}{2} \int \text{d}\mathbf{x} |\nabla_x h|^2\right].
    \label{seq:boltzdistexpabs}
\end{equation}
We now express the discrete Fourier series expansion of $h(\mathbf{x})$ as:
\begin{equation}
    h(\mathbf{x}) = \sum_{\mathbf{k}} h_{k}\text{exp}\left[-i\mathbf{k}\cdot\mathbf{x}\right].
    \label{seq:interfacefourierseries}
\end{equation}
Then the gradient of $h(\mathbf{x})$ is:
\begin{equation}
    \nabla_xh(\mathbf{x}) = -i\sum_{\mathbf{k}}h_k\mathbf{k}\text{exp}\left[-i\mathbf{k}\cdot\mathbf{x}\right].
    \label{seq:interfacefourierseriesgrad}
\end{equation}
We can then solve for square gradient of $h(\mathbf{x})$ as:
\begin{equation}
    |\nabla_xh(\mathbf{x})|^2 = -\sum_{\mathbf{k}}\sum_{\mathbf{k}'}h_kh_{k'}\mathbf{k}\cdot\mathbf{k}'\text{exp}\left[-i\left(\mathbf{k}+\mathbf{k}'\right)\cdot\mathbf{x}\right].
    \label{seq:interfacefourierseriesgradsquared}
\end{equation}
Then our distribution is proportional to:
\begin{equation}
    P\left[h\right] \sim \text{exp}\left[\frac{C}{2} \int \text{d}\mathbf{x}\sum_{\mathbf{k}}\sum_{\mathbf{k}'}h_kh_{k'}\mathbf{k}\cdot\mathbf{k}'\text{exp}\left[-i\left(\mathbf{k}+\mathbf{k}'\right)\cdot\mathbf{x}\right]\right].
    \label{seq:boltzdistsub}
\end{equation}
Evaluating the integral allows us to only consider the cases where $\mathbf{k} = -\mathbf{k}'$, and the symmetry of $h_{k} = h_{-k}$ for the even components and $h_{k} = -h_{-k}$ for the odd components allows us to identify:
\begin{equation}
    P[h_{k}] \sim \text{exp}\left[-\frac{C}{2}|h_{k}|^2|\mathbf{k}|^2 \right].
    \label{seq:boltzdistsubsimp}
\end{equation}
We now wish to determine whether the linearized interfacial dynamics derived in Section~\ref{sec:heightfieldevo} exhibits this distribution at steady-state.
We rewrite Eq.~\eqref{seq:interfacelangevin}:
\begin{equation}
    \zeta_{\rm eff}(k)\frac{\partial h}{\partial t} = -k^3\upgamma_{\rm cw}(k)h + \chi(k,t).
\end{equation}
For small $k$, the tension and effective drag become constants:
\begin{equation}
    \frac{\partial h}{\partial t} = -k^3\frac{\upgamma_{\rm cw}}{\zeta_{\rm eff}} h + \chi'(k,t),
\end{equation}
where the noise $\chi'$ has zero average and variance:
\begin{equation}
    \langle \chi'(k,t)\chi'(k',t')\rangle = \left(2k \frac{\left(k_BT^{\rm act}\right)}{\zeta_{\rm eff}} + \mathcal{A}(k)\right)\delta(t-t')\delta(k+k')(2\pi)^{d-1}.
\end{equation}
Here $\mathcal{A}(k)$ is the component noise variance originating from the anisotropic noise, 
\begin{equation}
    \mathcal{A}(k) = \frac{\zeta_{\rm{eff}}(C(k) + D(k))}{(d-1)\rho^{\rm surf}B(k)} 
    \label{seq:mathcalAdef}.
\end{equation}
The Fokker-Planck equation describing the distribution of $P\left[h\right]$ is then straightforwardly found:
\begin{equation}
    \frac{\partial}{\partial t}P\left[h\right] = -\frac{\partial}{\partial h}\left[-k^3\frac{\upgamma_{\rm cw}}{\zeta_{\rm eff}}h P[h]\right] +\frac{1}{2}L^{d-1}\left(2k \frac{\left(k_BT^{\rm act}\right)}{\zeta_{\rm eff}} + \mathcal{A}(k)\right)\frac{\partial^2}{\partial h^2}\left[P\left[h\right]\right].
    \label{seq:interfacefokkerplancksub}
\end{equation}
Solving for the steady state of Eq.~\eqref{seq:interfacefokkerplancksub} results in:
\begin{equation}
    0 = \frac{\partial}{\partial h} \left[k^3\frac{\upgamma_{\rm cw}}{\zeta_{\rm eff}}h P[h] + \frac{1}{2}L^{d-1}\left(2k \frac{\left(k_BT^{\rm act}\right)}{\zeta_{\rm eff}} + \mathcal{A}(k)\right)\frac{\partial}{\partial h}\left[P\left[h\right]\right]\right].
    \label{seq:interfacefokkerplancksub2}
\end{equation}
After rearranging Eq.~\eqref{seq:interfacefokkerplancksub2} and integrating we find:
\begin{equation}
    P[h] \sim \text{exp}\left[-2L^{1-d}k^2\upgamma_{\rm cw} h^2\left(2 \left(k_BT^{\rm act}\right) + \frac{1}{k}\mathcal{A}(k)\right)^{-1}\right].
\end{equation}
At small wavevectors, $\mathcal{A}(k)$ is negligible and we find:
\begin{equation}
    P[h] \sim \text{exp}\left[-2k^2 h^2 \frac{L^{1-d}\upgamma_{\rm cw}}{k_BT^{\rm act}}\right],
\end{equation}
which is indeed a Boltzmann distribution weighted by the interfacial area and the interfacial stiffness $\upgamma_{\rm cw}/{k_BT^{\rm act}}$. 
However, if $\mathcal{A}(k)$ is not negligible, then this distribution will only be recovered if $\mathcal{A}(k)$ \textit{scales linearly with $k$}.
Based on inspection of Eqs.~\eqref{seq:beekayynum},~\eqref{seq:ceekayynum}, and~\eqref{seq:deekayynum}, $\mathcal{A}(k)$ scales as $k^2$ in the low-$k$ limit.
This implies that when the anisotropic contributions to the noise are non-negligible, a Boltzmann distribution with respect to an effective free energy proportional to the interfacial area is not recovered.
 \newpage
\section{\label{sec:eq}Langevin Dynamics of an Equilibrium Interface}

In this Section we show that an equilibrium theory constructed in the same manner as Sections~\ref{sec:fluctuating_hydro} and~\ref{sec:interface_langevin} is equivalent to equilibrium capillary-wave theory when taken to a macroscopic limit. For brevity we will skip many steps of the derivation as this procedure will be entirely analogous to those outlined by Sections~\ref{sec:fluctuating_hydro} and~\ref{sec:interface_langevin}. 

\subsection{\label{sec:denseheighteq}Equilibrium Density and Height Field Dynamics} 
We begin with the fluctuating hydrodynamics for an equilibrium system. 
We consider particles which follow an overdamped Langevin equation with an additional translational Brownian stochastic noise that satisfies the FDT.
The overdamped Langevin equation for the $i$th particle of this system is:
\begin{align}
    \Dot{\mathbf{r}}_i = \frac{1}{\zeta}\sum_{j\neq i}^{N}\mathbf{F}_{ij} + \bm{\eta}_i(t),\label{seq:eqparticleequations}
\end{align}
where $\bm{\eta}_i$ is the translational Brownian noise with zero average and variance $\langle \bm{\eta}_i(t)\bm{\eta}_j(t') = 2(k_BT/\zeta)\delta(t-t')\delta_{ij}\mathbf{I}\rangle$.
We assume that the interparticle forces, overall system density, and temperature are such that the system is phase separated. 
Following Ref.~\cite{Dean1996LangevinProcesses}, we start from Eq.~\eqref{seq:eqparticleequations} and derive the following fluctuating hydrodynamics:
\begin{subequations}
\label{seq:eqhydro}
\begin{align}
    \frac{\partial\rho}{\partial t} &= -\boldsymbol{\nabla} \cdot \mathbf{J},\  \label{seq:continuityeq} \\ \mathbf{J} &= \frac{1}{\zeta}\boldsymbol{\nabla}\cdot \left(\bm{\sigma}^C + \bm{\sigma}^B\right) + \bm{\eta}^{\rm eq},\ \\ 
    \bm{\sigma}^C &= \left[-p_C(\rho) + \rho\kappa(\rho)\nabla^2\rho + \frac{\left(\kappa(\rho)+ \rho\kappa'(\rho)\right)}{2}|\boldsymbol{\nabla}\rho|^2\right]\mathbf{I}  - \kappa(\rho)\boldsymbol{\nabla}\rho\boldsymbol{\nabla}\rho,\  \label{eq:cstress} \\
    \bm{\sigma}^B &= -k_BT\rho\mathbf{I},\  \label{seq:idealgasstress}
\end{align}
\end{subequations}
where $\bm{\eta}^{\rm eq}$ is a stochastic flux with statistics
\begin{align}
     \langle \bm{\eta}^{\rm eq}(\mathbf{r},t)\bm{\eta}^{\rm eq}(\mathbf{r}',t') \rangle &= 2D_T\rho\mathbf{I}\delta(t-t')\delta(\mathbf{r}-\mathbf{r}') \ \label{seq:etaeqvar},
\end{align}
and $\bm{\sigma}^B$ is the ``ideal gas'' stress generated by the stochastic translational Brownian force (absent in our treatment of athermal active matter).
We then use Bray's ansatz~\cite{Bray1994TheoryKinetics,Bray2002InterfaceShear}, i.e. $\rho = \upvarphi(z-h(\mathbf{x},t))$, and proceed in the same manner as Section~\ref{sec:interface_langevin}, except now the identity of our pseudovariable is $\mathcal{E}\sim 1/\rho$~\cite{Omar2023b}. This results in the following equation of motion for the height field:
\begin{align}
    \zeta_{\rm{eff}} \frac{\partial h}{\partial t}  &= -k^3\upgamma_{\rm{cw}} h + \chi(k,t), 
\end{align}
where we have defined an effective drag coefficient,
\begin{align}
    \zeta_{\rm{eff}} = \frac{\zeta A^2(k)}{2\rho^{\rm surf}B(k)}.
\end{align}
$\upgamma_{\rm{cw}}$ is the capillary wave tension defined as
\begin{align}
    \upgamma_{\rm{cw}}(k) = \frac{2\zeta_{\rm{eff}}}{
    \zeta A(k)}\biggl[\int_{-\infty}^{\infty}\text{d}u \frac{\partial \mathcal{E}}{\partial z} \upvarphi' \upvarphi\kappa(\upvarphi)  - \int_{-\infty}^{\infty}\text{d}u \frac{\partial \mathcal{E}}{\partial z}\int_{-\infty}^{\infty} \text{d}z'\text{sgn}(u-z')e^{-k|u-z'|}\kappa(\upvarphi)(\upvarphi')^2 \biggr],
    \label{eq:capwavetension}
\end{align}
$\chi$ is a noise defined as
\begin{align}
    \chi(k,t) = \frac{A(k)\zeta }{B(k)(\rho^{\rm{liq}}+\rho^{\rm{gas}})} \int \text{d}u \frac{\partial \mathcal{E}}{\partial z}\int \text{d}z' e^{-k|u-z'|}\mathcal{F}\left[\nabla\cdot\bm{\eta}^{\rm eq}\right],
\end{align}
$A(k)$ is a function defined as
\begin{align}
    A(k) = \int \int \text{d}u\text{d}z e^{-k|u-z|}\frac{\partial \mathcal{E}}{\partial z}(u)\upvarphi'(z),
\end{align}
and $B(k)$ is a function defined as
\begin{align}
    B(k) = \int \int \text{d}u\text{d}u' e^{-k|u-u'|}\frac{\partial \mathcal{E}}{\partial z}(u)\frac{\partial \mathcal{E}}{\partial z}(u').
\end{align}
Following the same procedure as in Section~\ref{sec:noise_stat}, the statistics of $\chi(\mathbf{k},t)$ can be solved for as
\begin{align}
        \langle \chi(\mathbf{k},t)\chi(\mathbf{k}',t')\rangle = 2k(k_BT)\zeta_{\rm{eff}}\delta(t-t')\delta(\mathbf{k}+\mathbf{k}')(2\pi)^{(d-1)}\label{eq:chivariso}
\end{align}
We next argue that this Langevin equation for the height field agrees with equilibrium capillary wave theory in a macroscopic limit. This macroscopic limit implies two conditions:
\begin{itemize}
    \item The magnitude of the wave vector goes to zero, i.e. $k\to0$
    \item The width of the interfacial region becomes negligibly small compared to system dimensions, i.e. the density profile $\upvarphi$ becomes infinitely sharp.
\end{itemize}
In this macroscopic limit we will be able to argue that the capillary-wave tension is equivalent to the mechanical surface tension and that the interfacial fluctuations are given by the mechanical surface tension.
\subsection{\label{sec:macroscopic}Macroscopic Agreement With Equilibrium Capillary-Wave Theory}
The mechanical surface tension as defined by Kirkwood and Buff~\cite{Kirkwood1949,Rowlinson1982MolecularCapillarity} can be found by integrating the difference in the normal and tangential components of the stress tensor, i.e.
\begin{align}
    \upgamma_{\rm{mech}} = -\int \text{d}z \left[\sigma^C_{zz} - \frac{1}{2}\left(\sigma^C_{xx} + \sigma^C_{yy}\right)\right]\label{seq:passivemechtension},
\end{align}
where contributions to Eq.~\eqref{seq:passivemechtension} from $\bm{\sigma}^B$ will vanish because the ideal gas stress is isotropic.
We now substitute our definition for $\bm{\sigma}^C$ into our expression for the mechanical surface tension, and take the limit as $k$ goes to zero
\begin{align}
    \lim_{k\to 0}\upgamma_{\rm{mech}} = \int \text{d}z\kappa(\upvarphi)(\upvarphi')^2.
    \label{seq:mechsteq}
\end{align}
From our macroscopic assumptions listed in Section~\ref{sec:denseheighteq}, we can show that the ratio $A(k)/B(k)$ for an equilibrium system goes to 
\begin{align}
    \lim_{k\to 0}\frac{A(k)}{B(k)} \approx \frac{\left(\rho^{\rm surf}\right)^2}{\mathcal{C}},
\end{align}
where again, $\mathcal{C}$ is an arbitrary constant of integration associated with the pseudodvariable.
Then the low-$k$ capillary tension is given by:
\begin{align}
    \lim_{k\to0}\upgamma_{\rm{cw}}(k) = &\rho^{\rm surf}\biggl[\int_{-\infty}^{\infty}\text{d}u \frac{\upvarphi'}{\upvarphi^2} \upvarphi' \upvarphi \kappa(\upvarphi)\nonumber \\ &- \int_{-\infty}^{\infty}\text{d}u \frac{\upvarphi'}{(\upvarphi)^2}\int_{-\infty}^{\infty} \text{d}z'\text{sgn}(u-z')\kappa(\upvarphi)(\upvarphi')^2 \biggr].
\end{align}
The second integral can be split into regions where $z' > u$ and $z' < u$, giving
\begin{align}
    \lim_{k\to0}\upgamma_{\rm{cw}}(k) = &\rho^{\rm surf}\biggl[\int_{-\infty}^{\infty}\text{d}u \frac{\upvarphi'}{\upvarphi^2} \upvarphi' \upvarphi \kappa(\upvarphi) \nonumber \\ & - \int_{-\infty}^{\infty}\text{d}u \frac{\upvarphi'}{(\upvarphi)^2}\left(\int_{-\infty}^{u} \text{d}z'\kappa(\upvarphi)(\upvarphi')^2 - \int_{u}^{\infty} \text{d}z'\kappa(\upvarphi)(\upvarphi')^2\right)\biggr].
\end{align}
These integrals across $z'$ are equal to the mechanical surface tension if the integral goes across the interface and zero otherwise since, in the macroscopic limit, the interfaces are infinitely sharp.
Then,
\begin{align}
    \lim_{k\to0}\upgamma_{\rm{cw}}(k) = &\rho^{\rm surf}\biggl[\int_{-\infty}^{\infty}\text{d}u \frac{\upvarphi'}{\upvarphi^2} \upvarphi' \upvarphi \kappa(\upvarphi) \nonumber \\ & - \int_{-\infty}^{\infty}\text{d}u \frac{\upvarphi'}{(\upvarphi)^2}\biggl(\upgamma_{\rm{mech}}H(u) - \upgamma_{\rm{mech}}(1 - H(u))\biggr)\biggr],
\end{align}
where $H$ is the Heaviside step function.
From the identity $2H(u) - 1 = \text{sgn}(u)$, it follows that:
\begin{align}
    \lim_{k\to0}\upgamma_{\rm{cw}}(k) = \rho^{\rm surf}\biggl[\int_{-\infty}^{\infty}\text{d}u \frac{\upvarphi'}{\upvarphi^2} \upvarphi' \upvarphi \kappa(\upvarphi) - \upgamma_{\rm{mech}}\int_{-\infty}^{\infty}\text{d}u \frac{\upvarphi'}{(\upvarphi)^2}\text{sgn}(u)\biggr].
\end{align}
For an infinitely sharp profile, $\upvarphi'$ is nonzero only exactly at interface. 
We therefore approximate the $1/(\upvarphi)^2$ terms as constants evaluated at $u=0$, leaving us with
\begin{align}
    \lim_{k\to0}\upgamma_{\rm{cw}}(k) = \int_{-\infty}^{\infty}\text{d}u \kappa(\upvarphi)(\upvarphi')^2 = \lim_{k\to 0}\upgamma_{\rm{mech}}.
\end{align}
This derivation demonstrates that at equilibrium, the capillary-wave tension equals the mechanical surface tension in the macroscopic limit. 
By using the equivalence of the capillary-wave tension and mechanical surface tension, and using the same lines of argument as Section~\ref{sec:capillarystats}, we can show that
\begin{align}
    \langle h(\mathbf{k})h(-\mathbf{k}) \rangle = \frac{L^{(d-1)}k_BT}{\upgamma_{\rm{mech}}k^2}.
    \label{seq:capfluctslowkeq}
\end{align} \newpage
\section{\label{sec:numerics}Numerical Details}
In this Section, we list the details of all numerical calculations and simulations. 
We start by providing the equations of state necessary for calculating the quantities derived in Sections~\ref{sec:fluctuating_hydro} and~\ref{sec:interface_langevin}. 
We then summarize the procedure used to calculate the stationary noise-averaged density profile, $\upvarphi(z)$. 
Next, we give the details on computing the capillary-wave tension, interfacial height fluctuations, relaxation times, and mechanical surface tension. 
We then calculate the variance of the anisotropic noise $\chi^{\rm aniso}$ relative to the total noise in order to identify the conditions under which our interfacial Langevin equation effectively obeys the fluctuation dissipation theorem. 
A comparison of the above quantities calculated with and without the pseudovariable is then presented. 
This comparison strengthens our argument that the capillary-wave tension encapsulates the effects of nematic flows. 
Finally, we list the details of the Brownian dynamics simulations and statistical uncertainty of the interfacial fluctuations measured from simulation.

\subsection{\label{sec:eos}Equations of State}
In order to numerically solve for the quantities derived in this supplemental material, we use the equations of state reported Ref.~\cite{Omar2023b} for athermal active hard spheres.
The two contributions to the bulk dynamic pressure were found to be well described by:
\begin{subequations}
    \begin{align}
        \frac{p_{\rm act}}{\zeta U_0 / ( d_{\rm{hs}}^2)} &= \phi \left(\frac{\ell_0}{d_{\rm{hs}}}\right) \frac{1}{\pi}\overline{U}\nonumber \\ &= \phi \left(\frac{\ell_0}{\mathcal{D}}\right)\frac{1}{\pi} \Biggl[1+\biggl(1 - \exp\left[-2^{7/6} \left(\frac{\ell_0}{d_{\rm{hs}}}\right)\right] \biggr)\frac{\phi}{1 - \phi / \phi_{\rm max}} \Biggr]^{-1} \  \\
        \frac{p_{\rm C}}{\zeta U_0 / ( d_{\rm{hs}}^2)} &=  6\times 2^{-7/6} \frac{1}{\pi}\frac{\phi^2}{\sqrt{1-\phi/\phi_{\rm max}}} \ , 
\end{align}
\end{subequations}
where  $p_{\rm act}$ is the active pressure, $\phi = \rho\pi d_{\rm hs}^3/6$ is the volume fraction, and $\phi_{\rm max} = 0.645$ is the maximum volume fraction of disordered hard spheres.
The active pressure above is consistent with a dimensionless active speed given by:
\begin{align}
    \overline{U} = \Biggl[1+\biggl(1 - \exp\left[-2^{7/6} \left(\frac{\ell_0}{d_{\rm hs}}\right)\right] \biggr)\frac{\phi}{1 - \phi / \phi_{\rm max}} \Biggr]^{-1}.
\end{align}
Although these equations were empirically fit to simulations conducted in the homogeneous region of the phase diagram ($\lambda < 0$), we assume that they remain valid in the phase separated region. 
From these equations of state one has all information needed to numerically evaluate the quantities reported in the main text.
\subsection{\label{sec:profiles}Density Profile Evaluation}
We solve for the noise-averaged steady-state profile $\upvarphi$ by invoking the static linear momentum balance, $\boldsymbol{\nabla}\cdot\bm{\Sigma} = 0$.
As $\upvarphi$ is only a function of $z$, we focus on the z-component of the momentum balance with:
\begin{align}
    \Sigma_{zz} = -\mathcal{P}_{\rm coexist} = \rm{const.}, \label{seq:sigmaequal}
\end{align}
where the dynamic pressure $\mathcal{P}$ is defined as the bulk contribution to the dynamic pressure 
$\mathcal{P}\equiv p_{C} + p_{\rm act}$.
From Section~\ref{sec:summary} we can also express $\Sigma_{zz}$ as:
\begin{align}
    \Sigma_{zz} = -\mathcal{P}(\rho) + a(\rho)\frac{\partial^2\rho}{\partial z^2} + b(\rho)\left(\frac{\partial \rho}{\partial z}\right)^2.\label{seq:sigmazz} 
\end{align}
Substitution of Eq.~\eqref{seq:sigmaequal} into Eq.~\eqref{seq:sigmazz} and rearranging results in:
\begin{align}
    \mathcal{P}(\rho) - \mathcal{P}_{\rm coexist}  = a(\rho)\frac{\partial^2\rho}{\partial z^2} + b(\rho)\left(\frac{\partial \rho}{\partial z}\right)^2.\label{seq:sigmazzcoexist} 
\end{align}
We now integrate Eq.~\eqref{seq:sigmazzcoexist} from the liquid phase to the gas phase with respect to the pseudovariable $\mathcal{E}$:
\begin{align}
    \int_{\mathcal{E}^{\rm liq}}^{\mathcal{E}^{\rm gas}}\left(\mathcal{P}(\rho) - \mathcal{P}_{\rm coexist} \right)\text{d}\mathcal{E} = \int_{\mathcal{E}^{\rm liq}}^{\mathcal{E}^{\rm gas}}\left(a(\rho)\frac{\partial^2\rho}{\partial z^2} + b(\rho)\left(\frac{\partial \rho}{\partial z}\right)^2\right)\text{d}\mathcal{E},\label{seq:constructionsetup} 
\end{align}
where all functions dependent on density are implicitly dependent on $\mathcal{E}$.
Eq.~\eqref{seq:constructionsetup} can be rewritten in terms of an integral with respect to density:
\begin{align}
    \int_{\rho^{\rm liq}}^{\rho^{\rm gas}}\left(\mathcal{P}(\rho) - \mathcal{P}_{\rm coexist} \right)\frac{\partial \mathcal{E}}{\partial \rho}\text{d}\rho = \int_{\rho^{\rm liq}}^{\rho^{\rm gas}}\left(a(\rho)\frac{\partial^2\rho}{\partial z^2} + b(\rho)\left(\frac{\partial \rho}{\partial z}\right)^2\right)\frac{\partial \mathcal{E}}{\partial \rho}\text{d}\rho.\label{seq:constructionsetuptransform} 
\end{align}
Integrating the right hand side of Eq.~\eqref{seq:constructionsetuptransform} by parts and invoking Eq.~\eqref{seq:pseudodef} then results in:
\begin{align}
    \int_{\rho^{\rm liq}}^{\rho^{\rm gas}}\left(\mathcal{P}(\rho) - \mathcal{P}_{\rm coexist} \right)\frac{\partial \mathcal{E}}{\partial \rho}\text{d}\rho = \left[a(\rho)\frac{\partial \mathcal{E}}{\partial \rho}\left(\frac{\partial \rho}{\partial z}\right)^2\right]\biggr|^{\rho^{\rm gas}}_{\rho^{\rm liq}}.\label{seq:constructionsetupboundaries} 
\end{align}
Note that because the density profile is spatially constant at the binodal densities, the right hand side of Eq.~\eqref{seq:constructionsetupboundaries} is equal to zero (the pseudovariable is defined precisely such that this is the case, as discussed in Section~\ref{sec:heightfieldevo} and Refs.~\cite{Aifantis1983TheRule, Solon2018GeneralizedMatter, Omar2023b}).
Then, recalling that $\mathcal{E}\sim p_C(\rho)$ for ABPs, we identify the coexistence criteria for the binodal densities~\cite{Omar2023b}:
\begin{subequations}
    \begin{align}
        &\mathcal{P}(\rho^{\rm{liq}}) = \mathcal{P}(\rho^{\rm{gas}}), \label{seq:coexist1} \\ \int_{\rho^{\rm{liq}}}^{\rho^{\rm{gas}}}\biggl(&\mathcal{P}(\rho) - \mathcal{P}_{\rm{coexist}}\biggr)\frac{\partial p_C}{\partial\rho} \text{d}\rho = 0.
        \label{seq:coexist2}
    \end{align}
\end{subequations}
The binodal densities corresponding to each value of $\ell_o$ were evaluated by simultaneous solution of Eqs.~\eqref{seq:coexist1}~and~\eqref{seq:coexist2}. 

To solve for the full spatial density profile (rather than just the binodal densities), we integrate Eq.~\eqref{seq:sigmazzcoexist} from one of the binodal densities to an intermediate density between the two binodal densities.
In this case Eq.~\eqref{seq:constructionsetupboundaries} would have the form:
\begin{align}
    \int_{\rho^{\rm liq}}^{\rho}\left(\mathcal{P}(\rho) - \mathcal{P}_{\rm coexist} \right)\frac{\partial \mathcal{E}}{\partial \rho}\text{d}\rho = a(\rho)\frac{\partial \mathcal{E}}{\partial \rho}\left(\frac{\partial \rho}{\partial z}\right)^2.\label{seq:constructionsetupboundariesarb} 
\end{align}
We use Eq.~\eqref{seq:beta} to identify $a(\rho)$ and recall that $\mathcal{E}\sim p_C(\rho)$ for ABPs in order to map any value of density between the binodals to its derivative in the $z$-direction by rearranging Eq.~\eqref{seq:constructionsetupboundariesarb}:
\begin{align}
    \frac{\partial\rho}{\partial z} = \pm f(\rho) =\pm \frac{\sqrt{\frac{4d(d-1)(d+2)}{3}\int_{\rho^{\rm{liq}}}^{\rho}\bigl(\mathcal{P}(\rho') - \mathcal{P}_{\rm{coexist}}\bigr)\frac{\partial p_C}{\partial\rho'} \text{d}\rho'}}{\overline{U}\frac{\partial p_C}{\partial\rho}}.
    \label{seq:drhodz}
\end{align}
Then all densities between the binodals can be mapped to a $z$-coordinate using
\begin{align}
    z(\rho) = \int_{\rho^{\rm{liq}}}^{\rho}\frac{1}{f(\rho')}\text{d}\rho'.
\end{align}
We identify this steady-state mapping between density and $z$-coordinate as $\upvarphi(z)$.
\subsection{\label{sec:cwtension}Capillary-Wave Tension and Fluctuations}
\begin{figure}
    \centering
    \includegraphics[width = 0.75\textwidth]{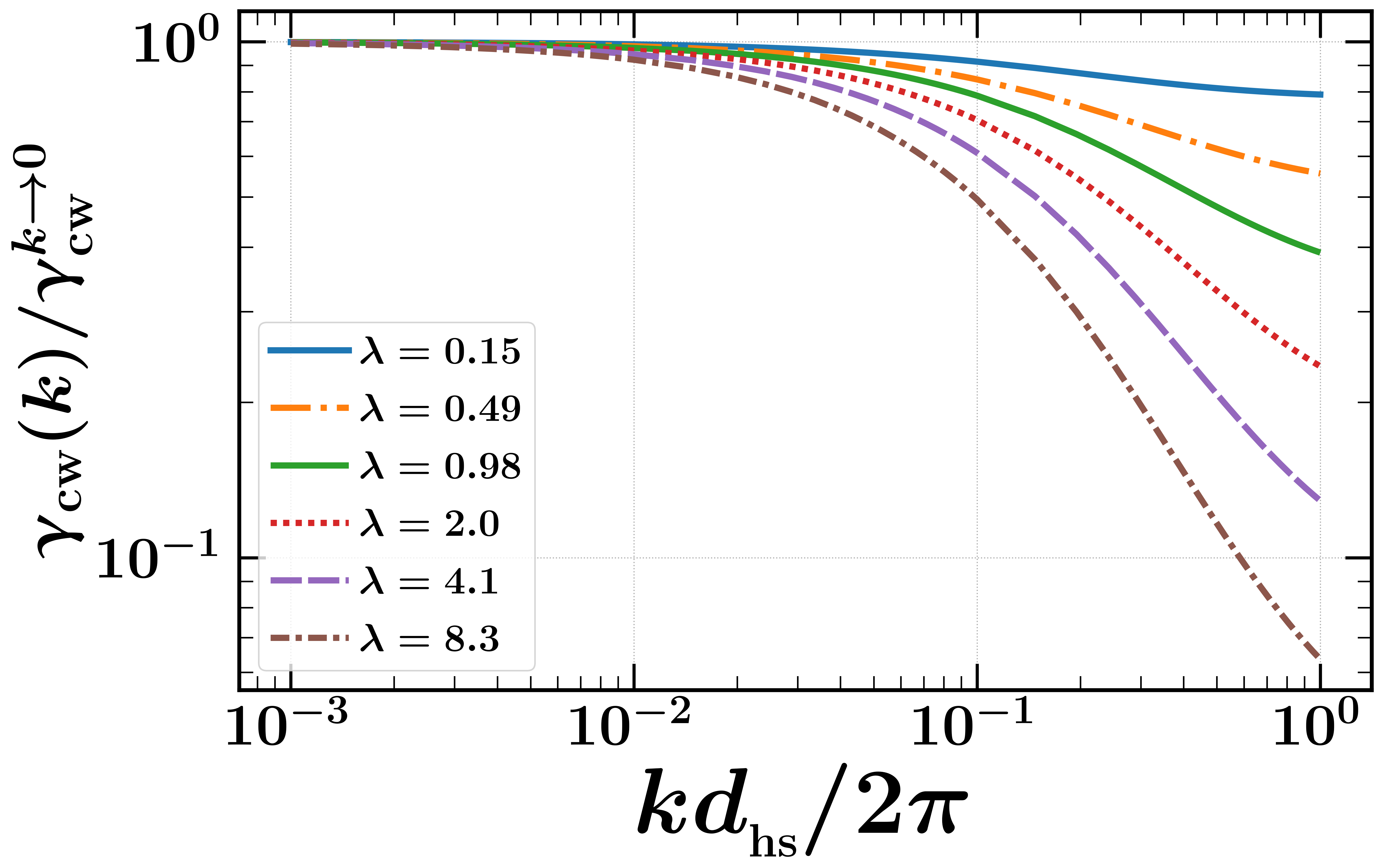}
    \caption{$\upgamma_{\rm cw}(k)$ normalized by its $k\to 0$ limit for selected values of $\lambda\equiv \ell_o/\ell_o^c - 1$}
    \label{fig:tensionmacro}
\end{figure}

Substituting the solution for $\mathcal{E}$ into the definition of capillary wave tension gives
\begin{align}
    \upgamma_{\rm{cw}}(k)/\mathcal{C} = &\frac{A(k)}{B(k)\rho^{\rm surf}}\biggl[\int_{-\infty}^{\infty}\text{d}u  \upvarphi' \upvarphi'\left(p'_C\overline{U}\right)^2\nonumber \\ & + \int_{-\infty}^{\infty}\text{d}u \frac{\left(p'_C\overline{U}\right)^2}{a(\upvarphi)}\upvarphi'\int_{-\infty}^{\infty} \text{d}z'\text{sgn}(u-z')e^{-k|u-z'|}b(\upvarphi)(\upvarphi')^2 \biggr].\ 
\end{align}
Similarly, we find that the functions $A(k)$, $B(k)$, $C(k)$, $D(k)$ are given by
\begin{align}
    A(k) = \mathcal{C}\int \text{d}z_1 \upvarphi'\int \text{d}z_2 \frac{\partial p_C}{\partial \rho}\upvarphi'e^{-k|z_1 - z_2|},
    \label{seq:ayykayynum}
\end{align}
\begin{align}
    B(k) = \mathcal{C}^2\int \text{d}z_1 \upvarphi'\frac{\partial p_C}{\partial \rho}\int \text{d}z_2 \frac{\partial p_C}{\partial \rho}\upvarphi'e^{-k|z_1 - z_2|},
    \label{seq:beekayynum}
\end{align}
\begin{align}
    C(k) = &\mathcal{C}^2\int \text{d}u \upvarphi'\frac{\partial p_C}{\partial \rho} \int \text{d}u' \upvarphi'\frac{\partial p_C}{\partial \rho} \int \text{d}z' e^{-k|u-z'|}e^{-k|u'-z'|}k^2\frac{a(\upvarphi)}{\overline{U}(\upvarphi)}\upvarphi'' \nonumber \\ &\times \left(1 + \text{sgn}\left((u-z')(u'-z')\right)\right),
    \label{seq:ceekayynum}
\end{align}
and
\begin{align}
    D(k) = &\mathcal{C}^2\int \text{d}u \upvarphi'\frac{\partial p_C}{\partial \rho} \int \text{d}u' \upvarphi'\frac{\partial p_C}{\partial \rho} \int \text{d}z'  e^{-k|u-z'|}e^{-k|u'-z'|} \nonumber \\ &\times k^2\frac{b(\upvarphi)}{\overline{U}(\upvarphi)}\upvarphi'\upvarphi'\text{sgn}\left((u-z')(u'-z')\right).
    \label{seq:deekayynum}
\end{align}
All of the above quantities were evaluated using trapezoidal integration with density profile vectors of size $200001$. 
For double (e.g. Eq.~\ref{seq:ayykayynum}) and triple (e.g. Eq.~\ref{seq:ceekayynum}) integrals the size $200001$ vector density profiles were too fine in resolution to compute in reasonable time. 
Compressed vectors using every $20$th entry were used to compute the double integrals and the innermost integral of the triple integrals. 
The two outer integrals of the triple integrals were computed using every $200$th entry of the full sized density profiles.
In order to avoid numerical instabilities associated with the $\text{sgn}(z)$ functions, they were replaced with $\tanh(yz)$ functions, where $y=1\times10^{6}~d_{\rm hs}^{-1}$ was a constant large enough to make the $\tanh$ function closely resemble a sign function for our data. 
The results of calculating the above equations as a function of activity for selected values of $k$ are plotted in the main text.
The results of calculating the above equations as a function of $k$ for selected values of activity are plotted in Fig.~\ref{fig:tensionmacro}.

\subsection{\label{sec:relaxmechtension}Relaxation Times and Mechanical Surface Tension}
As derived in Section~\ref{sec:capillarystats}, the characteristic timescale for the relaxation of a capillary wave is set by:
\begin{align}
    \tau = \frac{\zeta_{\rm{eff}}}{k^3\upgamma_{\rm{cw}}}.
\end{align}
Details on calculating all dependencies of $\tau$ have been discussed in Section~\ref{sec:cwtension}.
We plot this quantity as a function of activity for several selected values of $k$.
\begin{figure}
    \centering
    \includegraphics[width = 0.75\textwidth]{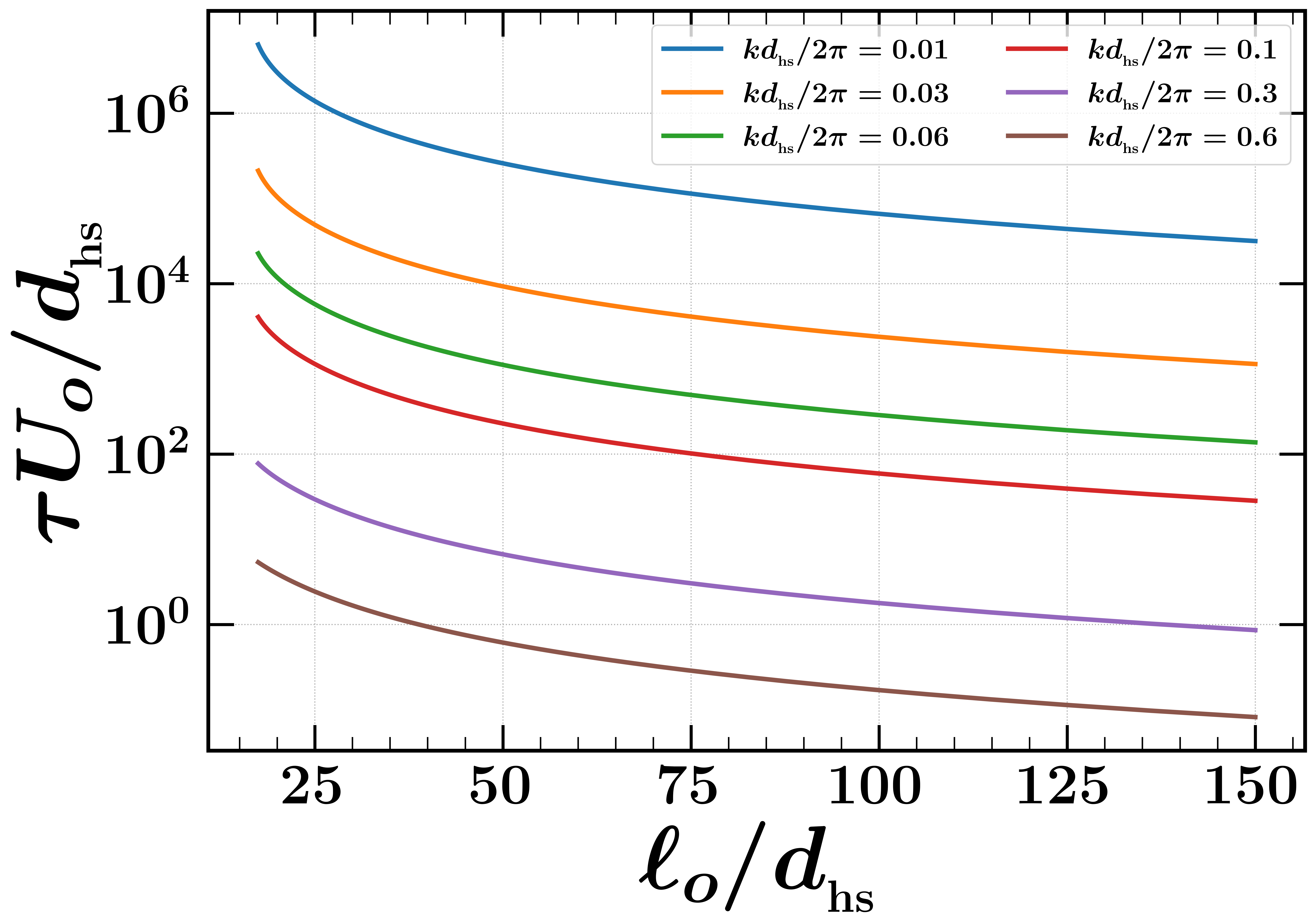}
    \caption{Activity dependence of relaxation time $\tau$ for selected values of $k$.}
    \label{fig:relaxation}
\end{figure}

\begin{figure}
    \centering
    \includegraphics[width = 0.75\textwidth]{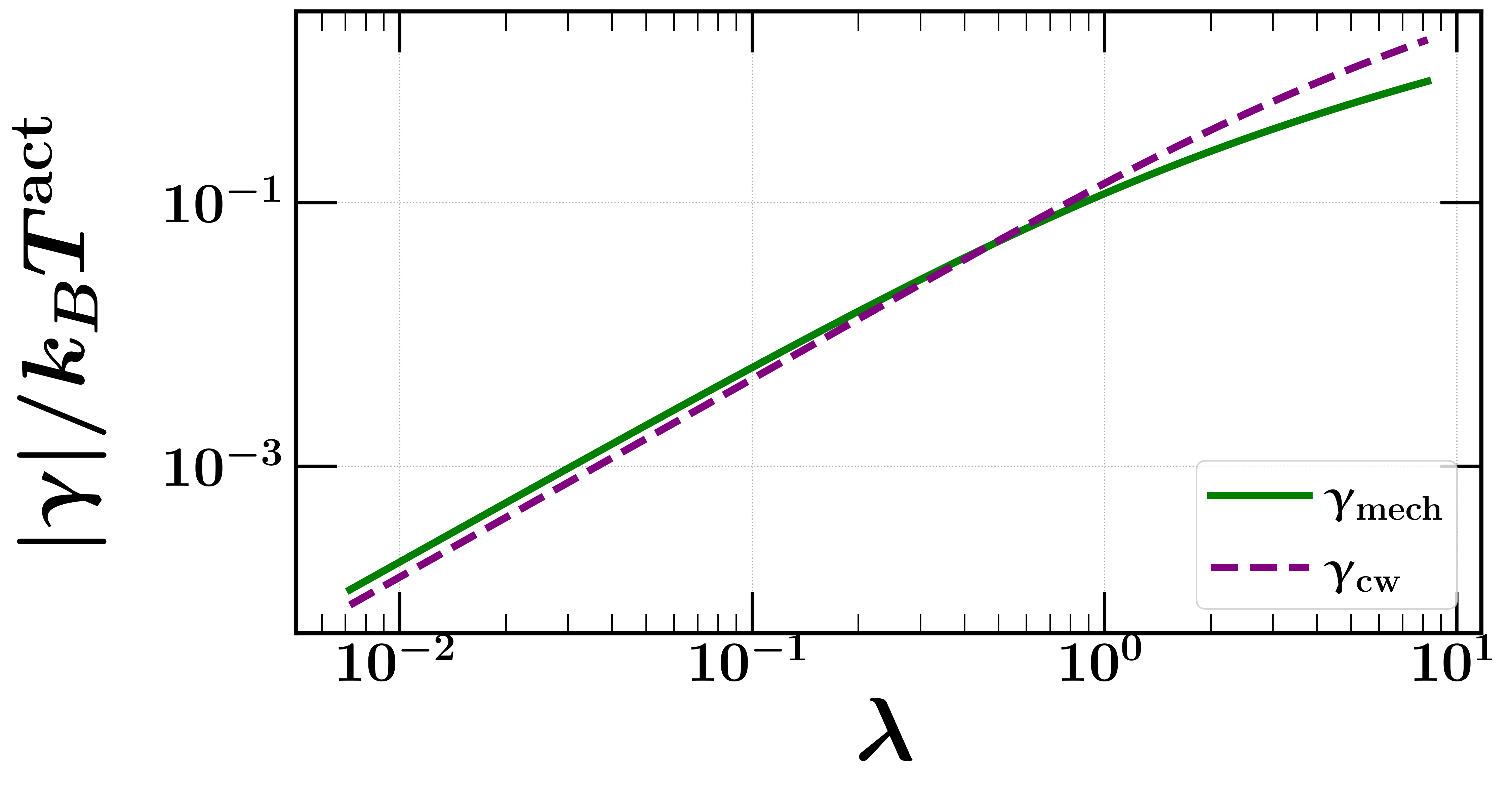}
    \caption{Absolute value of the mechanical surface tension as defined by Eq.~\eqref{seq:mechst} and $k\to0$ limit of the capillary-wave tension. Both quantities are normalized by $k_BT^{\rm act}$. Vertical axis has units of $d_{\rm hs}^{-2}$. Mechanical surface tension is negative for all run lengths above the critical point.}
    \label{fig:smmech}
\end{figure}

We now demonstrate that within our theory, the mechanical surface tension defined by Bialk\'e~\textit{et al.}~\cite{Bialke2015} is indeed distinct from the capillary-wave tension and negative. 
This mechanical surface tension can be found by integrating the difference in the normal and tangential components of the dynamic stress tensor Eq.~\eqref{seq:dynstressagain}:
\begin{align}
    \upgamma_{\rm{mech}} = -\int \text{d}z \left[\Sigma_{zz} - \frac{1}{2}\left(\Sigma_{xx} + \Sigma_{yy}\right)\right].
    \label{seq:mechtensiondef}
\end{align}
We now substitute Eq.~\eqref{seq:dynstressagain} into Eq.~\eqref{seq:mechtensiondef} and make the approximation that the variation of the stochastic density (and therefore the stress) in the $x,y$-directions are far less than that in the $z$-direction,
\begin{align}
    \upgamma_{\rm{mech}} = -\int \text{d}z b(\upvarphi)(\upvarphi')^2.
    \label{seq:mechst}
\end{align}
We calculate quantity as a function of run length in the same manner described by Section~\ref{sec:cwtension}.
The mechanical surface tension is \emph{negative} for all run lengths above the critical point.
We plot the magnitude of both the mechanical and capillary-wave tension (relative to $k_BT^{\rm act}$) in Fig.~\ref{fig:smmech}.
Intriguingly, when comparing the \emph{magnitudes} of the two surface tensions, we find that there is little qualitative distinction. Both surface tensions encapsulate the effects of nematic flows — while the sign of the mechanical surface tension suggests that nematic flows are destabilizing, our derived capillary tension reveals that these same flows suppress interfacial fluctuations, reflected in a positive capillary tension.
\subsection{\label{sec:isoerror}Isotropic Approximation Error}
\begin{figure}
    \centering
    \includegraphics[width = 0.75\textwidth]{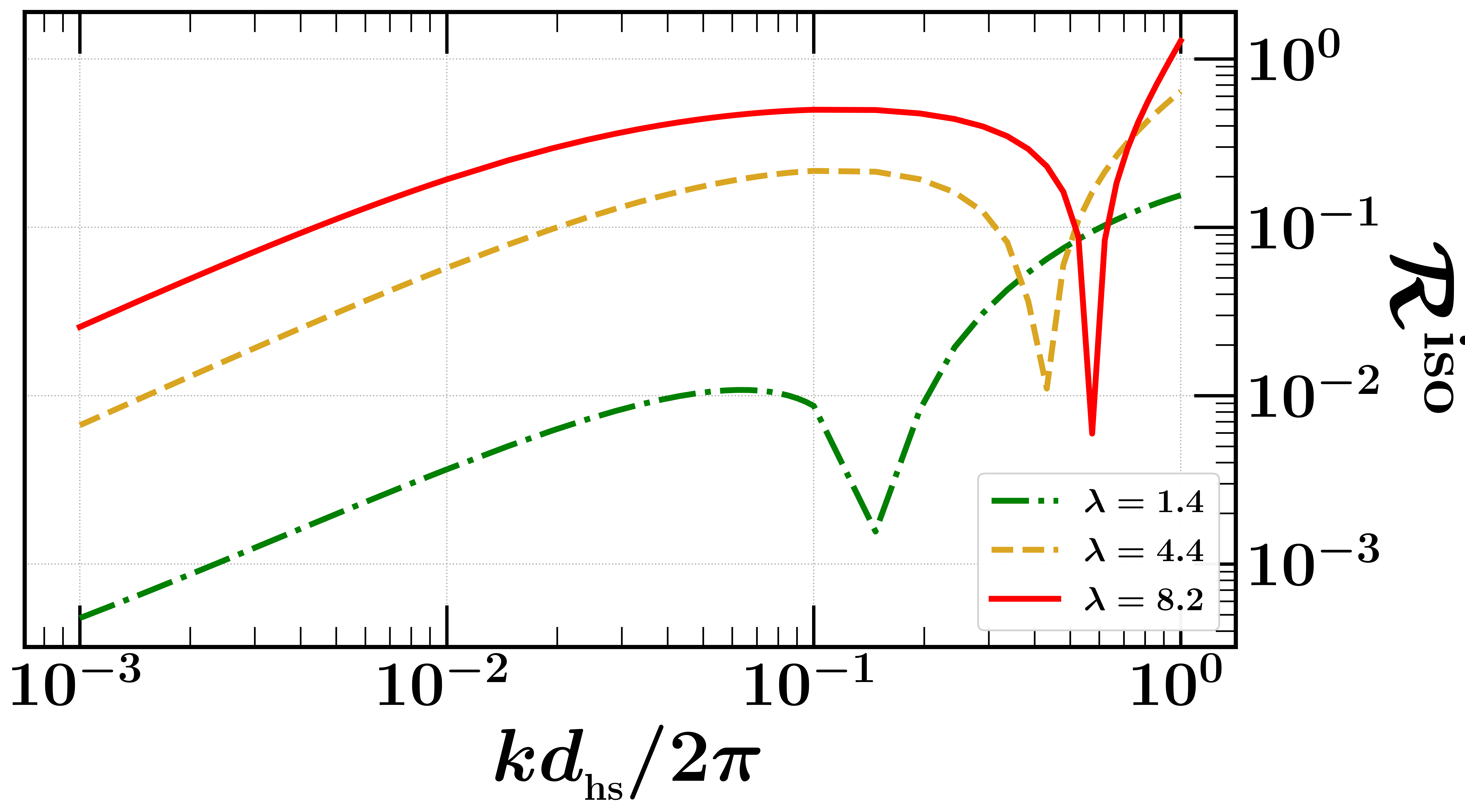}
    \caption{Relative error of the isotropic fluctuations as a function of $k$ for three selected activities.}
    \label{fig:smisoerrork}
\end{figure}
\begin{figure}
    \centering
    \includegraphics[width = 0.75\textwidth]{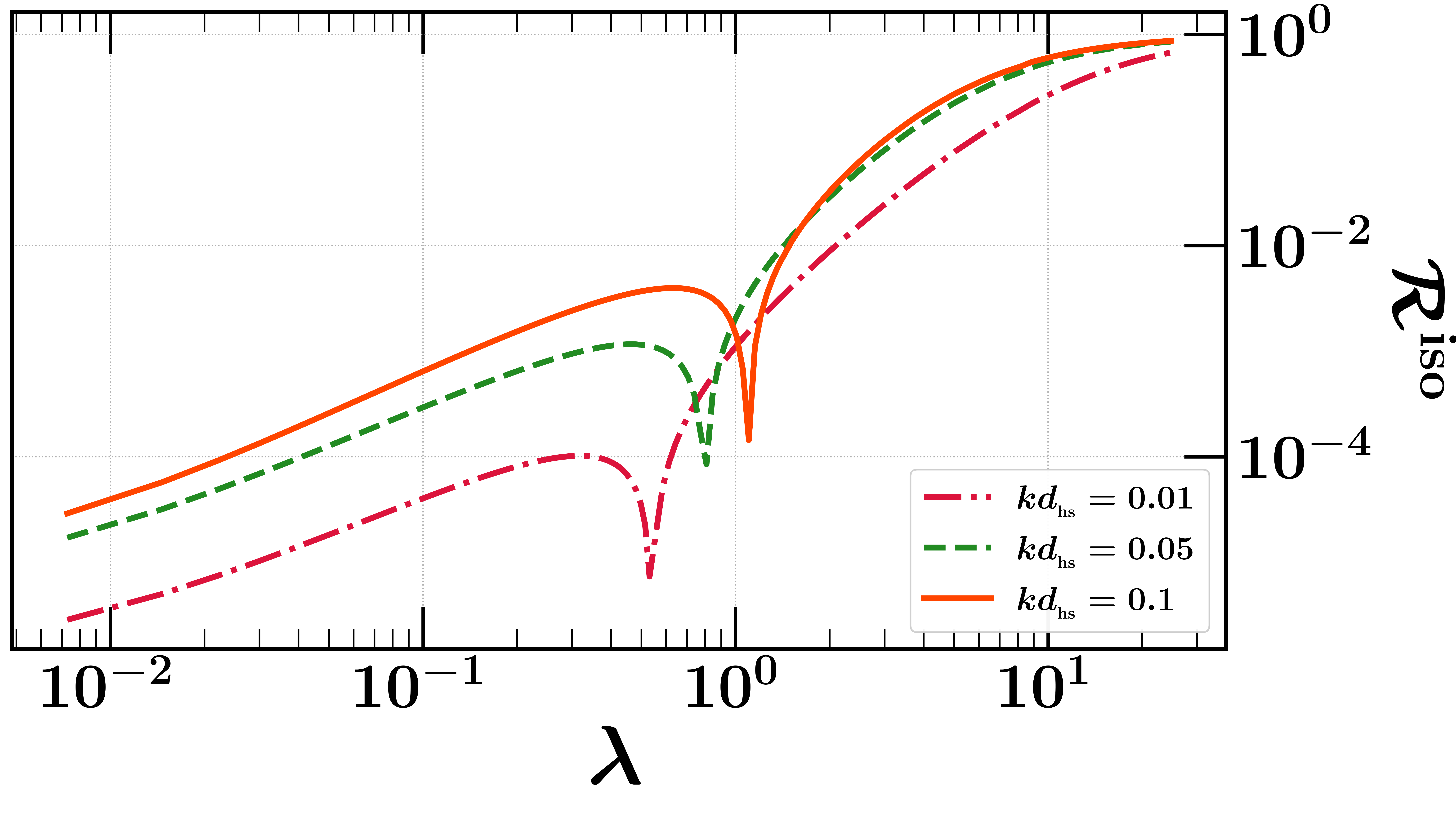}
    \caption{Relative error of the isotropic fluctuations as a function of $\lambda$ for three selected values of $k$.}
    \label{fig:smisoerrortau}
\end{figure}
\begin{figure}
    \centering
    \includegraphics[width = 0.75\textwidth]{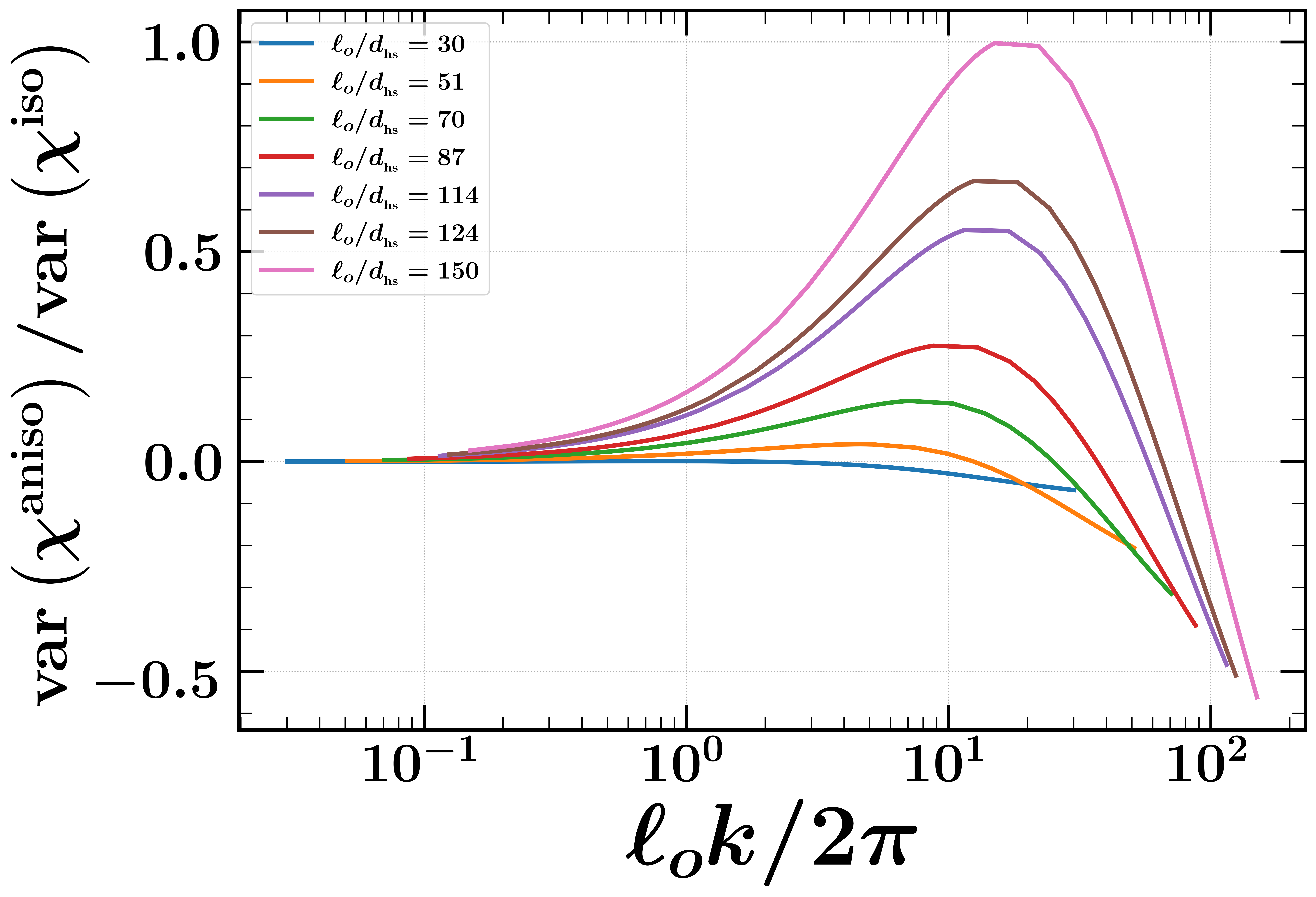}
    \caption{Variance of $\chi^{\rm{aniso}}$ divided by the variance of $\chi^{\rm{iso}}$ as a function of $\ell_o k$.}
    \label{fig:anisooveriso}
\end{figure}
We plot the relative error of the height field fluctuations calculated from only isotropic noise Eq.~\eqref{seq:chivariso} as compared to calculated from both Eq.~\eqref{seq:chivariso} and Eq.~\eqref{seq:chivaraniso} as a function of wave vector magnitude $k$ and as a function of reduced run length $\lambda$. 
We define the relative error of the fluctuations as
\begin{align}
    \mathcal{R}^{\rm iso} = \left|\frac{\langle|h(k)|^2 \rangle^{\rm iso} - \langle|h(k)|^2 \rangle}{\langle|h(k)|^2 \rangle}\right|.
\end{align}
The results of this calculation are shown in Fig.~\ref{fig:smisoerrork} and Fig.~\ref{fig:smisoerrortau}. 
At low activities and low values of $k$, the anisotropic components of the noise can be safely discarded.
However, the error associated with ignoring the anisotropic noise components becomes more significant as the wavelength decreases and activity increases. 
We remark that the kinks present in Fig.~\ref{fig:smisoerrork} and Fig.~\ref{fig:smisoerrortau} are due to a transition from the isotropic approximation underestimating to overestimating the fluctuations. 
This can be seen in Fig.~\ref{fig:anisooveriso} crossing from positive to negative values.

Fig.~\ref{fig:anisooveriso} demonstrates that as the run length approaches the wavelength of a capillary fluctuation, i.e. $\ell_o k$ approaches unity, the magnitude of the anisotropic noise becomes significant as compared to the isotropic noise. 
On the other hand, when $\ell_o k << 1$ the anisotropic noise can be safely ignored and the interfacial Langevin equation will recover area minimizing Boltzmann statistics [See Sec.~\ref{sec:fluctdiss}]~\cite{Bray2002InterfaceShear}.

\subsection{\label{sec:pseudoeffect}Effect of Pseudovariable}
\begin{figure}
    \centering
    \includegraphics[width = 0.75\textwidth]{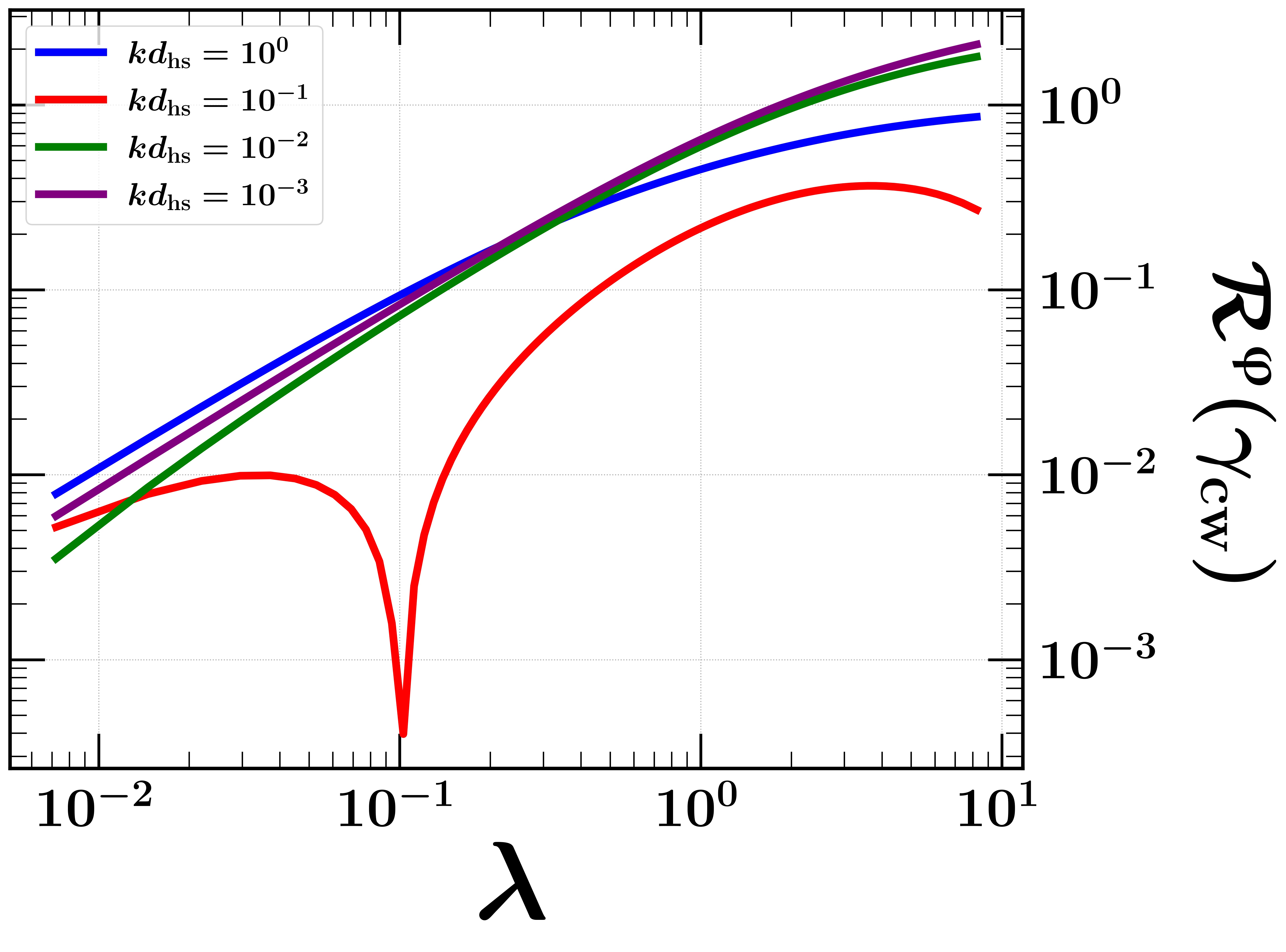}
    \caption{Relative difference of the capillary-wave tension as a function of $\lambda$ for four selected values of $k$.}
    \label{fig:pseudocomparetension}
\end{figure}

\begin{figure}
    \centering
    \includegraphics[width = 0.75\textwidth]{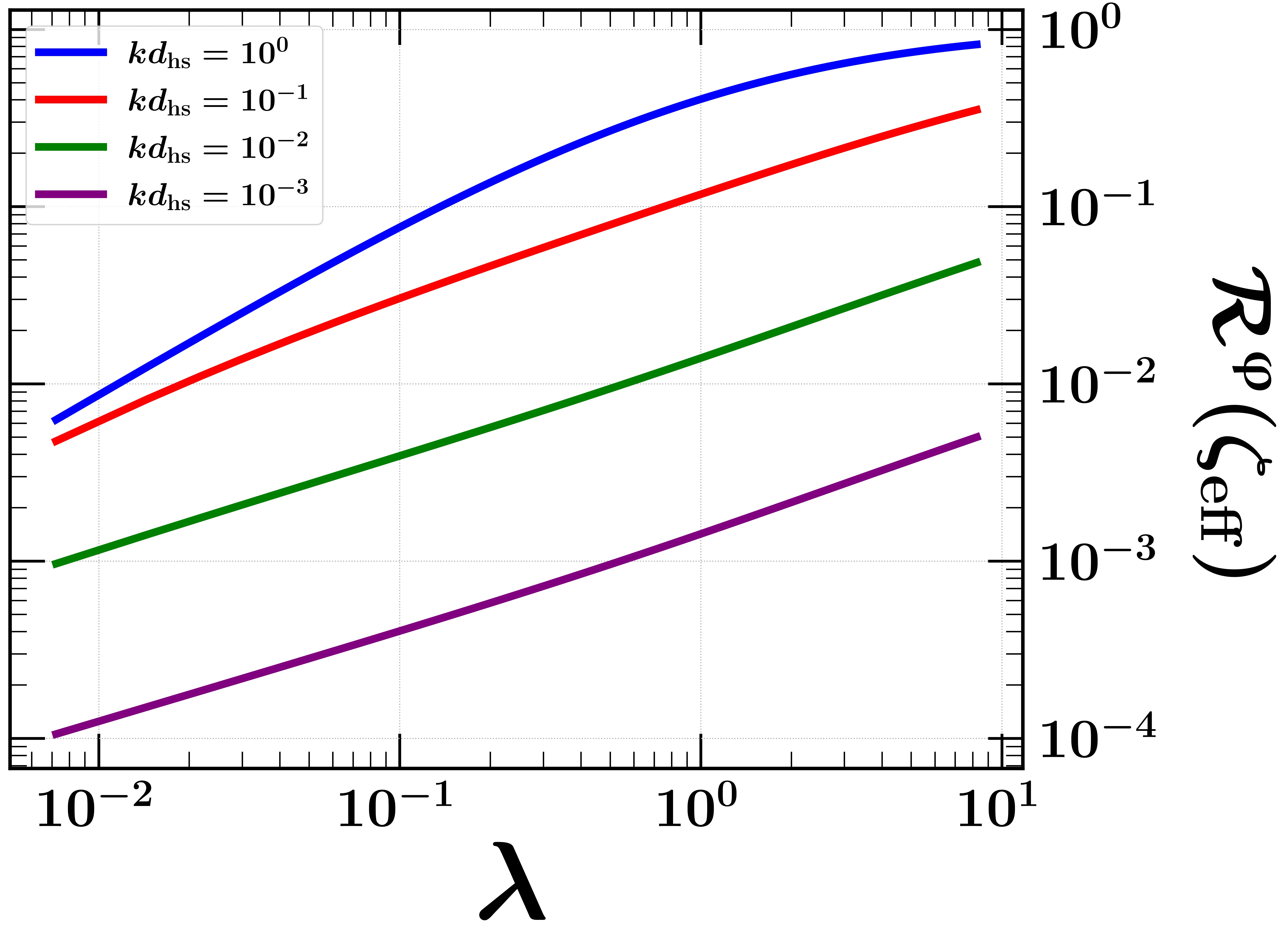}
    \caption{Relative difference of the effective drag as a function of $\lambda$ for four selected values of $k$.}
    \label{fig:pseudocomparezeta}
\end{figure}

\begin{figure}
    \centering
    \includegraphics[width = 0.75\textwidth]{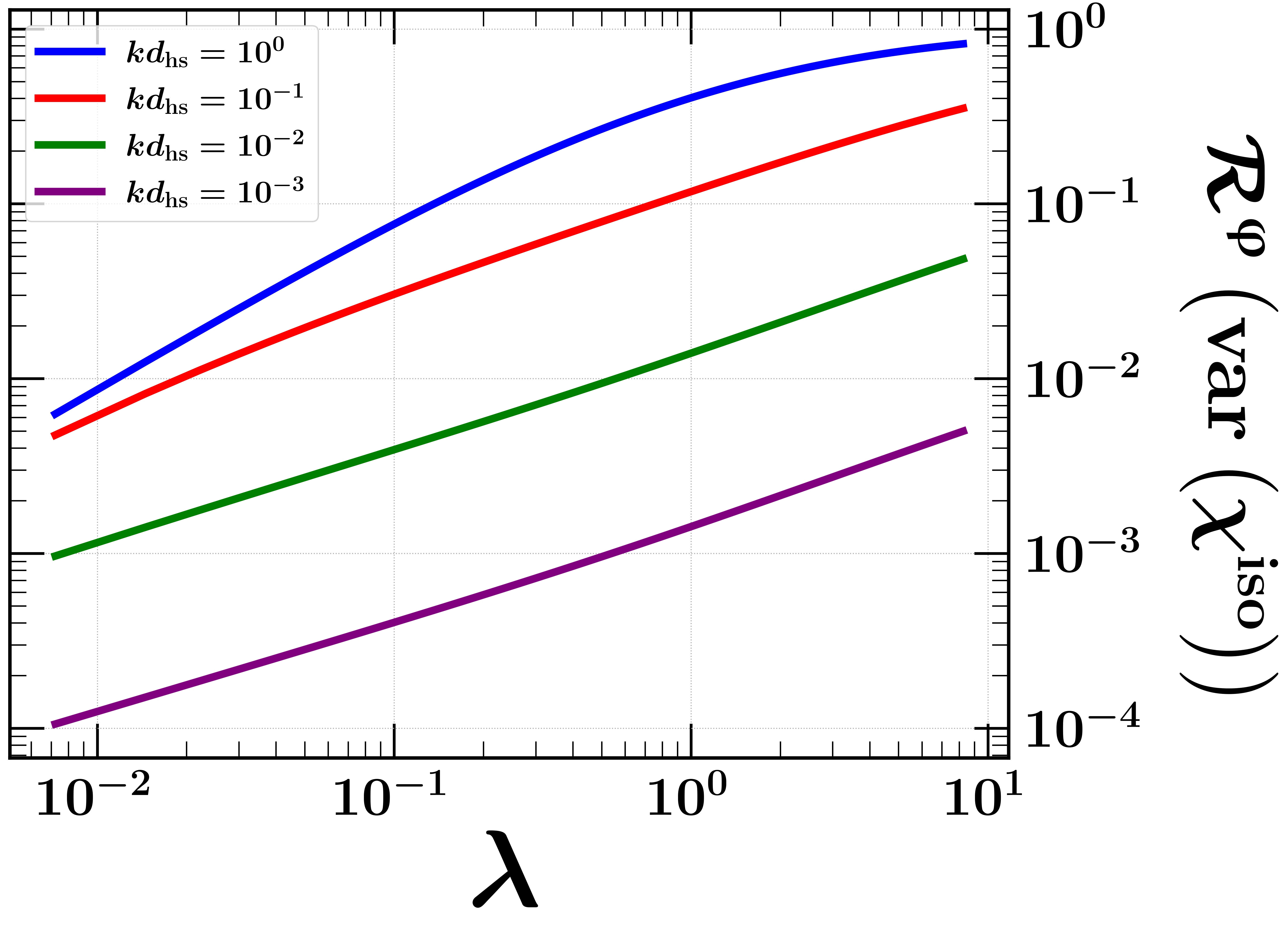}
    \caption{Relative difference of the isotropic noise variance as a function of $\lambda$ for four selected values of $k$.}
    \label{fig:pseudocompareiso}
\end{figure}

\begin{figure}
    \centering
    \includegraphics[width = 0.75\textwidth]{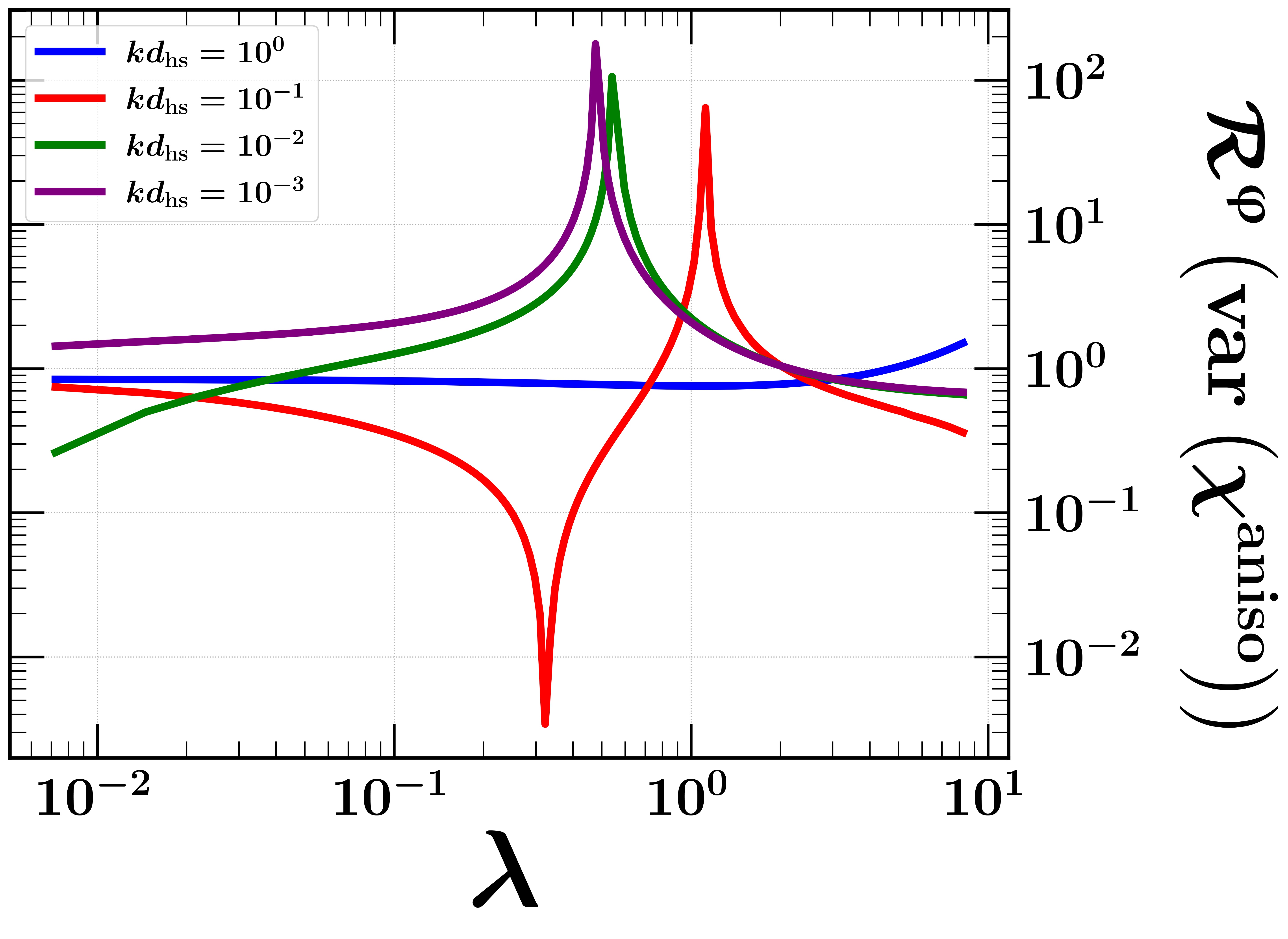}
    \caption{Relative difference of the anisotropic noise variance as a function of $\lambda$ for four selected values of $k$.}
    \label{fig:pseudocompareaniso}
\end{figure}

In the main text we have argued that $\upgamma_{\rm cw}$ includes the effect of nematic flows on the height field dynamics due, in part, to the use of the pseudovariable $\mathcal{E}$. 
Here we provide numerical analysis supporting this claim. 
Suppose Eq.~\eqref{seq:hevogf} was multiplied by $\upvarphi'$ (as Bray~\cite{Bray2002InterfaceShear} proposed in the context of passive systems) rather than $\partial\mathcal{E}/\partial z$ and integrated across all $u$.
Doing so would result in a final interface Langevin equation: 
\begin{align}
    \zeta^{\upvarphi}_{\rm{eff}} \frac{\partial h}{\partial t}  = -k^3\upgamma^{\upvarphi}_{\rm{cw}} h + k\nu \mathcal{F}\left[|\nabla_xh|^2\right] + \chi(k,t)^{\rm{iso,\upvarphi}} + \chi(k,t)^{\rm{aniso,\upvarphi}}.\ \label{eq:interfacelangevinp}  
\end{align}
We note that while this equation contains a nonlinearity, the nonlinearity is outside of the scope of Bray's ansatz: although this equation may capture a contribution to the nonlinear interfacial height evolution, there may be terms other than $\nu$ that would be found if a nonlinear ansatz was used.
The coefficients in the above equation with superscript $\upvarphi$ are defined differently than above, given by:
\begin{align}
    \zeta^{\upvarphi}_{\rm{eff}} = \frac{\zeta (A^{\upvarphi})^2(k)}{2\rho^{\rm surf}B^{\upvarphi}(k)},
\end{align}
\begin{align}
    \upgamma^{\upvarphi}_{\rm{cw}}(k) = \frac{2\zeta^{\upvarphi}_{\rm{eff}}}{
    \zeta A^{\upvarphi}(k)}\biggl[\int_{-\infty}^{\infty}\text{d}u (\upvarphi')^2  a(\upvarphi) + \int_{-\infty}^{\infty}\text{d}u \upvarphi'\int_{-\infty}^{\infty} \text{d}z'\text{sgn}(u-z')e^{-k|u-z'|}b(\upvarphi)(\upvarphi')^2 \biggr],\ \label{seq:capwavetensionp}
\end{align}
\begin{subequations}
    \begin{align}
        \langle \chi^{\rm{iso,\upvarphi}}(\mathbf{k},t)\chi^{\rm{iso,\upvarphi}}(\mathbf{k}',t')\rangle = &2k(k_BT)^{\rm{act}}\zeta^{\upvarphi}_{\rm{eff}}\delta(t-t')\delta(\mathbf{k}+\mathbf{k}')(2\pi)^{(d-1)}, \\ \langle \chi^{\rm{aniso,\upvarphi}}(\mathbf{k},t)\chi^{\rm{aniso,\upvarphi}}(\mathbf{k}',t')\rangle = &\frac{2\zeta^{\upvarphi}_{\rm{eff}}(C^{\upvarphi}(k) + D^{\upvarphi}(k))}{(d-1)2\rho^{\rm surf}B^{\upvarphi}(k)} \delta(t-t')\delta(\mathbf{k}+\mathbf{k}')(2\pi)^{(d-1)},
    \end{align}
\end{subequations}
\begin{align}
    A^{\upvarphi}(k) = B^{\upvarphi}(k) = \int \int \text{d}u\text{d}z e^{-k|u-z|}\upvarphi'(u)\upvarphi'(z),
\end{align}
\begin{align}
    C^{\upvarphi}(k) = &\int \int \text{d}u \text{d}u' \upvarphi'(u)\upvarphi'(u')\int\int \text{d}z' \text{d}z''\times  e^{-k|u-z'|}e^{-k|u'-z''|}\nonumber \\ & \times\biggl(k^2 \frac{a(\rho)}{ \overline{U}} \upvarphi''(z) \delta(z-z') - \frac{\partial}{\partial z}\left( \frac{a(\rho)}{ \overline{U}} \upvarphi''(z) \frac{\partial}{\partial z}\delta(z-z')\right)\biggr),
    \label{seq:ceekayyp}
\end{align}
and
\begin{align}
    D^{\upvarphi}(k) = &\int\int \text{d}u \text{d}u'\upvarphi'(u)\upvarphi'(u')\int\int \text{d}z'\text{d}z''\nonumber \\ &\times e^{-k|u-z'|}e^{-k|u'-z''|}  \frac{\partial}{\partial z'}\left[\frac{b(\rho)}{\overline{U}}(\upvarphi'(z'))^2\frac{\partial}{\partial z'}\delta(z-z')\right].
    \label{seq:deekayyp}
\end{align}
We numerically evaluate $\zeta^{\upvarphi}_{\rm{eff}}$, $\upgamma^{\upvarphi}_{\rm{cw}}$, and the compare them to the analagous quantities that are obtained using the pseudovariable.
We define this comparison as
\begin{equation*}
    \mathcal{R}^{\upvarphi}\left(f\right) = \left|\frac{f - f^{\upvarphi}}{f^{\upvarphi}}\right|,
\end{equation*}
where $f$ is an arbitrary quantity that we have calculated with and without the use of a pseudovariable.
These numerical results are plotted in Figs. \ref{fig:pseudocomparetension}-\ref{fig:pseudocompareaniso}.
The kinks in the plots correspond to the quantities in the absolute values switching from positive to negative. 
At low $k$ the numerics offer an attractive physical interpretation. 
In this limit the effective drag and isotropic noise variance defined by the pseudovariable become nearly identical to those defined by $\upvarphi'$.
As identified in Section~\ref{sec:isoerror}, the anisotropic noise variance is negligible in this limit, so the only difference in parameters for our interfacial Langevin equation are in the capillary-wave tension $\upgamma_{\rm{cw}}$ and driving coefficient $\nu$.
From Eq.~\eqref{seq:tracenemconst} one can identify the coefficients of the traceless nematic order to be proportional to the nonlinearity in Eq.~\eqref{seq:hevogf} (up to a 1d approximation), and because $\partial\mathcal{E}/\partial z$ is orthogonal to this nonlinearity, $\nu$ vanishes when the pseudovariable is used. 
In contrast, the magnitude of $\upgamma_{\rm{cw}}$ increases by up to three times (depending on activity) when the pseudovariable is used.
Because the only identified changes are in the disappearance of the nonlinearity proportional to nematic ordering and the increase of $\upgamma_{\rm{cw}}$, we interpret the effect of the pseudovariable as encoding the stabilizing effects of tangential flows into the capillary-wave tension.

\subsection{\label{sec:bddetails}Brownian Dynamics Simulation}
We take the interparticle force $\mathbf{F}_{ij}$ to result from a Weeks-Chandler-Anderson (WCA) potential~\cite{Weeks1971}, characterized by an an energy scale $\varepsilon$ and Lennard-Jones diameter $d_{\rm LJ}$.
Despite our use of a continuous potential, the finite and constant amplitude of the active force along with the use of a stiff WCA allow us to achieve hard-sphere statistics.  
A stiffness of $\mathcal{S}\equiv \varepsilon/\zeta U_o d_{\rm LJ} = 50$ was found to be sufficient to achieve effective hard-sphere interactions with a hard-sphere diameter of $d_{\rm{hs}} = 2^{1/6}d_{\rm LJ}$. 
In this hard-sphere limit, the system state is independent of the active force magnitude and is fully described by two parameters: the density and the intrinsic run length $\ell_0$. 
We will focus on the case of three-dimensional ABPs for the reasons outlined in the main text. 

Before choosing the specific details of our simulations, we consider the computational scaling. 
As demonstrated in Section~\ref{sec:capillarystats}, the \textit{number of timesteps} required to observe the relaxation of a capillary wave is derived to scale as $k^{-3}$. 
In addition, obtaining lower $k$ requires expansion of a simulation cell which, at constant density, requires additional particles. 
An isotropic expansion of the simulation cell will thus result in the scaling $1/k \sim N^d$. 
The computational cost of a single timestep scales linearly with $N$. 
Therefore, in order to relax capillary waves of increasingly small $k$, the computational cost of every time step scales as $(1/k)^{d}$ while the number of timesteps to obtain statistically significant data scales as $(1/k)^{3}$. 
As a result, for a square interface in 3d the total computational cost of relaxing a capillary-wave scales as $(1/k)^6$, placing intense resource limitations on sampling a low $k$ limit. 
In addition, because of the vanishing capillary-wave tension at low $\lambda$ found in Section~\ref{sec:capillarystats}, we expect simulations at lower activities to have significantly longer relaxation times and therefore higher statistical uncertainty for the same number of simulation time steps as higher activities.

Brownian dynamics simulations of the above system were then performed using $631444$ particles with intrinsic run lengths of $\ell_0/ d_{\rm{hs}} = \{40.09, 63.70, 89.09, 148.3\}$ for a length of $89000~d_{\rm{hs}}/U_o$ using \texttt{HOOMD-blue}~\cite{Anderson2020}. 
Rectangular simulations with dimensions of $L_z/d_{\rm{hs}} = 221.1$, $L_x/d_{\rm{hs}} = 196.7$, $L_y/d_{\rm{hs}} = 19.2$ were employed. 
This combination of cell dimensions and number of particles corresponds to an overall volume fraction of $\phi = 0.397$.

\begin{figure}
    \centering
    \includegraphics[width = 0.75\textwidth]{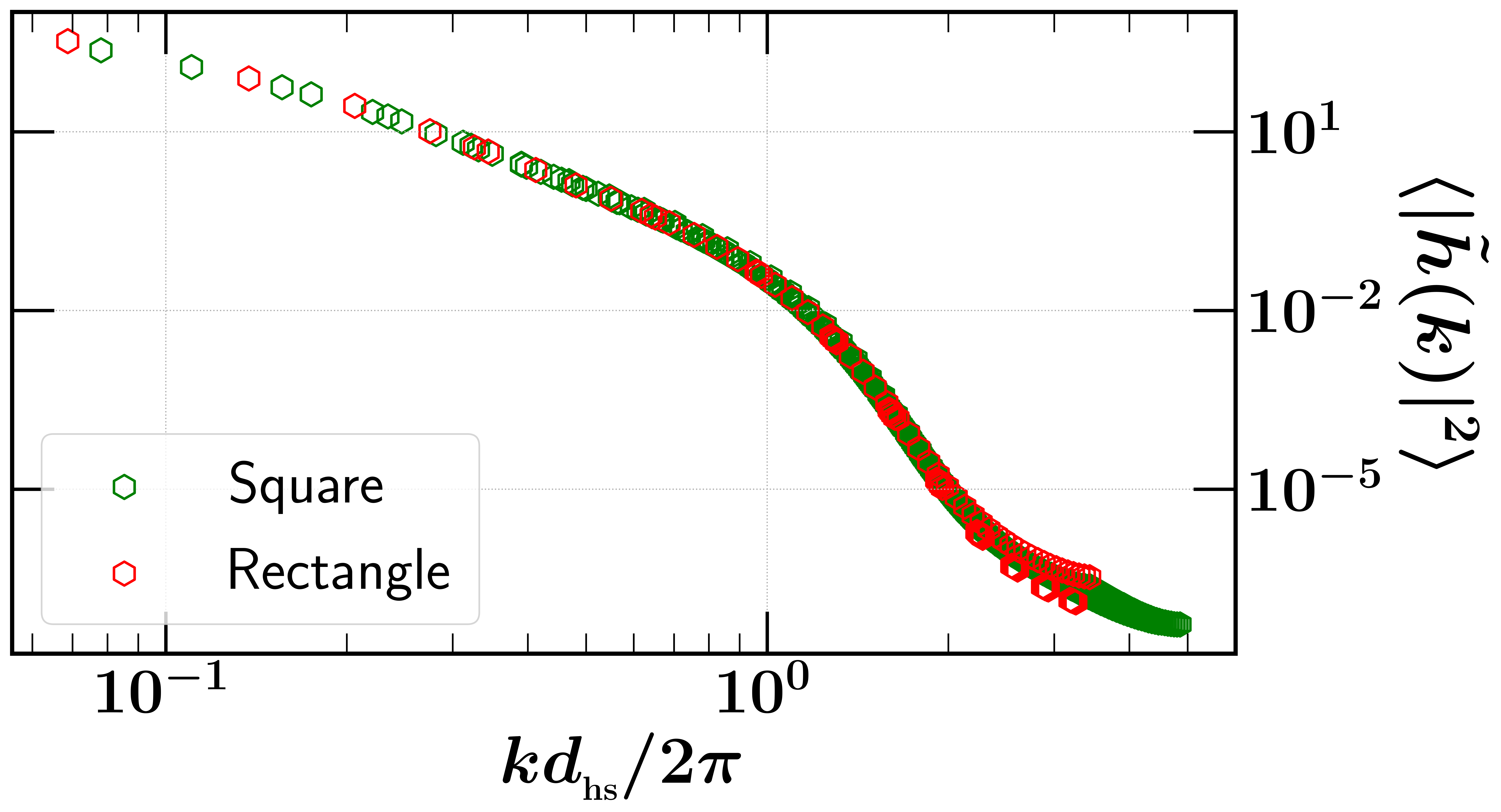}
    \caption{Interfacial fluctuation spectrum for square ($80.5\times 80.5~d_{\rm hs}^2$) and rectangular ($91.4\times 19.7~d_{\rm hs}^2$) interfaces. Simulations were conducted at an activity of $\ell_o/d_{\rm hs} = 89$.}
    \label{fig:dimensionscompare}
\end{figure}

Simulation cell dimensions were unequal in the directions tangential to the interface in order to maximize the length of one dimension (and therefore access lower wave vector fluctuations) without incurring additional computational expense.
For our 3d system, this allows the cost of each time step to scale as $(1/k)^2$ rather than $(1/k)^3$.
\emph{A priori}, it was unclear whether introducing unequal dimensions would introduce artefacts into the measured height fluctuation spectra. 
We therefore measured the height fluctuations of (smaller) systems with square and rectangular interfaces at an activity of $\ell_o/d_{\rm{hs}} = 89.09$. 
The dimensions of the square and rectangular interfaces were $L_x/d_{\rm{hs}} = L_y/d_{\rm{hs}} = 80.5$ and $L_x/d_{\rm{hs}}= 91.4,\quad L_y/d_{\rm{hs}} = 19.7$, respectively. 
Their height fluctuation spectra are plotted in Fig.~\ref{fig:dimensionscompare}. 
The fluctuation spectrum from the systems with equal and unequal dimensions are indistinguishable until wave vector far outside the scope of a low-$k$ theory, justifying the unequal simulation cell dimensions used in our production runs.

The instantaneous location of interfaces was identified using the algorithm proposed by Willard and Chandler~\cite{Willard2010}.
The coarse grained density field was defined by specifying the kernel $\Delta(\mathbf{r}-\mathbf{r}_i)$ to be a Gaussian,
\begin{align}
    \rho(\mathbf{r},t) = \left(2\pi\xi^2\right)^{-d/2}\sum_{i=1}^{N}\text{exp}\left[-\frac{|\mathbf{r}-\mathbf{r}_i(t)|}{2\xi^2}\right],
\end{align}
where $\xi$ is a coarse-graining length. 
The value of this field was calculated every $4.45~d_{\rm{hs}}/U_0$ with $\xi/d_{\rm{hs}} = 1.78$ on a cubic grid of points with spacing $0.89~d_{\rm{hs}}$.
The $(d-1)$ dimensional surface at which the density field was equal $\rho^{\rm surf}$ was then determined through linear interpolation.  

The statistical uncertainty of the height fluctuation spectra (as well as the exponent $\omega$, stiffness $K_s$, and later power spectra fits) was determined by calculating dividing the trajectory into five equally spaced periods in space, calculating the average fluctuation spectrum of each period independently, and taking the standard deviation between the periods.
\begin{figure}
    \centering
    \includegraphics[width = 0.75\textwidth]{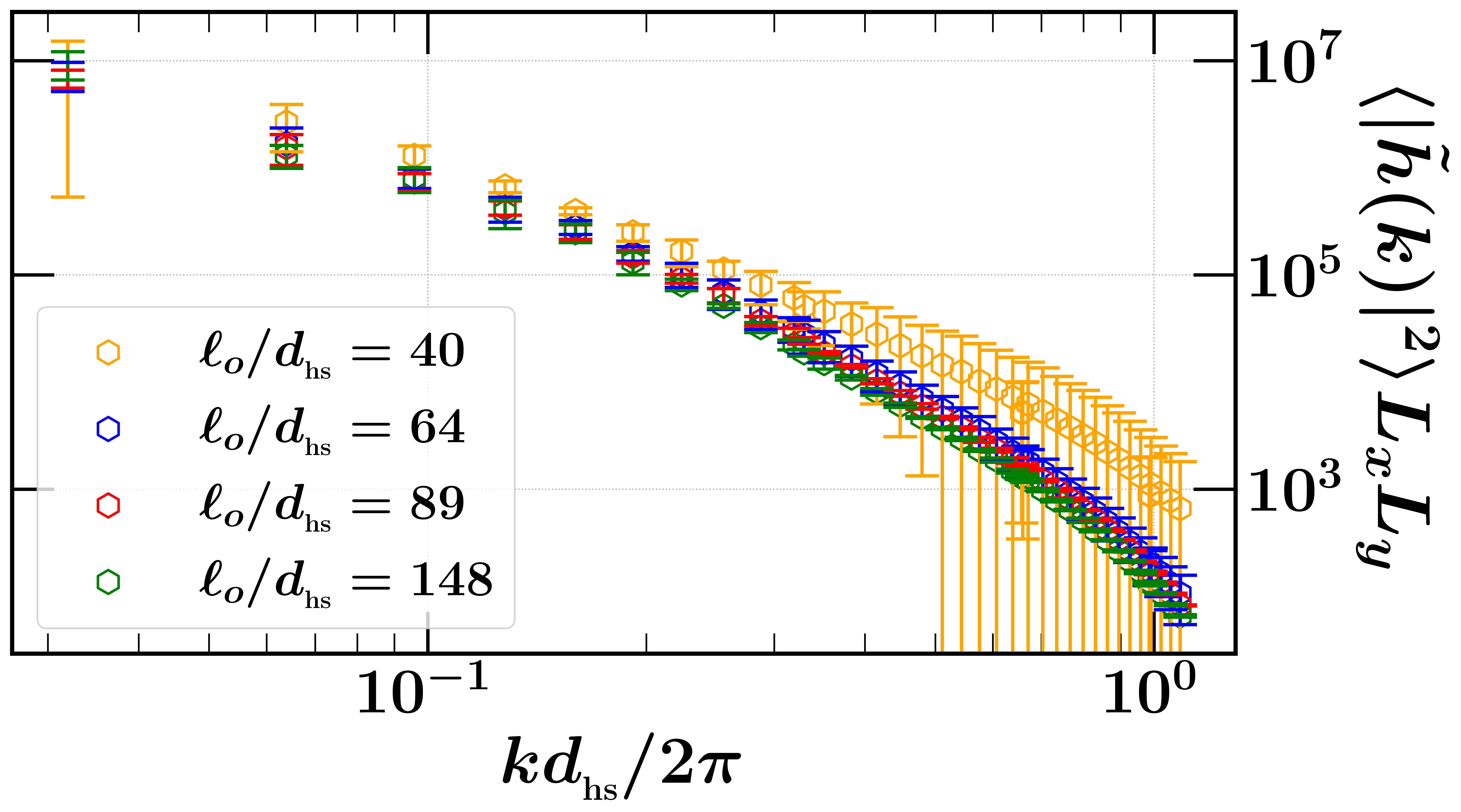}
    \caption{Interfacial fluctuation spectrum for production runs with error bars included.}
    \label{fig:flucterror}
\end{figure}
The spread on the data corresponding to $\ell_o/d_{\rm{hs}} = 40.09$ was significantly higher than the other data points. 
This is because the relaxation time of capillary excitations at this run length was up to an order of magnitude higher than the other activities simulated for any given wave vector, as shown by Eq.~\eqref{seq:relaxtime} and plotted in Fig.~\ref{fig:relaxation}.
We also attempt to collapse the height fluctuation spectra by multiplying each curve by $\ell_o$. 
This collapse was first observed by Patch \emph{et al.}~\cite{Patch2018} for 2d ABPs and a theoretical justification for this collapse is discussed in the main text. 
This curve collapse should, in principle, apply to 3d ABPs at sufficiently low $k$ but it is 
unclear whether our sampling reached low enough values of $k$ for Eq.~\eqref{seq:capfluctslowk} to hold.
In addition, the statistical uncertainty found in Fig.~\ref{fig:flucterror} may be concealing any potential curve collapse.
\begin{figure}
    \centering
    \includegraphics[width = 0.75\textwidth]{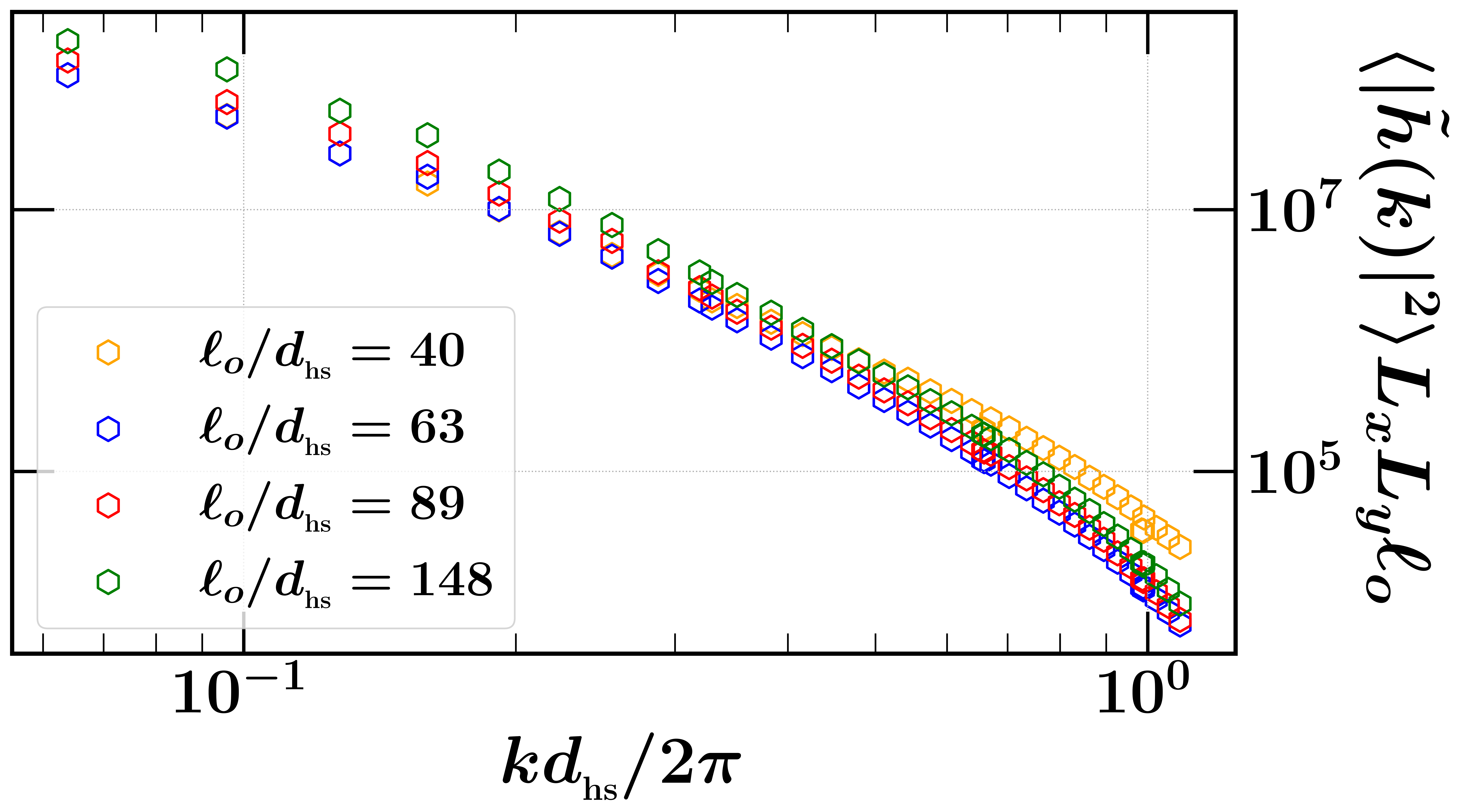}
    \caption{Interfacial fluctuation spectrum multiplied by $\ell_o$. At $k$ low enough to be described by Eq.~\eqref{seq:capfluctslowk}, all plots should collapse to a single point.}
    \label{fig:collapse}
\end{figure}

\begin{figure}
    \centering
    \includegraphics[width = 0.75\textwidth]{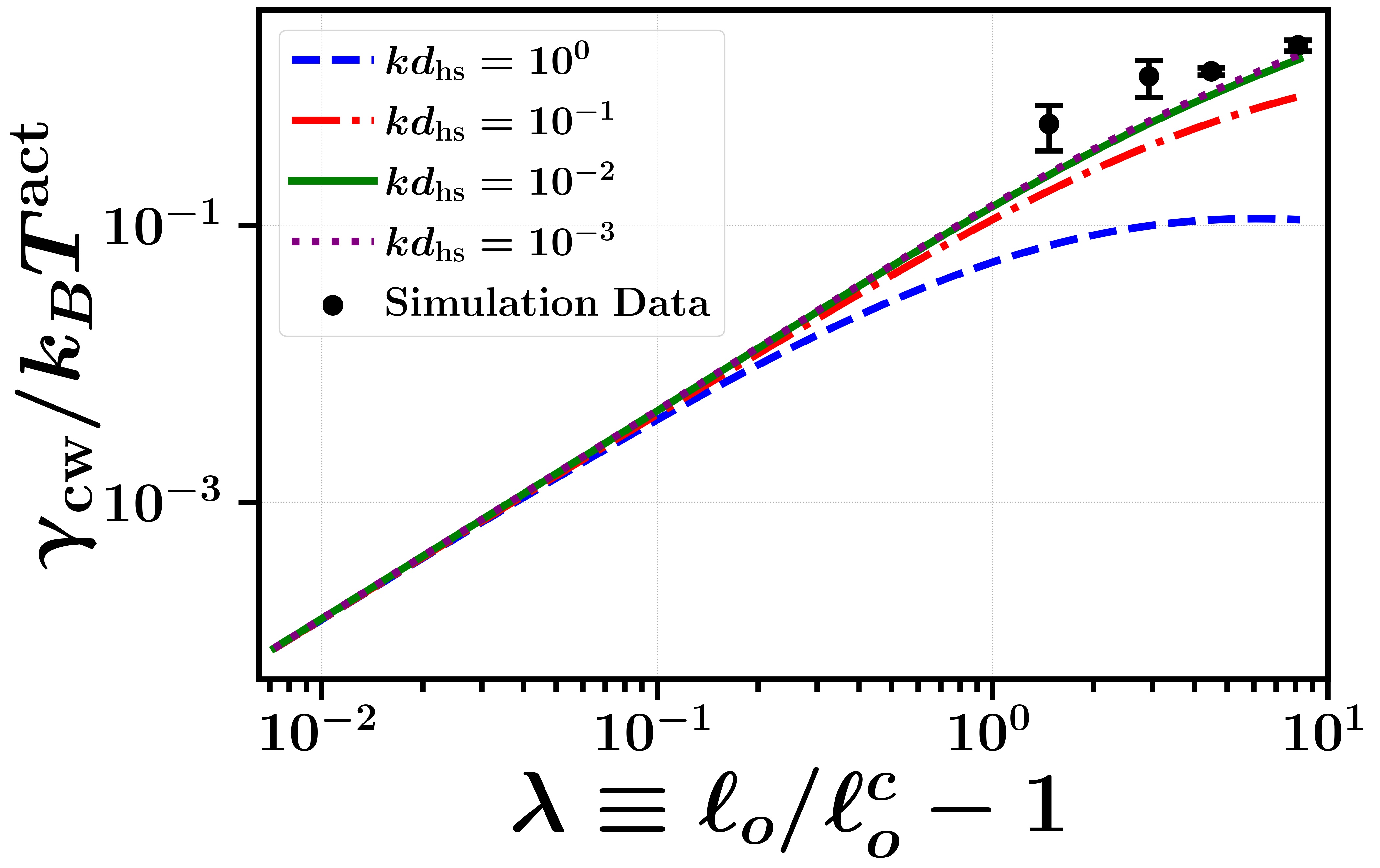}
    \caption{Interfacial stiffness $\upgamma_{\rm cw} / k_BT^{\rm act}$ as predicted by theory (lines) and as measured via simulation data (points). The stiffness is presented in units of $d_{\rm hs}^{-2}$. The simulation data points were fit from the low-$k$ ($kd_{\rm hs}/2\pi < 0.5$) data.}
    \label{fig:stiffnessdata}
\end{figure}

From a fit of the height spectrum data to $\langle |\tilde{h}(k)|^2\rangle = K_s k^{\omega}$ we obtain the interfacial stiffness $K_s^{-1}$. We find that the interfacial stiffness measured from simulation is within an order of magnitude agreement to the predictions of the theory, as plotted by Fig.~\ref{fig:stiffnessdata}.

\begin{figure}
    \centering
    \includegraphics[width = 0.75\textwidth]{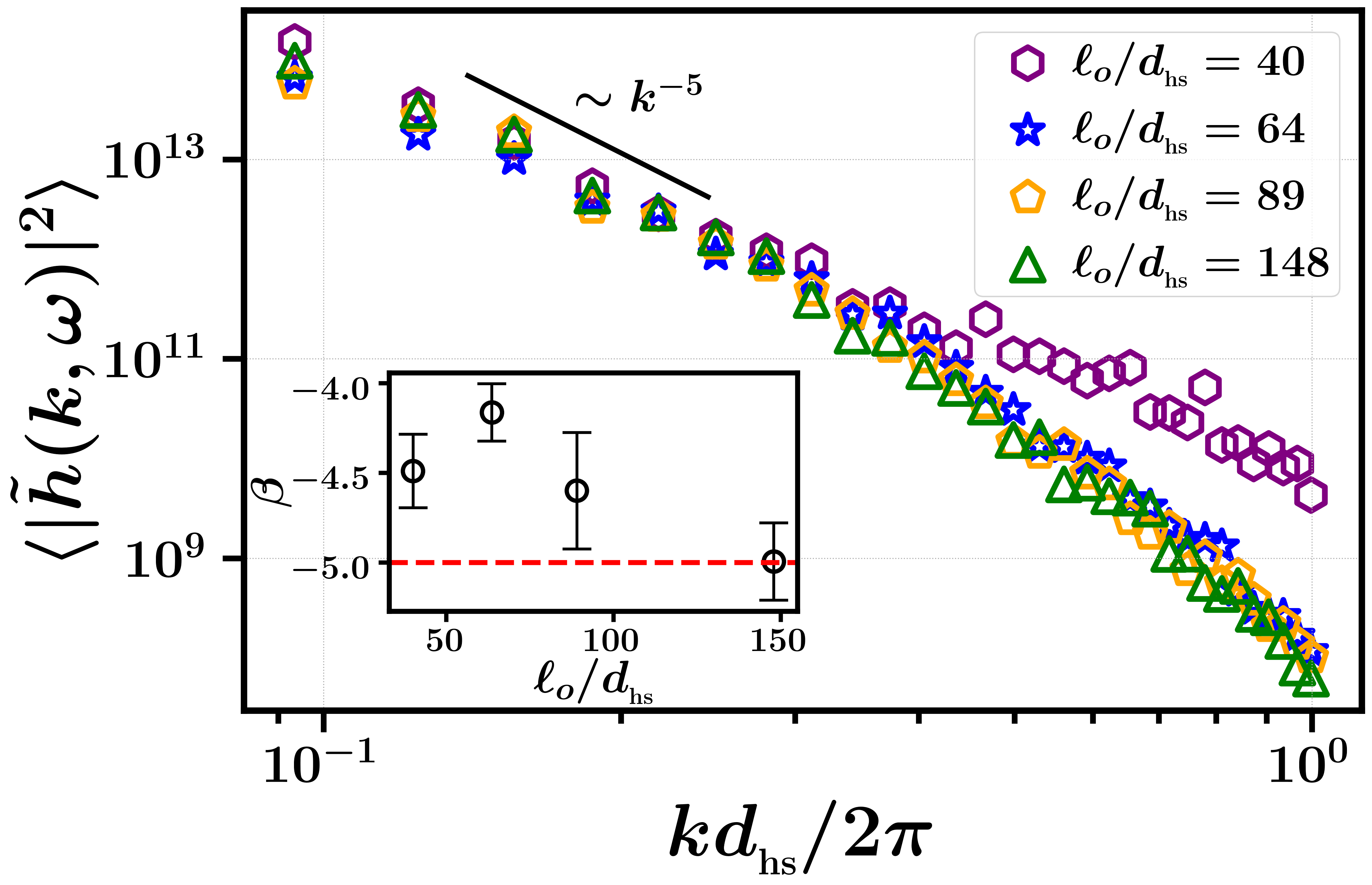}
    \caption{\protect\small{ Power spectra $\langle |h(k,\omega)|^2 \rangle$ at each simulated activity as a function of $k$ with $\omega d_{\rm hs}/ 2\pi U_o = 2.82\times 10^{-4}$ fixed. (Inset) Power law scaling obtained from logarithmic fits to $\langle |h(k,\omega)|^2 \rangle \sim k^{\beta}$, considering low $k$ such that $kd_{\rm hs}/2\pi < 0.5$. Lowest two values of $k$ not included in plot or fit. }}
    \label{fig:powerspec}
\end{figure}
The power spectra data with $\omega d_{\rm hs}/ 2\pi U_o = 2.82\times 10^{-4}$ fixed is plotted in Fig.~\ref{fig:powerspec}. 
The lowest twelve points of the power spectra was fit to $\langle |h(k,\omega)|^2 \rangle \sim k^{\beta}$ and values of $\beta$ are included in the inset of Fig.~\ref{fig:powerspec}.
This plot is interested in probing the $\omega \to 0$ limit as closely as possible, so the lowest two values of $k$ were not included in the plot or fit as $\tau(k)$ as predicted by Fig.~\ref{fig:relaxation} was on the order of $\omega^{-1}$
 \newpage
\section{\label{sec:media}Supplemental Media}
Each video listed below displays the instantaneous interface~\cite{Willard2010} determined from a Brownian dynamics simulation of ABP phase separation at a specified activity and simulation geometry. 
In each video, the instantaneous interface was computed using a cubic grid of points with a uniform spacing of $0.89~d_{\rm_{hs}}$ and a coarse-graining length of $\xi/d_{\rm{hs}} = 1.78$. 
Each frame is separated by a duration of $4.45~d_{\rm hs}/U_o$ and the videos are played at a rate of $25$ frames per second.
All videos are publicly available at the following URL:
\newline
\url{https://berkeley.box.com/s/09q5ccbbhf1h14d93gbsw70qjcdv4ufe}
\newline
\begin{itemize}
    \item \text{[}interfacedynamics\_40square.mp4\text{]}:\newline 
    Instantaneous interface dynamics with an activity of $40~\ell_o/d_{\rm hs}$. 
    The simulation box has dimensions $L_x/d_{\rm{hs}} = L_y/d_{\rm{hs}} = 80.5,\quad L_z/d_{\rm{hs}} = 144.0$. 
    \item \text{[}interfacedynamics\_89square.mp4\text{]}:\newline 
    Instantaneous interface dynamics with an activity of $89~\ell_o/d_{\rm hs}$. 
    The simulation box has dimensions $L_x/d_{\rm{hs}} = L_y/d_{\rm{hs}} = 80.5,\quad L_z/d_{\rm{hs}} = 144.0$. 
    \item \text{[}interfacedynamics\_40rectangle.mp4\text{]}:\newline 
    Instantaneous interface dynamics with an activity of $40~\ell_o/d_{\rm hs}$. 
    The simulation box has dimensions $L_x/d_{\rm{hs}} = 196.7,\quad L_y/d_{\rm{hs}} = 19.2,\quad L_z/d_{\rm{hs}} = 221.1$. 
    \item \text{[}interfacedynamics\_89rectangle.mp4\text{]}:\newline 
    Instantaneous interface dynamics with an activity of $89~\ell_o/d_{\rm hs}$. 
    The simulation box has dimensions $L_x/d_{\rm{hs}} = 196.7,\quad L_y/d_{\rm{hs}} = 19.2,\quad L_z/d_{\rm{hs}} = 221.1$. 
    \item \text{[}interfacedynamics\_148rectangle.mp4\text{]}:\newline 
    Instantaneous interface dynamics with an activity of $148~\ell_o/d_{\rm hs}$. 
    The simulation box has dimensions $L_x/d_{\rm{hs}} = 196.7,\quad L_y/d_{\rm{hs}} = 19.2,\quad L_z/d_{\rm{hs}} = 221.1$. 
\end{itemize}